\documentclass{preprint}

\usepackage{times}
\usepackage{booktabs}
\usepackage{enumitem}
\usepackage{tikz}
\usetikzlibrary{arrows.meta,positioning,calc}

\newcommand{\E}{\mathbb{E}}
\newcommand{\eu}{\mathrm{e}}
\newcommand{\upd}[3]{#1[#2\mapsto #3]}
\newcommand{\tm}{t_m}
\newcommand{\Dist}{D}
\newcommand{\Pk}{\Pi}
\newcommand{\cV}{\mathcal{V}}
\newcommand{\cZ}{\mathcal{Z}}
\newcommand{\cN}{\mathcal{N}}
\newcommand{\cU}{\mathcal{U}}
\newcommand{\cE}{\mathcal{E}}
\newcommand{\one}{\mathbf{1}}
\newcommand{\Id}{\mathrm{Id}}
\newcommand{\sdots}{\cdots}
\newcommand{\ren}{\mathrm{ren}}
\newcommand{\occ}{\mathrm{occ}}
\newcommand{\ov}{\mathrm{ov}}
\newtheorem{defn}{Definition}
\makeatletter\renewcommand*{\@titlefoot}{}\makeatother
\renewenvironment{proof}[1][\proofname]{\par\noindent{\bfseries\upshape #1\ }\ignorespaces}{\jmlrQED}

\title[How Far Must a Test Explore?]{How Far Must a Test Explore? Certifying Composable Read--Write Interfaces from Their Own Random Edits}

\author{%
 \Name{Jiaxuan Guo} \Email{guojx@stanford.edu}\\
 \addr Stanford University
 \AND
 \Name{Xingyu Ren}\\
 \addr The Chinese University of Hong Kong
 \AND
 \Name{Jingxin Yang}\\
 \addr Stanford University
 \AND
 \Name{Youran Sun}\\
 \Name{Haizhao Yang}\\
 \addr University of Maryland, College Park
}

\makeatletter\renewcommand\ps@jmlrtps{\let\@mkboth\@gobbletwo\let\@oddhead\@empty\let\@evenhead\@empty\def\@oddfoot{\hfill\small\rmfamily\thepage\hfill}\let\@evenfoot\@oddfoot}\makeatother
\hypersetup{pdftitle={How Far Must a Test Explore? Certifying Composable Read-Write Interfaces from Their Own Random Edits},pdfauthor={Jiaxuan Guo, Xingyu Ren, Jingxin Yang, Youran Sun, Haizhao Yang}}

\begin{document}

\maketitle

\begin{abstract}%
Random tests of stateful code check laws, like reading back a written value, at states that random writes reach. The laws hold at every state exactly when the code behaves like a record with hidden data, so a pass is taken as evidence. This can fail: the laws can hold at every sample while the code is far from any record. The laws are local, but behaving like a record is global. Self-testing and stability theorems bridge this gap when the query distribution is controlled by the tester or preserved by the tested maps, but here it is generated by the tested writes, which need not be invertible.

We ask how far a test must explore. An interface has $m$ binary variables, each with a read and two writes; in a record, reads return their bits and writes set only their bits. Samples are the ends of geometric runs of mean $\ell$ of uniformly random writes from a start distribution; a tester adds its own write chains of depth $d$. The distance is the least, over records with hidden data, of the largest probability that a read or write differs from it at a sample.

From one start state, law checks at $O(m^2\log m/\epsilon)$ samples certify, at fixed confidence, any threshold $\epsilon$ after geometric runs of every positive mean, while runs of fixed positive length can pass at distance $1/2$. For the worst start distribution, take $\epsilon=m^{-\beta}$ with fixed $0<\beta\le1$. For $d\le cm$ with fixed $c<1/2$, a polynomial budget of samples and writes exists exactly when $\ell=\Omega(m)$, where law checks suffice. With no runs, the least depth for this budget is $(1/2+o(1))m$. Counting samples alone, in the worst case at $\epsilon=1/(32m)$ and $d\le m/16$, the critical mean for a polynomial budget is $\Theta(m\log(m/d)/\log m)$, by exhaustive search. So one start state needs only law checks, and the worst case at these thresholds needs exploration linear in $m$ if writes are counted.%
\end{abstract}


\section{Introduction}\label{sec:intro}

Random tests check stateful code in the way \citet{claessen2002testing} check queues: they run random sequences of operations from a start state and check laws or a model at the reached states. For $m$ binary variables, each with a read and two writes, the laws say that a written value is read back (\emph{success}), that writing back the current value changes nothing (\emph{write-back}), that of two writes to one variable the second wins (\emph{overwrite}), that a write to one variable keeps the others (\emph{preservation}), and that writes to different variables commute (\emph{commutation}) \citep{foster2016unifying}. For one variable, success, write-back and overwrite are the PutGet, GetPut and PutPut laws of a very well-behaved total lens \citep{foster2007combinators}, which the Haskell lens library checks with QuickCheck \citep{lens536properties}.

The laws hold at every state exactly when the code behaves like a record of $m$ bits with \emph{hidden data}, which no read reveals and no write changes (Proposition~\ref{prop:exact}), so a pass is taken as evidence of this behavior. A one-bit record may have the states $A0$, $A1$, $B0$ and $B1$: reads and writes act on the bit, and the letter is hidden data. But a pass checks only the reached states: each law is local, checked at one state with at most two writes in a row, while behaving like a record is a global property of how all states fit together.

Self-testing and stability theorems justify this interpretation in other settings. There, rare failures of local laws imply closeness to an exact solution of the laws \citep{blum1993self,rubinfeld1996robust,becker2021abelian,becker2023stability}. Their analyses control the distribution at which the laws are checked. The homomorphism test of \citet{blum1993self} checks $f(xy)=f(x)f(y)$ at a uniformly random pair of group elements $x,y$, so each of $x$, $y$ and $xy$ is uniform. In stability theorems, the tested permutations preserve the counting measure.

In random tests of stateful software, even in a record, setting a variable to $0$ maps two states that differ only in this variable to the same state, so the image of a sample under a write need not be distributed like a sample. The writes under test also generate the sample distribution.

We ask how far beyond its samples a test must explore before a pass certifies that an interface, an implementation of these reads and writes, is close to a record, and we call such a record a \emph{composable} interface \citep[as in database view updates,][]{bancilhon1981update}. A sample is the end of a \emph{run}: a geometric number of uniformly random writes from a state drawn from a \emph{start distribution}. We measure exploration by the mean $\ell$ of the runs and the \emph{depth} $d$ of the tester's own chains of writes from each sample, and closeness by the distance of Section~\ref{sec:model}.

\begin{figure}[!h]
\centering
\begin{tikzpicture}[font=\scriptsize,
  lab/.style={inner sep=1.5pt,align=center},
  ax/.style={-{Stealth[length=5pt]}},
  title/.style={font=\small,anchor=south},
  cert/.style={black!6},
  tick/.style={inner sep=2pt}]
\begin{scope}
\node[title] at (2.0,3.3) {one start state};
\fill[cert] (0,0) rectangle (4.0,2.95);
\draw[ax] (0,0) -- (4.2,0) node[right]{$d$};
\draw[ax] (0,0) -- (0,3.15) node[above]{$\ell$};
\node[lab] at (2.0,1.5) {law checks certify\\at every $\ell>0$ (Thm.~\ref*{thm:onestate})};
\end{scope}
\begin{scope}[xshift=5.6cm]
\node[title] at (4.3,3.3) {worst start distribution, $\epsilon=m^{-\beta}$};
\fill[cert] (0,2.35) rectangle (8.6,2.95);
\fill[cert] (5.5,0) rectangle (8.6,2.95);
\fill[black!20] (0,0) rectangle (3.9,2.35);
\draw (0,2.35) -- (8.6,2.35);
\draw (5.5,0) -- (5.5,2.95);
\draw[ax] (0,0) -- (8.8,0) node[right]{$d$};
\draw[ax] (0,0) -- (0,3.15) node[above]{$\ell$};
\draw[densely dashed,thick] (0.06,2.3) .. controls (0.15,1.35) and (0.5,0.8) .. (1.25,0.62);
\node[lab,anchor=west] at (1.27,0.62) {$\ell^*_m$};
\draw[densely dotted] (1.25,0.62) -- (1.25,0);
\node[lab] at (2.35,1.5) {$\ell=o(m)$, $d\le cm$:\\no polynomial\\budget (Thm.~\ref*{thm:budget})};
\node[lab] at (2.75,2.65) {$\ell\ge\alpha m$: law checks certify (Thm.~\ref*{thm:budget})};
\node[lab] at (7.05,1.18) {audit certifies\\(Thm.~\ref*{thm:audit})};
\draw (1.25,0.06) -- (1.25,-0.06) node[below,tick]{$\frac m{16}$};
\draw (3.9,0.06) -- (3.9,-0.06) node[below,tick]{$cm$};
\draw (5.5,0.06) -- (5.5,-0.06) node[below,tick]{$\frac m2+O(\sqrt{m\log m})$};
\draw (0.06,2.35) -- (-0.06,2.35) node[left,tick]{$\Theta(m)$};
\end{scope}
\end{tikzpicture}
\caption{The depth $d\ge2$ and the run mean $\ell$, schematic and not to scale, for fixed $0<\beta\le1$, $\alpha>0$ and $1/16<c<1/2$; Theorem~\ref*{thm:budget} holds for every $0<c<1/2$. Budgets count samples and writes. The dashed curve $\ell^*_m=m\log(m/d)/\log m$ counts samples alone, at $\epsilon=1/(32m)$: polynomially many samples do not suffice below its order and suffice at its order (Thm.~\ref*{thm:samplesonly}). The boundary for $d-m/2=o(m)$ and $0<\ell=o(m)$ is open.}\label{fig:phase}
\end{figure}
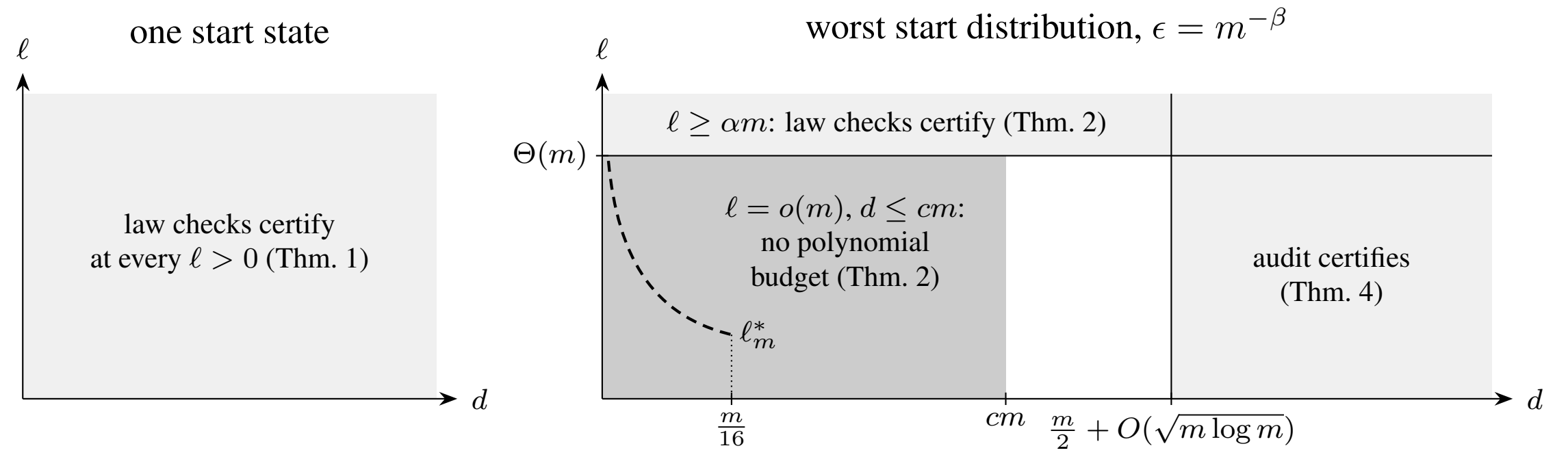

From one start state, law checks at the ends of geometric runs certify closeness at every positive mean, while for $m\ge2$ runs of a fixed positive length can pass at distance $1/2$ (Theorem~\ref{thm:onestate}). A geometric run can stop after any of its writes, so the sample distribution puts weight on every prefix of a run.

For the worst start distribution at polynomially small thresholds, exploration linear in $m$ is necessary. At each threshold $\epsilon=m^{-\beta}$ with fixed $0<\beta\le1$, a polynomial budget of samples and writes forces $\ell=\Omega(m)$ as long as $d\le cm$ for a fixed $c<1/2$ (Theorem~\ref{thm:budget}). Either kind of exploration suffices. Law checks after runs of linear mean meet the budget (Theorem~\ref{thm:budget}). So does an \emph{audit} with no runs, which walks from each sample to a target label and checks that flips rejoin this walk, at depth about $m/2$, the least depth at these thresholds (Theorem~\ref{thm:audit}).

At $\epsilon=1/(32m)$ and $d\le m/16$, counting samples alone lowers the necessary mean by a factor of at most $\log m$. A tester that exhausts the ball of radius $d$ around each sample needs only $\ell=\Theta(m\log(m/d)/\log m)$, at a cost in writes exponential in $d$ (Theorem~\ref{thm:samplesonly}). So the two budgets separate: at $d=\lfloor m/16\rfloor$ and $\ell=m/\log m$, polynomially many samples certify at every threshold $m^{-\beta}$, and polynomially many samples and writes do not. Figure~\ref{fig:phase} draws these regions.

\subsection{Related work}\label{sec:related}

\paragraph{Self-testing and stability.} In a composable interface, the flips, which write the opposite value of one variable, commute with each other, each flip undoes itself, and flipping a nonempty set of variables changes every state (Proposition~\ref{prop:exact}). So testing composability resembles testing stability for the group $\{0,1\}^m$ \citep{chapman2025stability}. \citet{chapman2026efficiently} ask how far unitary matrices $U_1,\dots,U_m$ that nearly satisfy the relations of the flips, $U_i^2=I$ and $U_iU_j=U_jU_i$, are from unitary matrices $V_1,\dots,V_m$ of the same size that satisfy them exactly. For $n\times n$ matrices, they take as squared distance the sum of the squared moduli of the entries of the difference, divided by $n$. Suppose that, for a relation drawn as an equal mixture of a uniform $U_i^2=I$ and a uniform $U_iU_j=U_jU_i$ with $i\ne j$, its two sides are at mean squared distance at most $\eta$. Then some exact solution has $U_i$ and $V_i$ at mean squared distance $O(\min\{1,m\eta\})$ for a uniform $i$, and this order is best possible. Their lower bound uses permutations that flip an extra bit conditionally on another bit, close to the twisted interface of Section~\ref{sec:calls}. Our setting differs in two ways. A flip of a general interface need not be a bijection, and the writes under test generate the distribution under which failures are counted.

\paragraph{Testing from samples.} Distribution-free testers measure the distance under an unknown distribution from which they sample, and they may also query the tested function at any point \citep{halevy2003distribution}. In the vertex-distribution-free model of \citet{goldreich2019testing}, testers may be assumed, at a constant-factor cost, to query only vertices that they have sampled or received as answers. Our testers obey the same restriction for states. In the huge-object model of \citet{goldreich2023testing}, a tester draws samples of long strings and probes each at locations of its choice, and the tester with the fewest samples need not be the one with the fewest queries. Our two budgets are these two measures for probes that are writes, and Section~\ref{sec:samples} separates them at the thresholds $m^{-\beta}$ for the worst start distribution. Local exploration from randomly selected vertices is a standard method in bounded-degree graph testing \citep{goldreich2002bounded,bender2002directed}, and our depth is such an exploration radius, bounded as a function of $m$ alone. Identity testers that decide from a single trajectory whether a symmetric Markov chain equals a known reference chain also face states that the trajectory rarely visits \citep{daskalakis2018testing,cherapanamjeri2019testing}; here the chain is generated by the interface under test.

\paragraph{Conformance testing.} Conformance testing of finite-state machines decides from input and output sequences whether a black box behaves like a given specification machine, and the length of the sequences is a central quantity, bounded in terms of the number of states \citep{vasilevskii1973failure,chow1978testing,lee1996principles}. Here the laws are the only specification, and exploration is measured against the number of variables, not of states.
\section{Interfaces, laws, runs and distances}\label{sec:model}

\paragraph{Interfaces and laws.} An \emph{interface} $I=(r,T)$ supplies readers and setters on a finite set $S$ of states, for $m\ge1$ binary variables. A \emph{reader} $r_i\colon S\to\{0,1\}$ reports the value of variable $i$; a \emph{setter} $T_{i,a}\colon S\to S$ is the operation intended to write $a\in\{0,1\}$ to it. The \emph{label} $r(x)=(r_i(x))_{i\in[m]}$ collects all readings at $x$. The \emph{flip} of variable $i$ is $F_i(x)=T_{i,1-r_i(x)}x$, the write of the opposite value.

A \emph{composable} interface $\tilde I$ acts like a record with hidden data. Formally, a bijection $\phi\colon S\to\{0,1\}^m\times\Xi$ gives $\tilde r_i(x)=u_i$ and $\phi(\tilde T_{i,a}x)=(\upd{u}{i}{a},z)$ when $\phi(x)=(u,z)$. Here $\upd{u}{i}{a}$ replaces coordinate $i$ of $u$ by $a$. Fixing $z$ gives a \emph{cube}, the $2^m$ states $\phi^{-1}(\{0,1\}^m\times\{z\})$, one of each label, such as $\{A0,A1\}$ in the one-bit record of Section~\ref{sec:intro}. A composable interface on $S$ exists exactly when $2^m$ divides $|S|$, with $M=|S|/2^m$ cubes.

Fixing the variables and values in a law gives a \emph{law instance}. The following laws hold at every state $x$ of a composable interface, for all $i\ne j$ and $a,b$, with maps composed from right to left.
\begin{itemize}[nosep,leftmargin=1.5em]
\item \emph{success}, $r_i(T_{i,a}x)=a$; \emph{write-back}, $T_{i,r_i(x)}x=x$; \emph{overwrite}, $T_{i,a}T_{i,b}x=T_{i,a}x$;
\item \emph{preservation}, $r_j(T_{i,a}x)=r_j(x)$; \emph{commutation}, $T_{i,a}T_{j,b}x=T_{j,b}T_{i,a}x$.
\end{itemize}
Conversely, satisfying all five laws everywhere makes an interface composable (Proposition~\ref{prop:exact}; \citealp{power2004comodels}).

\paragraph{Runs and samples.} A \emph{random edit} applies a uniformly random setter. It sends a distribution $\mu$ to $\mu P$, where $P=(2m)^{-1}\sum_{i,a}T_{i,a}$. The \emph{start distribution} $\nu$ specifies where runs start: a \emph{run} draws its \emph{start state} from $\nu$, makes $L$ independent random edits, and returns its final state as a sample. For geometric run lengths of mean $\ell>0$, put $q=\ell/(1+\ell)$ and $\Pr(L=s)=(1-q)q^s$, $s\ge0$. The \emph{sample distribution} is
\begin{equation}\label{eq:access}
\mu_\ell=(1-q)\sum\nolimits_{s\ge0}q^s\,\nu P^s ,
\end{equation}
and $\mu_0=\nu$. For measures, $\mu\le\mu'$ means $\mu(X)\le\mu'(X)$ for every $X\subseteq S$. The measure $\mu_\ell$ satisfies $\mu_\ell=(1-q)\nu+q\,\mu_\ell P$, hence $\mu_\ell P\le(1+1/\ell)\mu_\ell$. More generally, $\mu$ is \emph{$\lambda$-subinvariant} if $\mu P\le\lambda\mu$, for $\lambda\ge1$: one random edit increases the probability of any set by at most a factor $\lambda$.

\paragraph{Violation rate and distances.} Fix a probability measure $\mu$ on $S$. Let $\cN$ be the set of states at which some law instance fails. The \emph{violation rate} is $\gamma_\mu(I)=\mu(\cN)$, the probability that checking all law instances at one sample rejects. The \emph{disagreement} of a candidate $\tilde I$ with $I$ is the largest probability that a reader or setter differs,
\begin{equation}\label{eq:distance}
R_\mu(I,\tilde I)=\max_{i,a}\max\big\{\mu\{r_i\ne\tilde r_i\},\,\mu\{T_{i,a}\ne\tilde T_{i,a}\}\big\}.
\end{equation}
The \emph{distance} $\Dist_\mu(I)$ is the least disagreement over composable interfaces on $S$, which exist when $2^m$ divides $|S|$. The distance $\Dist^\infty_\mu(I)$ is the infimum of the disagreement over composable interfaces on finite sets $S'\supseteq S$, measured under $\mu$ on $S$ with an output outside $S$ counted as a disagreement, so $\Dist^\infty_\mu\le\Dist_\mu$. The theorems below hold for both distances: an upper bound speaks of $\Dist_\mu$ only when $2^m$ divides $|S|$, and every lower bound holds for both. For $\Dist_\mu$, the audit of Section~\ref{sec:calls} adds a \emph{bridge check}: it draws a uniform label and, when the depth tests allow, attempts to reach a state with that label and checks, through intermediate states, that the audit's walk sends this state to the same state as the sample (Appendix~\ref{app:auditbridge}). The distance on $S$ needs this check because a composable interface on $S$ itself cannot add states to supply missing labels. Subscripts $\ell$ mean evaluation at $\mu_\ell$: we write $\gamma_\ell$, $\Dist_\ell$ and $\Dist^\infty_\ell$. All logarithms are natural.

\begin{defn}[Testers]\label{def:tester}
\begin{enumerate}[label=(\roman*),nosep,leftmargin=*,beginpenalty=10000]
\item \emph{Access.} A \emph{tester} holds states only as opaque handles: the samples of $\mu$ that it draws and the outputs of its \emph{setter calls}, also called writes. On any held state it may call readers and setters, and it may compare two held states for equality. Its \emph{transcript} lists its draws, calls and comparisons, with the reader outputs and comparison outcomes, and numbers each handle by its first appearance.
\item \emph{Depth.} A held state has \emph{depth} $j$ if the shortest chain of the tester's setter calls that leads from a sample to this state has $j$ calls. A setter call has depth one more than its argument, determined when the call is made. A tester has depth $d$ on an input if all its setter calls on this input have depth at most $d$.
\item \emph{Law checks.} A \emph{law-check tester} evaluates law instances at samples and rejects only when one of them fails. It may decide how many samples to draw and which instances to evaluate from the outcomes so far. It calls readers and setters only on samples and on outputs of the setter calls of these evaluations, so its depth is $2$.
\item \emph{One-sided error.} A tester has \emph{one-sided error} if it accepts every composable interface with probability~$1$.
\end{enumerate}
\end{defn}

\noindent Depth measures how far calls reach from a sample, not how many calls there are. A call has depth one more than its argument even if it returns a state that the tester already holds, such as its argument. In a composable interface, writing the current value of a variable is such a call. Later calls use the shortest chain of calls that has reached the returned state. For instance, take a sample $x$ with $r_1(x)=0$ and call $T_{1,1}$ on $x$, then $T_{1,0}$, then any setter, each on the previous output. In a composable interface, the first output is a new state of depth $1$ and the second is $x$ again, so the three calls have depths $1$, $2$ and $1$. If the first two outputs are new states, the calls have depths $1$, $2$ and $3$.

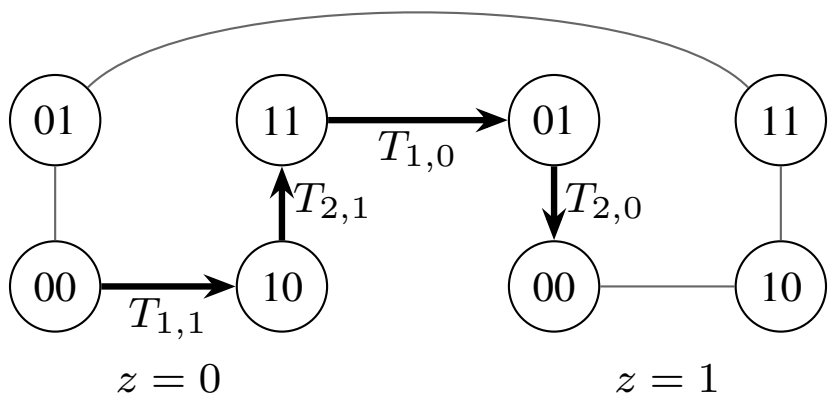
\begin{figure}[!h]
\centering
\begin{tikzpicture}[x=1.5cm,y=1.1cm,
  st/.style={circle,draw,inner sep=0pt,minimum size=17pt,font=\scriptsize},
  key/.style={-{Stealth[length=6pt]},very thick},
  oth/.style={black!60},
  lab/.style={font=\scriptsize,inner sep=1.5pt}]
\node[st] (a00) at (0,0) {00};
\node[st] (a10) at (1,0) {10};
\node[st] (a01) at (0,1) {01};
\node[st] (a11) at (1,1) {11};
\node[st] (b00) at (2.2,0) {00};
\node[st] (b10) at (3.2,0) {10};
\node[st] (b01) at (2.2,1) {01};
\node[st] (b11) at (3.2,1) {11};
\node[lab] at (0.5,-0.55) {$z=0$};
\node[lab] at (2.7,-0.55) {$z=1$};
\draw[oth] (a00) -- (a01);
\draw[oth] (b00) -- (b10);
\draw[oth] (b10) -- (b11);
\draw[oth] (a01) to[out=45,in=135,looseness=0.55] (b11);
\draw[key] (a00) -- node[lab,below]{$T_{1,1}$} (a10);
\draw[key] (a10) -- node[lab,right]{$T_{2,1}$} (a11);
\draw[key] (a11) -- node[lab,below]{$T_{1,0}$} (b01);
\draw[key] (b01) -- node[lab,right]{$T_{2,0}$} (b00);
\end{tikzpicture}
\caption{The eight states $(u,z)$ of the interface defined below, each shown by its label $u_1u_2$, in two layers by the hidden bit $z$. Lines join states that differ by one flip. The thick path is $(00,0)\to(10,0)\to(11,0)\to(01,1)\to(00,1)$.}\label{fig:eight}
\end{figure}

\paragraph{Certification and budgets.} A tester \emph{certifies} $\Dist_\mu<\epsilon$ with confidence $1-\delta$ if it accepts every composable interface with probability at least $1-\delta$ and rejects every interface with $\Dist_\mu\ge\epsilon$ with probability at least $1-\delta$; define this for $\Dist^\infty_\mu$ in the same way. Each theorem below certifies the distance under the sample distribution $\mu_\ell$ of its own mean, and Section~\ref{sec:scope} compares the measures of different means.

A budget counts samples, or samples and setter calls. It does not count the writes that generate a sample, $\ell$ in expectation (Lemma~\ref{lem:coverage}), and lower bounds leave reader calls, comparisons and computation free. For sequences $\ell_m\ge0$ and $d_m\ge2$ and thresholds $\epsilon_m$, a \emph{polynomial budget} is a constant $A$ and, for all large $m$, a tester of depth at most $d_m$ on samples of $\mu_{\ell_m}$ that certifies at $\epsilon_m$ with confidence $2/3$ for all interfaces and start distributions, and uses at most $m^A$ counted operations in expectation on every composable interface. A quantity is \emph{superpolynomial} if, for each $A$, it exceeds $m^A$ for all large $m$. A bound $\Dist_\mu\le K\gamma_\mu$ yields a law-check tester from $\lceil K\log(1/\delta)/\epsilon\rceil$ samples (Appendix~\ref{app:lawcheck}).

\paragraph{Reads and state identity.} Success and preservation speak only about reads, while write-back, overwrite and commutation compare states. The interface of Figure~\ref{fig:eight} separates the two. Its states are $S=\{0,1\}^2\times\{0,1\}$ with $r(u,z)=u$, and its setters are $T_{1,v}(u,z)=(\upd{u}{1}{v},z\oplus(v\oplus u_1)u_2)$ and $T_{2,v}(u,z)=(\upd{u}{2}{v},z)$. All its reads are correct: after any sequence of writes, the label is the one that a record predicts. But the thick path of the figure returns to label $00$ with the hidden bit toggled, and a composable interface that kept these four writes would return to the state where it started. So the distance is at least $1/8$ under the uniform measure, which is stationary (Proposition~\ref{prop:eight}).

For $\ell>0$, along independent uniform random writes from a sample of $\mu_\ell$, reads alone are controlled by the violation rate, with no factor of $m$ (Proposition~\ref{prop:reads}). So the exploration question below asks about \emph{state identity}: whether two write sequences from the same sampled state that predict the same label reach the same state. In the interface of Figure~\ref{fig:eight}, commutation fails at every state, so one law check finds the failure. The instance of Section~\ref{sec:calls} hides it.
\section{One start state}\label{sec:onestate}

\begin{theorem}[One start state]\label{thm:onestate}
Let $m\ge2$.
\begin{enumerate}[label=(\alph*),nosep,leftmargin=*]
\item Let the start distribution be a point mass $\nu=\delta_o$. There is a coefficient $K=O(m^2\log m)$, explicit in Appendix~\ref{app:os-assemble}, such that for every interface and every $\ell>0$, $\Dist_\ell\le\min\{1,K\gamma_\ell\}$ (with $\Dist^\infty_\ell$ in place of $\Dist_\ell$ if $2^m\nmid|S|$; Section~\ref{sec:model}). So, for $0<\epsilon,\delta<1$, law checks at $\lceil K\log(1/\delta)/\epsilon\rceil$ independent samples of $\mu_\ell$ certify $\Dist_\ell<\epsilon$ with confidence $1-\delta$.
\item For every integer $k\ge1$, there are a set $S$ of states with $2^m\mid|S|$, an interface on $S$ and a point mass $\nu$ such that the sample distribution $\nu P^k$ of runs of exactly $k$ writes has $\gamma_{\nu P^k}=0$ and $\Dist^\infty_{\nu P^k}=\Dist_{\nu P^k}=1/2$.
\end{enumerate}
\end{theorem}

\paragraph{Why part (b) holds.} Figure~\ref{fig:core} shows the core, for $m=2$, with samples uniform on $x$ and $w$. All law instances hold at $x$ and at $w$. Take a composable interface, even with additional states, that keeps the four bold entries $T_{1,1}x=y$, $T_{2,1}x=v'$, $T_{2,0}w=y$ and $T_{1,0}w=v$. A setter of a composable interface keeps the cube of its argument, changes only its own variable, and leaves its argument unchanged exactly when the variable already has the written value. So $x$, $y$ and $w$ lie in one cube, drawn in Figure~\ref{fig:core}(c). There $y=T_{1,1}x=T_{2,0}w$ gets the label $10$, by the values these setters write, and as $T_{1,1}$ changes $x$ and $T_{2,0}$ changes $w$, $x$ and $w$ get the labels $00$ and $11$. Then $v'=T_{2,1}x$ and $v=T_{1,0}w$ both get the label $01$, which is impossible as $v'\ne v$. So the distance is at least $1/2$. The law instances at $x$ never reach $v$, and those at $w$ never reach $v'$, so no law instance compares these two states. For $k=1$, add a start state $o$ with $T_{i,0}o=x$ and $T_{i,1}o=w$: every run of exactly one write ends at $x$ or $w$ (Appendix~\ref{app:fixed}).

\begin{figure}[!t]
\centering
\begin{tikzpicture}[yscale=0.5,
  st/.style={circle,draw,inner sep=0pt,minimum size=15pt,font=\small},
  atom/.style={st,fill=black!13,very thick},
  key/.style={-{Stealth[length=6pt]},very thick},
  oth/.style={-{Stealth[length=4pt]},black!50},
  lab/.style={font=\scriptsize,inner sep=1.5pt},
  ent/.style={font=\scriptsize,inner sep=1.5pt},
  gent/.style={ent,black!55},
  pan/.style={font=\scriptsize\bfseries,inner sep=0pt,anchor=north west}]
\node[atom] (x) at (0,0) {$x$};
\node[st] (y) at (3.4,0) {$y$};
\node[atom] (w) at (3.4,2.5) {$w$};
\node[st] (z) at (0,2.5) {$v'$};
\node[st] (v) at (1.7,3.7) {$v$};
\node[lab,left=2pt of x] {$00$};
\node[lab,right=2pt of y] {$10$};
\node[lab,right=2pt of w] {$11$};
\node[lab,left=2pt of z] {$01$};
\node[lab,left=2pt of v] {$01$};
\draw[key] (x) -- node[ent,below]{$T_{1,1}$} (y);
\draw[key] (x) -- node[ent,left]{$T_{2,1}$} (z);
\draw[key] (w) -- node[ent,right]{$T_{2,0}$} (y);
\draw[key] (w) -- node[ent,above right,pos=0.4]{$T_{1,0}$} (v);
\draw[oth] (v) -- node[gent,above left,pos=0.3]{$T_{2,0}$} (x);
\node[pan] at (-0.75,4.3) {(a)};
\node[font=\scriptsize,anchor=north west] (tab) at (4.75,4.0) {\setlength{\tabcolsep}{2.6pt}\begin{tabular}{@{}ccccccl@{}}
\toprule
state & $r_1r_2$ & $T_{1,0}$ & $T_{1,1}$ & $T_{2,0}$ & $T_{2,1}$ & laws failing\\
\midrule
$x$ & $00$ & $x$ & $\boldsymbol y$ & $x$ & $\boldsymbol{v'}$ & none\\
$w$ & $11$ & $\boldsymbol v$ & $w$ & $\boldsymbol y$ & $w$ & none\\
$y$ & $10$ & $x$ & $y$ & $y$ & $w$ & commutation\\
$v'$ & $01$ & $v'$ & $w$ & $x$ & $v'$ & overwrite\\
$v$ & $01$ & $v$ & $w$ & $x$ & $v$ & overwrite\\
\bottomrule
\end{tabular}};
\node[pan] at ($(tab.west |- 0,4.3)+(-0.38,0)$) {(b)};
\begin{scope}[shift={($(tab.east)+(0.5,-1.25)$)}]
\fill[black!13] (0,1.25) rectangle (1.2,2.5);
\draw (0,0) rectangle (2.4,2.5);
\draw (1.2,0) -- (1.2,2.5);
\draw (0,1.25) -- (2.4,1.25);
\node[ent] at (0.6,0.625) {$00$: $x$};
\node[ent] at (1.8,0.625) {$10$: $y$};
\node[ent] at (1.8,1.875) {$11$: $w$};
\node[ent] at (0.6,1.875) {$01$: $v'$, $v$};
\coordinate (boxnw) at (0,2.5);
\end{scope}
\node[pan] at (boxnw |- 0,4.3) {(c)};
\end{tikzpicture}
\caption{An interface with two variables whose square of labels has a split corner, reached at two different states $v'$ and $v$. (a)~Labels $r_1r_2$ of the states $x,y,v',v,w$; the samples are the shaded states $x$ and $w$. Thick arrows are the bold entries of~(b), and the gray arrow is $T_{2,0}v=x$. (b)~The images under each $T_{i,a}$ and the laws failing at each state. (c)~The cube that a composable interface keeping the thick arrows needs, each cell giving a label and its states; the shaded cell holds both $v'$ and $v$.}\label{fig:core}
\end{figure}

\paragraph{Why part (a) holds.} In a composable interface, a run from $o$ stays in one cube, and its end state is determined by the \emph{predicted label} $B$, obtained by applying the run's writes to $r(o)$ as coordinate writes. The proof shows an approximate version for a general interface in four steps (Appendix~\ref{app:onestate}).
\begin{itemize}[nosep,leftmargin=1.5em]
\item \emph{The sample distribution weights every prefix of a run.} A geometric run makes its $(s+1)$-st write with probability $q^{s+1}$, from a state distributed as $\nu P^s$. So by~\eqref{eq:access}, the expected number of writes that a run makes at states of a set $X$ is $\ell\mu_\ell(X)$, and the probability that a run makes a write at a state of $\cN$ is at most $\ell\gamma_\ell$. Runs of exactly $k$ writes have no such property: for $k=1$ above, write-back fails at $o$, but every run ends at $x$ or $w$.
\item \emph{Runs with one predicted label usually end at one state.} A run \emph{meets a failure with} a variable $j$ if a law instance that involves a setter of $j$ fails at one of its states. Otherwise the last write to $j$ can be moved to the front and the other writes to $j$ deleted, so these writes matter only through $B_j$. For instance, let a run make the writes $T_{1,1}$, $T_{2,0}$, $T_{1,0}$ and $T_{2,1}$ in this order, and take $j=1$. Commutation at $T_{1,1}o$ and overwrite at $o$ give $T_{1,0}T_{2,0}T_{1,1}o=T_{2,0}T_{1,0}T_{1,1}o=T_{2,0}T_{1,0}o$, so the run ends at $T_{2,1}T_{2,0}T_{1,0}o$. Now let writes arrive in continuous time at rate $1$, and stop the run at an independent exponential time $\iota$ of mean $\ell$, its \emph{stopping time}, which gives the same sample distribution. Given $B$ and $\iota$, the \emph{histories} of the variables, the times and values of the run's writes to each variable, are independent. So, by the Efron--Stein inequality \citep{vanhandel2016probability}, which bounds a variance by resampling one independent variable at a time, two independent runs with the same $B$ and $\iota$ end at different states with probability at most the sum over $j$ of the probability that resampling the history of $j$ alone, with the same $B_j$, changes the end state. The $j$-th term compares the run with the run whose history of $j$ is resampled, and these two end at different states only if one of them meets a failure with $j$.
\item \emph{The stopping time matters little.} A cube has one state of each label, so given $B$ the end state must also vary little with $\iota$. Given $B$, the conditional histories at different stopping times can be coupled so that the write counts of the variables are pure-birth processes with rates at most $2/m$. At a birth the history of that variable gains a write and may redraw its previous last value, and the end state is unchanged if the old run meets no failure with that variable (Appendix~\ref{app:os-birth}). Appendix~\ref{app:os-kernel} controls the law of $\iota$ given $B$ through an integral-kernel bound, which turns this rate into a bound on the variation over $\iota$.
\item \emph{One state per label forms a cube.} For each label, choose among the states that read this label the one where runs with this predicted label most often end, weighting a run by $1/(1-\eu^{-\iota/m})$, one over the probability, given $\iota$, that it writes a given variable, and choose $o$ for $r(o)$. These states have distinct labels, so they fit into one cube, and filling the other positions, including labels that no state reads, gives the composable interface. A setter $T_{j,a}$ at a sample $X$ acts like resampling the history of $j$ with last value $a$, and the weight offsets the factor by which this conditioning can raise the probability of a history (Appendix~\ref{app:os-weight}). If neither run meets a failure with $j$, then $T_{j,a}X$ is the end of the new run, whose predicted label is $\upd{B}{j}{a}$, so $T_{j,a}X$ is usually the chosen state of this label, which is the output of the composable setter.
\end{itemize}
\section{Many start states}\label{sec:calls}

\begin{theorem}[Worst start distribution]\label{thm:budget}
Fix $0<\beta\le1$ and $0<c<1/2$, and let $\epsilon_m=m^{-\beta}$ and $2\le d_m\le cm$. For both distances, a polynomial budget of samples and setter calls at $\epsilon_m$ exists if and only if $\ell_m=\Omega(m)$.
\begin{enumerate}[label=(\alph*),nosep,leftmargin=*]
\item \emph{Necessity.} If $\ell_m=o(m)$, every tester of depth at most $d_m$ that certifies at $\epsilon_m$ with confidence $2/3$ for all interfaces and start distributions uses a superpolynomial expected number of samples and setter calls on some composable interface and start distribution, even if it learns the length of each run.
\item \emph{Sufficiency.} Let $m\ge2$, $\alpha>0$ and $K_\alpha=4\alpha(2\eu)^{2/\alpha}m^{1+2/\alpha}$. If $\ell\ge\alpha m$, then for all $0<\epsilon,\delta<1$, law checks at $\lceil K_\alpha\log(1/\delta)/\epsilon\rceil$ samples of $\mu_\ell$ certify with confidence $1-\delta$, at depth $2$ and with at most $2m^2+2m$ setter calls per sample. Also $\Dist_\mu\le K_\alpha\gamma_\mu$ for every $(1+1/(\alpha m))$-subinvariant measure $\mu$.
\end{enumerate}
\end{theorem}

\noindent The necessity extends beyond these thresholds. For a fixed $C>1$, the conclusion of (a) holds for $1/(32m)\le\epsilon_m\le(\log m)^{-C}$ and $\ell_m=o(m\log(1/\epsilon_m)/\log m)$ (Proposition~\ref{prop:necessity}(b)); a matching upper bound at this scale is open. So if $(\log m)^{C}\le\ell_m=o(m)$, there is no polynomial budget at the threshold $1/\ell_m$ either.

\paragraph{Why the necessity holds.} The proof builds one family of interfaces at distance above $\epsilon_m$, and the worst start distribution comes from this family (Appendix~\ref{app:twist}).
\begin{itemize}[nosep,leftmargin=1.5em]
\item \emph{One twist.} Split $8J$ variables into two blocks of $4J$, with values $u'$ and $u''$. In the \emph{twisted interface}, a setter of the first block also toggles a hidden bit when it moves $|u'|$ across $2J$ while $|u''|\ge2J$. So commutation between the blocks fails only where both Hamming weights are near $2J$ (Figure~\ref{fig:blocks}).
\item \emph{No composable interface, even with additional states, can match the twist.} The loop of Figure~\ref{fig:blocks} writes each coordinate it touches once to $1$ and once back to $0$, and $|u'|$ crosses $2J$ upwards while $|u''|=J$ and downwards while $|u''|=3J$. So it returns to its start label with the hidden bit toggled, and a composable interface that kept every write would return to the start state. Averaging over such loops, for $\ell=o(m)$ and every composable interface, the expected number of setters of the two blocks that differ from it at a sample is at least a positive constant.

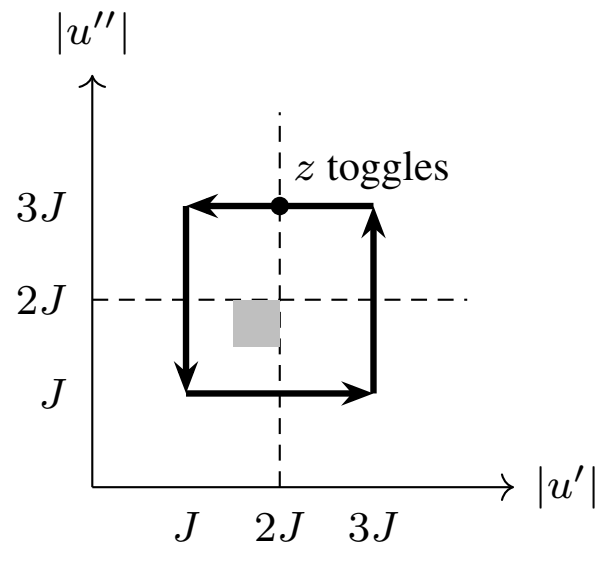
\begin{figure}[!h]
\centering
\begin{tikzpicture}[scale=0.62,every node/.style={font=\scriptsize}]
\draw[->] (0,0) -- (4.5,0) node[right]{$|u'|$};
\draw[->] (0,0) -- (0,4.4) node[above]{$|u''|$};
\draw[densely dashed] (2,0) -- (2,4);
\draw[densely dashed] (0,2) -- (4,2);
\fill[black!25] (1.5,1.5) rectangle (2,2);
\draw[very thick,-{Stealth[length=6pt]}] (1,1) -- (3,1);
\draw[very thick,-{Stealth[length=6pt]}] (3,1) -- (3,3);
\draw[very thick,-{Stealth[length=6pt]}] (3,3) -- (1,3);
\draw[very thick,-{Stealth[length=6pt]}] (1,3) -- (1,1);
\fill (2,3) circle (2.8pt);
\node[anchor=south west,inner sep=1.5pt] at (2.05,3.08) {$z$ toggles};
\foreach \v/\t in {1/J,2/2J,3/3J} {\node[anchor=north] at (\v,-0.05) {$\t$}; \node[anchor=east] at (-0.05,\v) {$\t$};}
\end{tikzpicture}
\caption{The Hamming weights of the two blocks of $4J$ variables of one group, with the thresholds $2J$ (dashed). Dark square, enlarged: $\{2J-1,2J\}^2$, where commutation between the blocks can fail. Thick: the start states, each with a frozen block of weight $J$ or $3J$, and the loop along them in the order bottom, right, top, left. The hidden bit $z$ toggles only where the top edge crosses the threshold (dot). Appendix~\ref{app:small} writes out $J=1$.}\label{fig:blocks}
\end{figure}

\item \emph{A frozen block hides the twist.} Each start state has a \emph{frozen block} of weight $J$ or $3J$. Suppose that its threshold function, $f_1=\one\{|u'|\ge2J\}$ or $f_2=\one\{|u''|\ge2J\}$, keeps its initial value at every label that the tester queries from a sample. If the first block is frozen, or the second with $f_2=0$, no setter toggles the hidden bit. If the second block is frozen with $f_2=1$, a setter of the first block toggles the hidden bit exactly when it changes $f_1$, so the hidden bit plus $f_1$ stays constant. Either way, up to this renaming the tester sees a composable interface (Lemma~\ref{lem:rename}).
\item \emph{The distance.} For one twist at a random place, dividing the expected number above among all $2m$ setters gives a disagreement lower bound of only order $1/m$. So the instance uses $N_g=\lfloor m/(8J)\rfloor$ disjoint \emph{groups} of $8J$ variables, each with its own twist, hidden bit and frozen block. Their expected numbers add up before the division, which gives a lower bound of order $N_g/m$, or $1/J$. With $J=\lfloor1/(512\epsilon_m)\rfloor$, this lower bound exceeds $\epsilon_m$ (Lemma~\ref{lem:rectdistance}). At thresholds above order $1/m$, this $J$ is much smaller than $m$, and the instance hides these small blocks.

\item \emph{Copies separate samples.} A state of the instance consists of a copy, a label in $\{0,1\}^m$ and one hidden bit for each group. Each of $2^m$ \emph{copies} of the states, closed under the setters, draws its own groups and blocks, and polynomially many samples lie in distinct copies except with exponentially small probability. On this event, no state is reached from two samples, and one sample says nothing about where the blocks of another lie.
\item \emph{A mask keeps the tester from aiming at a block.} Each copy also draws a \emph{mask}, a random vector added to the labels before the weights are computed. After averaging over the mask, the label of a sample is uniform and independent of the groups, the blocks, and the start label and edits relative to the mask, given the run length. On the composable interface the answers depend only on the labels of the samples and the setters called. So the coordinates that the tester flips are independent of the groups and the frozen blocks, however it chooses them (Lemma~\ref{lem:independence}).
\item \emph{The margin: half the dimension.} Flipping $d$ coordinates of a sample flips $4Jd/m$ coordinates of a frozen block in the mean. At a start state of weight $J$, a quarter of the coordinates of this block are ones, so each flip in this block raises its weight by $1/2$ in the mean, and the mean weight rises to $J+2Jd/m$. Call the distance from the mean weight to the threshold the \emph{margin}. Then flips multiply the margin by $1-2d/m$, for every $J$. This is a positive constant for $d\le cm$ with $c<1/2$, and the margin vanishes only when $d$ is about $m/2$.
\item \emph{Time: long runs, and where $\log(1/\epsilon)$ comes from.} A run lowers the margin as well. Each write to a group replaces a coordinate by a fair bit, so $\hat t$ such writes multiply the margin by $(1-1/(8J))^{\hat t}$, and the margin falls to the fluctuations of the weight after order $J\log J$ of them. The number of writes to one group in a run is geometric with mean $\ell_0=8J\ell/m$, so with $q_0=\ell_0/(1+\ell_0)$ it exceeds $\hat t$ with probability $q_0^{\hat t+1}\le\exp[-(\hat t+1)/(1+\ell_0)]$. For $\ell>0$ and $\hat t$ of order $J\log J$, this exponent has order $\min\{J\log J,m\log J/\ell\}$, so the tail is superpolynomially small for $\ell=o(m\log J/\log m)$ once $J\log J/\log m\to\infty$. As $J$ has order $1/\epsilon_m$, this is the scale stated after Theorem~\ref{thm:budget}, and at $\epsilon_m=m^{-\beta}$ it reads $\ell=o(m)$. These bounds hold state by state and group by group, so a union over them needs a polynomial budget (Lemma~\ref{lem:maskedtv}).
\end{itemize}

\paragraph{Why long runs suffice.} A \emph{word} is a sequence of setters. The \emph{path} of $x$ under $W=E_t\sdots E_1$ with $E_s=T_{i_s,a_s}$ is $x_0=x$, $x_s=E_sx_{s-1}$, and the \emph{fiber} of $x$ is $W^{-1}(Wx)$. In a composable interface, a word that writes every variable sends two states to the same state exactly when they lie in one cube, so fibers serve as candidate cubes. With the \emph{old value} $b_s=r_{i_s}(x_{s-1})$, step $s$ \emph{passes} if
\begin{equation}\label{eq:goodfull}
r(x_s)=\upd{r(x_{s-1})}{i_s}{a_s},\qquad T_{i_s,b_s}x_s=x_{s-1}.
\end{equation}
Where all steps pass, $x\mapsto(r(x),Wx)$ is injective, so a fiber fits into a cube at its labels.

Take a word of $t=\lceil m\log(2m)\rceil$ random setters, conditioned on writing every variable, an event of probability at least $1/2$. Before conditioning, the state after $s$ random edits has distribution $\mu P^s\le\lambda^s\mu$, so the path meets $\cN$ with probability at most $2\sum_{s<t}\lambda^s\gamma_\mu$. For $\mu_\ell$, $\lambda=1+1/\ell$, and $\lambda^t$ is polynomial exactly when $\ell\gtrsim m$. If the path avoids $\cN$, commutation moves any setter along it until the first write to its variable absorbs it, so every setter at $x$ stays in the fiber (Lemma~\ref{lem:insert}). With added states, each fiber takes a cube of its own, and this finishes the proof for $\Dist^\infty$.

On the same states there are only $M$ cubes, so the proof keeps the $M$ heaviest fibers and bounds the probability that this loses (Lemma~\ref{lem:weighted}). Here the $\mu$-mass of a fiber $W^{-1}(z)$ counts only its states whose path avoids $\cN$, and if weights $\kappa(z)\in[0,1]$ on the fibers of positive mass sum to at most $M$, this loses at most the mean of $1-\kappa(Wx)$ for $x\sim\mu$ besides the paths that meet $\cN$.
\begin{lemma}[Capacity]\label{lem:capacity}
A transition of $T_{i,a}$ at $x$ is \emph{kept} if it passes~\eqref{eq:goodfull}, with $T_{i,a}x=x$ when $a=r_i(x)$. The \emph{capacity} $\kappa^\circ(z)$ of a state $z$ is the probability that a second random word, drawn like the first, makes only kept transitions along the path of $z$. For every label $b$, the capacities of the states with label $b$ sum to at most $M$.
\end{lemma}
\noindent If the transition of $T_{i,a}$ from $x$ to $y\ne x$ is kept, then the transition of $T_{i,r_i(x)}$ from $y$ back to $x$ is kept. The random word has the same law as its reverse, as its setters are independent and uniform before the conditioning, and reversal keeps the event of writing every variable. So a walk along kept transitions leads from $x$ to $y$ with the same probability as from $y$ to $x$, and this symmetry replaces the invariance of the counting measure, which permutations have and general setters lack. By this symmetry, the sum of these capacities equals the sum over all states $x$ of the probability that such a walk from $x$ ends at label $b$, at most $2^{-m}$ each as the written label is uniform (Appendix~\ref{app:capacityfn}). Passing paths end at the label that $W$ writes, so Lemma~\ref{lem:weighted} applies, losing at most the mean of $1-\kappa^\circ(Wx)$ over the covering word $W$ and $x\sim\mu$. As a transition fails to be kept only where a law other than commutation fails, this mean is at most $2\lambda^t$ times the bound $2\sum_{s<t}\lambda^s\gamma_\mu$ on paths that meet $\cN$. Adding the earlier path loss gives $K_\alpha$ (Appendix~\ref{app:repaircoeff}).

\paragraph{An audit replaces the run by its own walk from each sample.} Fix a \emph{target label} $\hat a$ and put $W_{\hat a}=T_{m,\hat a_m}\sdots T_{1,\hat a_1}$. The \emph{path check} follows the path of a sample $x$ under $W_{\hat a}$ and rejects if a step fails~\eqref{eq:goodfull}. Each step that writes a new value first checks the label of its output and then calls the \emph{undo}, the setter of the old value, whose output must be the state before the step. The \emph{flip check} rejects if write-back fails at $x$, if a flip $y_i=F_ix$ has a wrong label, or if the path of $y_i$ fails~\eqref{eq:goodfull} or does not join the path of $x$ at step $i$. In a composable interface, step $i$ is the first write to variable $i$ and erases the flip, so the two paths meet there. Passing both checks puts every setter at $x$ in the fiber of $x$ with the updated label (Lemma~\ref{lem:checked}).

The calls of these checks have depth at most $|r(x)\oplus\hat a|+2$, the Hamming distance from the label of $x$ to the target label plus two (Lemma~\ref{lem:rootdepth}). A passed step that writes the current value returns its argument, so the path of $x$ gains depth only at the coordinates where $r(x)$ differs from $\hat a$. The flip adds one more, and the undo of a step adds one more. A uniform target label lies at distance about $m/2$ from the label of a sample, so depth about $m/2$ covers all but a small fraction of the rounds, and the other rounds accept without checking. Algorithm~\ref{alg:round} in Appendix~\ref{app:audit} lists the calls of one round.

On the interface of Figure~\ref{fig:core}, with samples uniform on $x$ and $w$, take $\hat a=00$ and $d=3$, so $W_{\hat a}=T_{2,0}T_{1,0}$. From $w$, step~$1$ calls $T_{1,0}w=v$, and its undo $T_{1,1}v=w$ passes. Step~$2$ calls $T_{2,0}v=x$, but its undo $T_{2,1}x=v'$ is not $v$, so the path check rejects $w$, although every law instance holds at $w$.

\begin{theorem}[Audit]\label{thm:audit}
Let $\mu$ be a probability measure on $S$ and $0<\epsilon,\delta<1$. Rounds of Algorithm~\ref{alg:round} have one-sided error.
\begin{enumerate}[label=(\alph*),nosep,leftmargin=*]
\item For $d=m+1$ and any fixed $\hat a$, a round has depth at most $m+1$, makes at most $m^2+5m$ setter calls and rejects with probability at least $\Dist^\infty_\mu$, so $\lceil\log(1/\delta)/\epsilon\rceil$ rounds certify. With the bridge check of Appendix~\ref{app:auditbridge}, the same holds for $\Dist_\mu$, with at most $2m^2+6m$ setter calls per round.
\item Let $2\le d\le m+1$ and $\tau_j=\Pr\{\mathrm{Bin}(m,1/2)>d-j\}$. For a uniform $\hat a$, a round has depth at most $d$ and rejects with a probability $p$ such that $\Dist^\infty_\mu\le p+\tau_2$; with the bridge check, $\Dist_\mu\le p+(m+2)\tau_2$. We call the term added to $p$ the \emph{tail term}; if it is at most $\epsilon/2$, $\lceil2\log(1/\delta)/\epsilon\rceil$ rounds certify.
\item Let $\ell=0$ and $\epsilon_m=m^{-\beta}$ with fixed $0<\beta\le1$. The least depth with a polynomial budget of samples and setter calls at $\epsilon_m$ is $(1/2+o(1))m$, for both distances.
\end{enumerate}
\end{theorem}

\noindent By Hoeffding's inequality, depth $m/2+O(\sqrt{m\log m})$ makes the tail term of (b) at most $\epsilon_m/2$. This is the upper bound in (c), and the lower bound is the instance of Theorem~\ref{thm:budget} at mean zero (Appendix~\ref{app:necessity}).
\section{Counting samples alone}\label{sec:samples}

\begin{theorem}[Samples alone]\label{thm:samplesonly}
Let $2\le d_m\le\lfloor m/16\rfloor$ and $\ell^*_m=m\log(m/d_m)/\log m$. For both distances, a polynomial budget of samples at $\epsilon_m=1/(32m)$ exists if and only if $\ell_m=\Omega(\ell^*_m)$.
\begin{enumerate}[label=(\alph*),nosep,leftmargin=*]
\item \emph{Necessity.} If $\ell_m=o(\ell^*_m)$, every tester of depth at most $d_m$ that certifies at $\epsilon_m=1/(32m)$ with confidence $2/3$ for all interfaces and start distributions uses a superpolynomial expected number of samples on some composable interface and start distribution, whatever its number of setter calls.
\item \emph{Sufficiency.} If $\ell_m\ge b\ell^*_m$ with a constant $b>0$, then for all $0<\epsilon,\delta<1$, evaluating all law instances at the ends of all words of length at most $d_m-2$ from each sample certifies from a number of samples polynomial in $m$, $1/\epsilon$ and $\log(1/\delta)$. The setter calls of this tester can be exponential in $d_m$. For $m\ge96$, $d=\lfloor m/16\rfloor$ and $\ell\ge\varsigma m/\log m$ with $\varsigma>0$, $\lceil140m^{2+8/\varsigma}\log(1/\delta)/\epsilon\rceil$ samples suffice.
\end{enumerate}
\end{theorem}

\paragraph{Why Theorem~\ref{thm:samplesonly} holds.} For sufficiency, a random word need not write every variable. If it misses at most $d-2$ of them, the exhaustive check at its end stands in for the missing writes, as all words of length at most $d-2$ on the missing variables behave like writes in a record (Theorem~\ref{thm:partial}). Writing all but $d-2$ variables takes about $t=2m\log(m/d)$ random writes, against $m\log m$ for all of them. The factor $\lambda^t=(1+1/\ell)^t$ that moves failures along the word to the samples is polynomial exactly when $t/\ell\lesssim\log m$, which is $\ell\gtrsim\ell^*_m$.

For necessity, take the instance of Theorem~\ref{thm:budget} with one group of $8\lfloor m/8\rfloor$ variables and no mask. The transcripts stay equal as long as the threshold function of the frozen block is constant on the whole ball of radius $d$ around a sample. For a run with at most $\hat t$ writes to the group, this holds with high probability if the margin $J(1-1/(8J))^{\hat t}$ exceeds $d$ by more than the fluctuations of the weight, which allows $\hat t$ of order $m\log(m/d)$. The writes to this group, which holds almost all variables, exceed $\hat t$ with probability at most $\exp[-\hat t/(1+\ell)]$, so polynomially many samples include such a run only if $\hat t/(1+\ell)\lesssim\log m$, the same scale $\ell\gtrsim\ell^*_m$ (Appendix~\ref{app:unmasked}).

The two theorems differ in how the tester's writes can lower the margin. Exhausting the ball reaches the labels that subtract $d$ from the margin, at a cost in setter calls exponential in $d$. Writes chosen without knowing the mask only multiply the margin by about $1-2d/m$, so with a polynomial budget the twist appears only once the margin falls to the fluctuations, after order $m\log m$ writes.

\paragraph{The two budgets separate.} Take $d=\lfloor m/16\rfloor$ and $\ell=m/\log m$. For $m\ge96$, part~(b) with $\varsigma=1$ says that $\lceil140m^{10}\log(1/\delta)/\epsilon\rceil$ samples certify at every $\epsilon$, so polynomially many samples certify at each $\epsilon_m=m^{-\beta}$. As $\ell=o(m)$, Theorem~\ref{thm:budget}(a) with $c=1/16$ says that at each such $\epsilon_m$ polynomially many samples and setter calls do not.
\section{Discussion}\label{sec:scope}

\paragraph{Which measure a certificate speaks about.} Law checks after runs certify $\mu_\ell$, and the audit with no runs certifies $\nu$, so these are two different certificates. A two-state interface shows the difference: with target label $0$ and depth $2$, every audit round passes, and $\Dist_\nu=0$, while after one random write the distance is $1/2$ (Proposition~\ref{prop:twostate}).

A tester that may change the mean of its runs can still certify the distance under the original $\mu_\ell$. With $L=\max\{\ell,m\}$, $\mu_\ell\le\frac{1+L}{1+\ell}\mu_L$, so law checks at mean $L$ certify the distance under $\mu_\ell$ from polynomially many samples (Proposition~\ref{prop:lengthen}). For $\ell>0$, a distance certified under $\mu_\ell$ bounds how often independent random edits from a sample leave the path of one composable interface (Corollary~\ref{cor:workload}).

\paragraph{How large the worst start distribution must be.} With added states and no runs, depth one and a budget polynomial in $m$ and $N$ suffice for a start distribution of support at most $N$, as a tester can compare states across samples (Proposition~\ref{prop:support}). So at mean zero, the necessity of Theorem~\ref{thm:budget} for $\Dist^\infty$ needs start distributions of superpolynomial support.

\paragraph{Larger thresholds.} At a fixed threshold, the question for law checks reduces to one quantity. Call a state \emph{locally exact to depth $d$} if every word of length at most $d$ from it ends at the label that coordinate writes predict, and two such words end at one state exactly when these labels agree. Let $a_d$ be the supremum of $\Dist^\infty_\nu(I)$ over finite interfaces $I$ of any number of variables and distributions $\nu$ on such states. Law checks at each fixed positive mean certify $\Dist^\infty$ at each fixed positive threshold from polynomially many samples if and only if $a_d\to0$ (Theorem~\ref{thm:modulus}). Without runs, a sample bound uniform over interfaces and start distributions requires depth $\Omega(\min\{m,1/\epsilon\})$ at $0<\epsilon\le1/512$ and confidence above $1/2$, unconditionally (Theorem~\ref{thm:depthsource}).

\paragraph{Tightness.} For $\epsilon\le1/16$ and $\delta\le1/3$, the number of samples of the audit is optimal up to a constant factor (Proposition~\ref{prop:auditlower}). At $\ell=\alpha m$ the coefficient of law checks is at least of order $m^{1/(2\alpha)}$ (Proposition~\ref{prop:coefflower}), so the degree of the law-check budget obtained from this coefficient grows with $1/\alpha$.

\paragraph{Scope.} The results assume binary variables, uniformly random writes in the runs, geometric run lengths, exact comparison of states, and a tester that is not given the start states or write sequences of the sampling runs (Definition~\ref{def:tester}(i)). The lower bounds apply to testers that stay local: the audit of depth $m+1$ is always available.

\paragraph{Open problems.} First, whether $a_d\to0$. Second, the condition on the mean for a polynomial budget at depths with $d_m-m/2=o(m)$. Third, the orders of the coefficients in Theorem~\ref{thm:onestate} and in law checks after long runs. Fourth, how the exploration needed at positive means varies between one start state and the worst start distribution, for instance at support polynomial in $m$.

\bibliography{references}

\appendix

\section{Preliminaries}\label{app:prelim}

\begin{lemma}[Properties of the sample distribution]\label{lem:resolvent}
For all $\ell>0$ and integers $s\ge0$, (i)~$\mu_\ell=(1-q)\nu+q\,\mu_\ell P$, (ii)~$\mu_\ell P^s\le q^{-s}\mu_\ell$ and (iii)~$\nu\le(1+\ell)\,\mu_\ell$.
\end{lemma}

Put $\tm=\lceil m\log(2m)\rceil$, and let $h_t$ be the probability that $t$ independent uniform draws cover $[m]$.

\begin{lemma}[Samples and coverage]\label{lem:coverage}
For every $\ell>0$, a sample as in Section~\ref{sec:model} has distribution $\mu_\ell$ and expected cost $\E L=\ell$ setter calls. Also $h_t\ge1-m(1-1/m)^t\ge1-m\eu^{-t/m}$, $h_t>0$ for $t\ge m$, and $h_{\tm}\ge1/2$. Finally $\tm\le m\log(2m)+1$, and $\tm\ge m$ for $m\ge2$.
\end{lemma}

\begin{proof}[Proof of Lemmas~\ref{lem:resolvent} and~\ref{lem:coverage}]
Write $\mu=\mu_\ell$. Given $L=s$, the returned state has distribution $\nu P^s$, so the sample has distribution~\eqref{eq:access}, and $\E L=\sum_{s\ge1}\Pr(L\ge s)=\sum_{s\ge1}q^s=q/(1-q)=\ell$. Reindexing the series gives~(i) of Lemma~\ref{lem:resolvent}, $q\mu P=(1-q)\sum_{s\ge0}q^{s+1}\nu P^{s+1}=\mu-(1-q)\nu$. Dropping either nonnegative term of~(i) gives~(iii), $\nu\le(1-q)^{-1}\mu=(1+\ell)\mu$, and $\mu P\le q^{-1}\mu$. By positivity of $P$, induction with $\mu P^{s+1}\le q^{-s}\mu P\le q^{-(s+1)}\mu$ gives~(ii). For coverage, a union bound over the variables absent from the $t$ draws gives the first inequality, and $1-1/m\le\eu^{-1/m}$ the second. The first $m$ draws are distinct with positive probability, so $h_t>0$ when $t\ge m$. For $m=1$ coverage is certain; for $m\ge2$ and $t=\tm\ge m\log(2m)$, $m\eu^{-t/m}\le1/2$. The ceiling gives $\tm\le m\log(2m)+1$, and $\log(2m)\ge1$ gives $\tm\ge m$ for $m\ge2$.
\end{proof}

\begin{lemma}[Subinvariant measures]\label{lem:subinv}
Let $\lambda\ge1$. If $\mu$ is $\lambda$-subinvariant, then $\mu P^s\le\lambda^s\mu$ for all $s\ge0$.
\end{lemma}

\begin{proof}
As $P$ is nonnegative, applying it to both sides of $\mu P^s\le\lambda^s\mu$ gives $\mu P^{s+1}\le\lambda^s\mu P\le\lambda^{s+1}\mu$, and induction gives the claim.
\end{proof}

\begin{proposition}[Characterization]\label{prop:exact}
An interface satisfies the five laws at every state if and only if it is composable.
\end{proposition}

\begin{proof}
A composable interface satisfies the five laws by direct evaluation. Conversely, suppose the laws hold everywhere. By success and preservation, the flip $F_i$ changes exactly reader $i$. Overwrite and write-back give $F_i^2x=T_{i,r_i(x)}T_{i,1-r_i(x)}x=T_{i,r_i(x)}x=x$. For $i\ne j$, preservation makes the values chosen by $F_i$ and $F_j$ in $F_iF_jx$ and $F_jF_ix$ equal to the values chosen at $x$, and commutation of the corresponding setters gives $F_iF_j=F_jF_i$. These involutions define an action of $\{0,1\}^m$ under bitwise addition. Applying $u\in\{0,1\}^m$ changes the label by $u$, so nonzero elements move every state and every orbit has $2^m$ points, one per label. With $\cZ$ the set of orbits, $x\mapsto(r(x),[x])$ is the required bijection; by write-back, $T_{i,a}x$ is $x$ if $a=r_i(x)$ and $F_ix$ otherwise.
\end{proof}

\paragraph{Law checks at samples.}\label{app:lawcheck} The five laws of Section~\ref{sec:model} have $4m^2+3m$ law instances: $2m$ for success, $m$ for write-back, $4m$ for overwrite, $2m(m-1)$ for preservation and $2m(m-1)$ for commutation, one for each pair $i<j$ and values $a,b$. Let $x$ be a state with label $u$, and put $y_i=T_{i,1-u_i}x$. The $2m^2+2m$ setter calls $T_{i,a}x$, $T_{i,a}y_i$ and $T_{i,a}y_j$, for all $i\ne j$ and $a$, give every law instance at $x$ as a reading or a comparison of their outputs, given write-back at $x$; for instance, overwrite of $i$ compares $T_{i,a}y_i$ with $T_{i,a}x$.

Let $K$ be a coefficient with $\Dist_\mu\le K\gamma_\mu$ for all $\mu$ in a class. Evaluating all law instances at $n=\lceil K\log(1/\delta)/\epsilon\rceil$ independent samples of $\mu$ and rejecting when one fails has one-sided error, by Proposition~\ref{prop:exact}. If $\Dist_\mu\ge\epsilon$, each sample shows a failure with probability $\gamma_\mu\ge\epsilon/K$, so all $n$ samples pass with probability at most $\eu^{-n\epsilon/K}\le\delta$. The same holds for $\Dist^\infty_\mu$. Conversely, a law-check tester rejects with probability at most $\gamma_\mu$ times its expected number of samples, as the decision to draw a sample does not depend on that sample.

\paragraph{What a distance bound gives.}\label{app:workload} A small distance controls sequences of random edits from the samples. Along a sequence, the same setters are applied under $I$ and under a closest composable interface.

\begin{corollary}[Sequences of random edits]\label{cor:workload}
Let $\ell>0$, let $\tilde I$ be a composable interface on $S$ with $R_{\mu_\ell}(I,\tilde I)=\Dist_\ell$, and let $k\ge1$ be an integer. Draw $x\sim\mu_\ell$ and independent uniformly random setters $(i_1,a_1),\dots,(i_k,a_k)$, independent of $x$, and put $X_0=\tilde X_0=x$, $X_s=T_{i_s,a_s}X_{s-1}$ and $\tilde X_s=\tilde T_{i_s,a_s}\tilde X_{s-1}$ for $s=1,\dots,k$. Then the probability that $X_s\ne\tilde X_s$ for some $s\le k$ is at most $\ell\,(q^{-k}-1)\,\Dist_\ell$, and the probability that this happens or $r(X_k)\ne\tilde r(\tilde X_k)$ is at most $(\ell\,(q^{-k}-1)+mq^{-k})\,\Dist_\ell$. For $k\le\ell$, these bounds are at most $\eu k\,\Dist_\ell$ and $\eu(k+m)\,\Dist_\ell$.
\end{corollary}

\begin{proof}
At the first $s$ with $X_s\ne\tilde X_s$, the setters $T_{i_s,a_s}$ and $\tilde T_{i_s,a_s}$ differ at $X_{s-1}=\tilde X_{s-1}$. Here $X_{s-1}$ has distribution $\mu_\ell P^{s-1}\le q^{-(s-1)}\mu_\ell$ by Lemma~\ref{lem:resolvent}(ii), and $(i_s,a_s)$ is uniform and independent of it, so averaging over the $2m$ setters bounds this probability by $q^{-(s-1)}\Dist_\ell$. Summing over $s$ with $\sum_{s<k}q^{-s}=\ell(q^{-k}-1)$ gives the first bound. If $X_s=\tilde X_s$ for every $s\le k$ but $r(X_k)\ne\tilde r(\tilde X_k)$, then $r_i(X_k)\ne\tilde r_i(X_k)$ for some $i$. As $X_k$ has distribution $\mu_\ell P^k\le q^{-k}\mu_\ell$, this has probability at most $q^{-k}\sum_i\mu_\ell\{r_i\ne\tilde r_i\}\le mq^{-k}\Dist_\ell$, which gives the second bound. For $k\le\ell$, $q^{-k}=(1+1/\ell)^k\le\eu^{k/\ell}\le\eu$ and $\ell(q^{-k}-1)\le\ell(\eu^{k/\ell}-1)\le k\eu^{k/\ell}\le\eu k$, by $\eu^y-1\le y\eu^y$ for $y\ge0$.
\end{proof}
\section{Completion and reversible words}\label{app:completion}

This appendix holds the tools behind the upper bounds of Theorems~\ref{thm:budget}, \ref{thm:audit} and~\ref{thm:samplesonly}. In a composable interface, a word that writes every variable sends two states to the same state exactly when they lie in one cube. So the fibers of such a word serve as candidate cubes in a general interface. With added states, each fiber can take a cube of its own. On the same states there are only $M$ cubes, so the proofs keep the $M$ heaviest fibers and bound the probability that this loses.

\begin{lemma}[Completion by heavy fibers]\label{lem:weighted}
Let $|S|=M2^m$, let $\mu$ be a probability measure on $S$, let $W$ be any map $S\to S$, let $G\subseteq S$ be a set on which $x\mapsto(r(x),Wx)$ is injective, and let $A\subseteq G$. Put $w_z=\mu(A\cap W^{-1}(z))$ for $z\in W(S)$, let $\kappa\colon W(S)\to[0,1]$ satisfy $\sum_{z:\,w_z>0}\kappa(z)\le M$, and put $\Delta=\sum_zw_z(1-\kappa(z))$. There are a composable interface $\tilde I$ on $S$ and a set $G'\subseteq G$ with $\mu(A\setminus G')\le\Delta$ such that $\tilde r=r$ on $G'$, and $\tilde T_{i,a}x=T_{i,a}x$ whenever $x\in G'$, $y=T_{i,a}x\in G$, $Wy=Wx$ and $r(y)=\upd{r(x)}{i}{a}$.
\end{lemma}

\begin{proof}
Let $\Omega_z=G\cap W^{-1}(z)$. By injectivity, the states of $\Omega_z$ have distinct labels. Let $Z_*$ be a set of at most $M$ points $z$ with nonempty $\Omega_z$ and largest $w_z$, containing all of them if there are at most $M$, and let $G'$ be the union of the sets $\Omega_z$ with $z\in Z_*$. If at most $M$ sets $\Omega_z$ are nonempty, then $G'=G\supseteq A$. Otherwise $|Z_*|=M$; let $w_*=\min_{z\in Z_*}w_z$. If $w_*=0$, then $w_z=0$ for $z\notin Z_*$. If $w_*>0$, then
\[
\sum_{z\in Z_*}w_z-\sum_zw_z\kappa(z)=\sum_{z\in Z_*}w_z(1-\kappa(z))-\sum_{z\notin Z_*}w_z\kappa(z)\ge w_*\Big(M-\sum_{z:\,w_z>0}\kappa(z)\Big)\ge0 .
\]
In both cases $\sum_{z\in Z_*}w_z\ge\sum_zw_z\kappa(z)$, so $\mu(A\setminus G')\le\sum_zw_z-\sum_{z\in Z_*}w_z\le\Delta$. Place each set $\Omega_z$ with $z\in Z_*$ in its own cube, at the positions given by the labels of its states. For $|Z_*|=M'\le M$, these cubes have $M'2^m-|G'|\le|S|-|G'|$ empty positions, which we fill with states outside $G'$. The remaining $(M-M')2^m$ states form the other cubes in any way, which defines $\tilde I$. At $x\in G'$ the readers agree with $\tilde I$. If $y=T_{i,a}x\in G$, $Wy=Wx$ and $r(y)=\upd{r(x)}{i}{a}$, then $y$ is the state of $\Omega_{Wx}$ with this label, which is the position of $\tilde T_{i,a}x$.
\end{proof}

The first choice of $\kappa$ is the fraction of labels that occur in a fiber, $\kappa(z)=|r(W^{-1}(z))|/2^m$. The fibers are disjoint and each has at least as many states as distinct labels, so these fractions sum to at most $M$ over all $z$. A uniformly random label occurs in the fiber $W^{-1}(z)$ with probability $\kappa(z)$, so $\Delta$ is the probability that $x\sim\mu$ lies in $A$ and a uniformly random label occurs nowhere in the fiber $W^{-1}(Wx)$. This fiber is a subset of $S$ and may contain states outside $G$.

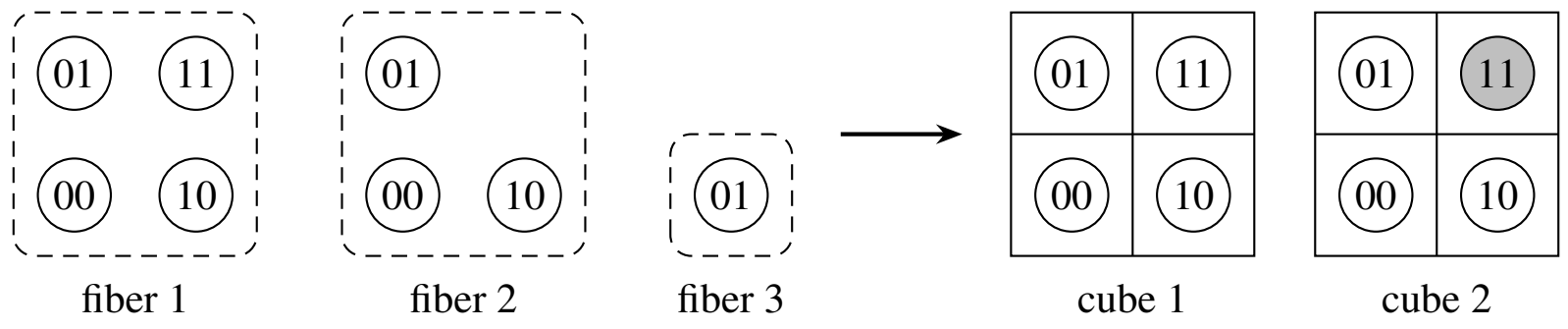
\begin{figure}[!t]
\centering
\begin{tikzpicture}[scale=0.82,
  st/.style={circle,draw,inner sep=0pt,minimum size=14pt,font=\scriptsize},
  gr/.style={circle,draw,fill=black!25,inner sep=0pt,minimum size=14pt},
  fib/.style={draw,densely dashed,rounded corners=4pt},
  lab/.style={font=\scriptsize}]
\draw[fib] (-0.5,-0.5) rectangle (1.5,1.5);
\node[st] at (0,0) {00}; \node[st] at (1,0) {10}; \node[st] at (0,1) {01}; \node[st] at (1,1) {11};
\draw[fib] (2.2,-0.5) rectangle (4.2,1.5);
\node[st] at (2.7,0) {00}; \node[st] at (3.7,0) {10}; \node[st] at (2.7,1) {01};
\draw[fib] (4.9,-0.5) rectangle (5.9,0.5);
\node[st] at (5.4,0) {01};
\node[lab] at (0.5,-0.85) {fiber 1}; \node[lab] at (3.2,-0.85) {fiber 2}; \node[lab] at (5.4,-0.85) {fiber 3};
\draw[-{Stealth[length=5pt]},thick] (6.3,0.5) -- (7.3,0.5);
\draw (7.7,-0.5) rectangle (9.7,1.5); \draw (8.7,-0.5) -- (8.7,1.5); \draw (7.7,0.5) -- (9.7,0.5);
\node[st] at (8.2,0) {00}; \node[st] at (9.2,0) {10}; \node[st] at (8.2,1) {01}; \node[st] at (9.2,1) {11};
\draw (10.2,-0.5) rectangle (12.2,1.5); \draw (11.2,-0.5) -- (11.2,1.5); \draw (10.2,0.5) -- (12.2,0.5);
\node[st] at (10.7,0) {00}; \node[st] at (11.7,0) {10}; \node[st] at (10.7,1) {01}; \node[gr,font=\scriptsize] at (11.7,1) {11};
\node[lab] at (8.7,-0.85) {cube 1}; \node[lab] at (11.2,-0.85) {cube 2};
\end{tikzpicture}
\caption{Completion by heavy fibers for $m=2$ and $M=2$ cubes. Left: three fibers of a word, with each state shown by its label. Right: the two cubes when the first two fibers carry the most probability. The state of the third fiber (gray), whose label was $01$, fills the empty position and reads $11$ there.}\label{fig:fibers}
\end{figure}

Figure~\ref{fig:fibers} shows how the sum bounds the loss. There the second fiber misses a quarter of the labels, the third misses three quarters, and the second carries at least as much probability as the third. So the average of $1-\kappa$ over the fibers, which counts a quarter of the probability of the second fiber and three quarters of that of the third, is at least the probability of the third. This is the part that the composable interface drops. Appendix~\ref{app:repair} uses another choice of $\kappa$.

\paragraph{Reversible words.} Let $W=E_t\sdots E_1$ be a word with $E_s=T_{i_s,a_s}$, put $U_s=E_s\sdots E_1$ and $U_0=\Id$, and let $x_s=U_sx$ be the path of $x$. The \emph{reversible set} $G_W$ is the set of states $x$ whose path passes~\eqref{eq:goodfull} at every step $s=1,\dots,t$.

\begin{lemma}[Reversible set]\label{lem:reversible}
The map $x\mapsto(r(x),Wx)$ is injective on $G_W$.
\end{lemma}

\begin{proof}
Let $x\in G_W$. The first equalities of~\eqref{eq:goodfull} determine each label $r(x_s)$, and so each old value $b_s$, from $r(x)$ and the word. Starting at $Wx$, the setters $T_{i_t,b_t},\dots,T_{i_1,b_1}$ undo the steps in reverse order by the second equalities and end at $x$.
\end{proof}

\begin{lemma}[Inserted setters]\label{lem:insert}
Let $W=E_t\sdots E_1$ write every variable, and let $x$ be a state with $x_j\notin\cN$ for $0\le j<t$. Then $x\in G_W$, and every setter $T_{i,a}$ satisfies $T_{i,a}x\in G_W$, $WT_{i,a}x=Wx$ and $r(T_{i,a}x)=\upd{r(x)}{i}{a}$.
\end{lemma}

\begin{proof}
A state $z\notin\cN$ and a setter $T_{k,e}$ pass~\eqref{eq:goodfull}: success and preservation give the label, and with $v=r_k(z)$, write-back handles $e=v$, while overwrite and write-back give $T_{k,v}T_{k,e}z=T_{k,v}z=z$ for $e\ne v$. So $x\in G_W$. Fix $T=T_{i,a}$ and let $\tau$ be the first position with $i_\tau=i$. The label claim is success and preservation at $x$. Let $y_s=U_sTx$.

For $j<\tau$, let $z=x_{j-1}$, $z'=x_j$ and $E_j=T_{k,e}$ with $k\ne i$. Both $z$ and $z'$ lie outside $\cN$, as $j<t$. Commutation at $z$ gives $E_jTz=Tz'$, so $y_j=Tx_j$ by induction. Success and preservation at $z$ and at $z'$ give $r(Tz)=\upd{r(z)}{i}{a}$ and $r(Tz')=\upd{r(z')}{i}{a}=\upd{r(Tz)}{k}{e}$, which is the first equality of~\eqref{eq:goodfull} for this step of the path of $Tx$. For the second, let $v=r_k(Tz)=r_k(z)$. If $e=v$, write-back at $z$ gives $z'=z$, so $T_{k,v}Tz'=E_jTz=Tz'=Tz$. If $e\ne v$, commutation at $z'$, overwrite at $z$ and write-back at $z$ give $T_{k,v}Tz'=TT_{k,v}T_{k,e}z=TT_{k,v}z=Tz$.

At $j=\tau$, let $z=x_{\tau-1}$ and $E_\tau=T_{i,e}$. Overwrite at $z$ gives $E_\tau Tz=T_{i,e}z=x_\tau$, so the two paths meet, and $T_{i,a}x_\tau=T_{i,a}T_{i,e}z=T_{i,a}z=Tz$, where $a=r_i(Tz)$ by success. Success and preservation at $z$ give $r(x_\tau)=\upd{r(Tz)}{i}{e}$. So this step passes~\eqref{eq:goodfull}. From step $\tau$ on, the path of $Tx$ is that of $x$, and $x\in G_W$. Hence $Tx\in G_W$ and $WTx=Wx$.
\end{proof}

\noindent For a word $W$ and $A\subseteq S$, let $\Delta_A$ be the probability that $x\sim\mu$ lies in $A$ and a uniformly random label, independent of $x$, occurs nowhere in the fiber $W^{-1}(Wx)$. This is the $\Delta$ of Lemma~\ref{lem:weighted} for the first choice of $\kappa$.

\begin{lemma}[Bounds from a reversible set]\label{lem:structure}
Let $W$ be a word, let $A\subseteq G_W$, and suppose every $x\in A$ and every setter $T_{i,a}$ satisfy $T_{i,a}x\in G_W$, $WT_{i,a}x=Wx$ and $r(T_{i,a}x)=\upd{r(x)}{i}{a}$. Then $\Dist^\infty_\mu\le\mu(S\setminus A)$, with no condition on $|S|$. If $2^m\mid|S|$, then $\Dist_\mu\le\mu(S\setminus A)+\Delta_A$. In both cases one composable interface agrees with all readers and setters at once outside an event of at most this probability.
\end{lemma}

\begin{proof}
For $\Dist^\infty_\mu$: by Lemma~\ref{lem:reversible}, the states of $G_W$ in one fiber have distinct labels. Additional states complete each such set, and each state of $S\setminus G_W$, to a cube of its own, with the states of $S$ at the positions of their labels. Coordinate setters on these cubes define a composable interface on a finite set containing $S$, whose readers agree with $r$ on $S$. For $x\in A$, $T_{i,a}x$ is the state of the cube of $x$ with the label $\upd{r(x)}{i}{a}$, which is the output of the composable setter. For $\Dist_\mu$: apply Lemma~\ref{lem:weighted} with $G=G_W$ and the first choice of $\kappa$, by Lemma~\ref{lem:reversible}. Its composable interface agrees with $I$ on $A\cap G'$.
\end{proof}
\section{Proof of Theorem~\ref{thm:audit}(a) and (b)}\label{app:audit}

Throughout, $\hat a$ is a target label, $W_{\hat a}=T_{m,\hat a_m}\sdots T_{1,\hat a_1}$, $U_j=T_{j,\hat a_j}\sdots T_{1,\hat a_1}$ and $U_0=\Id$. A \emph{checked step} from a state $z$ under $T_{j,\hat a_j}$ calls $z'=T_{j,\hat a_j}z$ and checks $r(z')=\upd{r(z)}{j}{\hat a_j}$. With $b=r_j(z)$, it then compares $z'$ with $z$ if $b=\hat a_j$, and otherwise calls $T_{j,b}z'$ and compares the output with $z$. Each checked step makes at most two calls, and a passed one satisfies~\eqref{eq:goodfull}. A round stops at its first failed check. On the path of $y_i$, step $i$ compares its output with the state $x_i=U_ix$ of the path of $x$ before it calls the undo. Part~(c) is proved in Appendix~\ref{app:necessity}.

\begin{algorithm2e}[!b]
\caption{One audit round with target label $\hat a$ and depth cap $d$. A failed check rejects.}\label{alg:round}
\DontPrintSemicolon\SetAlgoNoEnd
draw $x\sim\mu$; put $x_0=x$\;
\lIf(\tcp*[f]{first test}){$d\le m$ and $|r(x)\oplus\hat a|>d-2$}{accept the round}
\For(\tcp*[f]{path check}){$s=1,\dots,m$}{
put $b_s=r_s(x_{s-1})$; call $x_s=T_{s,\hat a_s}x_{s-1}$ and check its label; if $\hat a_s=b_s$, compare $x_s$ with $x_{s-1}$; otherwise call the undo $T_{s,b_s}x_s$ and compare its output with $x_{s-1}$\;
}
\For(\tcp*[f]{flip check}){$i=1,\dots,m$}{
check $T_{i,r_i(x)}x=x$; put $y_i=F_ix$; check $r(y_i)=\upd{r(x)}{i}{1-r_i(x)}$\;
\For{$s=1,\dots,i$}{
do step $s$ of the path of $y_i$ as above; if $s=i$, compare its output with $x_i$ before any undo\;
}
}
\end{algorithm2e}

\subsection{Path and flip checks}\label{app:auditroot}

\begin{lemma}[Checked setters]\label{lem:checked}
Let $x$ pass the path check and the flip checks of a round with target label $\hat a$. Then $x\in G_{W_{\hat a}}$, and every setter $T_{i,e}$ satisfies $T_{i,e}x\in G_{W_{\hat a}}$, $W_{\hat a}T_{i,e}x=W_{\hat a}x$ and $r(T_{i,e}x)=\upd{r(x)}{i}{e}$.
\end{lemma}

\begin{proof}
The path check gives $x\in G_{W_{\hat a}}$. If $e=r_i(x)$, the check of write-back gives $T_{i,e}x=x$, and the claims follow. Otherwise $T_{i,e}x=y_i$, and the check of its label gives the third claim. The path of $y_i$ satisfies~\eqref{eq:goodfull} at steps $1,\dots,i$ and reaches the same state as the path of $x$ at step $i$. From step $i$ on, the two paths coincide, so the path of $y_i$ satisfies~\eqref{eq:goodfull} at the later steps as well. Hence $y_i\in G_{W_{\hat a}}$ and $W_{\hat a}y_i=W_{\hat a}x$.
\end{proof}

\begin{lemma}[Depth of the path and flip checks]\label{lem:rootdepth}
Let $\varrho=|r(x)\oplus\hat a|$. The calls of the path and flip checks on $x$ have depth at most $\min\{\varrho+2,m+1\}$ when they are made, and there are at most $m^2+5m$ of them.
\end{lemma}

\begin{proof}
Let $\varrho_j$ be the number of $h\le j$ with $r_h(x)\ne\hat a_h$. On the path of $x$, a passed step that writes the old value is a self-loop that the tester has seen, so $x_j$ has depth at most $\varrho_j$ when step $j+1$ is made. Hence the calls of the path check have depth at most $\varrho+1\le m+1$. Let $j<i$. Before step $j$ of the path of $y_i$, the flip and the earlier passed steps give a chain of length at most $1+\varrho_{j-1}$, so this step and its undo have depth at most $\varrho_j+2$. As $\varrho_j\le\min\{\varrho,m-1\}$, this is at most $\min\{\varrho+2,m+1\}$. Step $i$ itself has depth at most $\varrho_{i-1}+2$. It is followed by a comparison with $x_i$, so its undo acts on a state of depth at most $\varrho_i$ and has depth at most $\varrho_i+1$. These bounds use only comparisons that passed before the call. The path check makes at most $2m$ calls, the write-backs and flips $2m$, and the paths of the states $y_i$ at most $\sum_i2i=m(m+1)$, in total $m^2+5m$.
\end{proof}

\subsection{The bridge check}\label{app:auditbridge}

On the same states, the $M$ heaviest fibers lose at most the probability that $x$ passes the checks above and a uniform label $u$ is missing from its fiber (Lemma~\ref{lem:structure}). The \emph{bridge check} pays for this loss by showing that a state of label $u$ lies in the fiber of $x$. It runs after the path and flip checks of Algorithm~\ref{alg:round}, as follows.
\begin{enumerate}[nosep,leftmargin=*]
\item Draw a uniform label $u$, and let $u^{(j)}=(\hat a_1,\dots,\hat a_j,u_{j+1},\dots,u_m)$ for $0\le j\le m$.
\item If some $|u^{(j)}\oplus r(x)|>d-1$, accept the round (the \emph{second test}).
\item For $j=0,\dots,m-1$: from $x$, write the values of $u^{(j)}$ where it differs from $r(x)$, to get $w_j$, and check $r(w_j)=u^{(j)}$.
\item Put $w_m=W_{\hat a}x$, and check $T_{j,\hat a_j}w_{j-1}=w_j$ for $j=1,\dots,m$.
\end{enumerate}
Applying $W_{\hat a}$ to $w_0$ directly could take depth $2m$. The bridge check instead reaches each intermediate state $w_j$ from $x$ by a word of length at most $m$ and checks one setter between neighbors, so its depth is at most $m+1$.

\begin{lemma}[Bridge check]\label{lem:bridge}
If the bridge check of a round passes, then $w_0$ reads $u$ and lies in the fiber of $x$ under $W_{\hat a}$. The bridge check makes at most $m^2+m$ calls, and with a depth cap $d$, its calls have depth at most $d$.
\end{lemma}

\begin{proof}
The label check gives $r(w_0)=u^{(0)}=u$, and the comparisons give $W_{\hat a}w_0=w_m=W_{\hat a}x$. Each $w_j$ with $j<m$ takes at most $m$ calls, and there are $m$ comparisons with a call each. The word that produces $w_j$ has length $|u^{(j)}\oplus r(x)|\le d-1$, so its calls have depth at most $d-1$, whatever their outputs are, and a bridge call on $w_{j-1}$ has depth at most $d$. For a depth cap $d=m+1$ the length is at most $m=d-1$ for every $u$ and $\hat a$.
\end{proof}

\subsection{Proof of parts (a) and (b)}\label{app:auditproof}

\begin{proof}[Proof of Theorem~\ref{thm:audit}(a) and (b)]
\emph{One-sided error.} In a composable interface every path satisfies~\eqref{eq:goodfull}, writing the current value fixes the state, and a flip changes only coordinate $i$. The paths of $x$ and $y_i$ reach the same state at step $i$, the first write to $i$. Each $w_j$ is the state of the cube of $x$ with the label $u^{(j)}$, and $T_{j,\hat a_j}$ takes $u^{(j-1)}$ to $u^{(j)}$. So every check passes, and a round that stops early accepts.

\emph{Depth and calls.} Lemma~\ref{lem:rootdepth} bounds the depth by $m+1$ without the first test of Algorithm~\ref{alg:round}, and by $d$ with it. Lemma~\ref{lem:bridge} bounds the bridge check. So a round has depth at most $d$, with at most $m^2+5m$ calls without the bridge check and at most $2m^2+6m$ with it.

\emph{Part (a).} Let $d=m+1$, fix $\hat a$, and let $A$ be the set of states that pass the path and flip checks. By Lemma~\ref{lem:checked}, Lemma~\ref{lem:structure} applies with $W=W_{\hat a}$. So $\Dist^\infty_\mu\le\mu(S\setminus A)$, which is the rejection probability of a round without the bridge check. For $\Dist_\mu$, the label $u$ is uniform and independent of $x$. If $x\in A$ and $u$ occurs nowhere in the fiber of $x$, the bridge check rejects, by Lemma~\ref{lem:bridge}. So a round rejects with probability at least $\mu(S\setminus A)+\Delta_A\ge\Dist_\mu$. If the distance is at least $\epsilon$, then $\lceil\log(1/\delta)/\epsilon\rceil$ independent rounds all pass with probability at most $\delta$.

\emph{Part (b).} Let $2\le d\le m$; for $d=m+1$ the bounds follow from part (a). For a fixed target label $\hat a$, let $A(\hat a)$ be the set of states that pass the first test of Algorithm~\ref{alg:round} and the path and flip checks. By Lemma~\ref{lem:structure}, $\Dist^\infty_\mu\le\mu(S\setminus A(\hat a))$. Averaged over the uniform target label, the probability that the first test ends the round is $\tau_2$, as $r(x)\oplus\hat a$ is uniform. So $\Dist^\infty_\mu\le p+\tau_2$. For $\Dist_\mu$, Lemma~\ref{lem:structure} gives $\Dist_\mu\le\mu(S\setminus A(\hat a))+\Delta_{A(\hat a)}$. If $x\in A(\hat a)$ and $u$ occurs nowhere in the fiber of $x$, then either the second test ends the round or the bridge check rejects, by Lemma~\ref{lem:bridge}. For fixed $x$ and $j$, the label $u^{(j)}$ is uniform, as $\hat a$ and $u$ are, so the second test ends the round with probability at most $(m+1)\tau_1$. Rejections by the two kinds of checks are disjoint events, so $\Dist_\mu\le p+\tau_2+(m+1)\tau_1\le p+(m+2)\tau_2$. If the tail is at most $\epsilon/2$ and the distance at least $\epsilon$, then $p\ge\epsilon/2$, and $\lceil2\log(1/\delta)/\epsilon\rceil$ independent rounds all pass with probability at most $\delta$.
\end{proof}

\subsection{Lower bounds for every tester}\label{app:lower}

\begin{proposition}[Every tester]\label{prop:auditlower}
Let $0<\epsilon\le1/16$, $0<\delta\le1/3$ and $N_{\delta,\epsilon}=\lceil\log(1/\delta)/(2\epsilon)\rceil+2$. Let a tester, whenever $\mu$ is the uniform measure on a set of $2^mN_{\delta,\epsilon}$ states, accept composable interfaces and reject interfaces with $\Dist^\infty_\mu>\epsilon$, each with probability at least $1-\delta$. Then it uses, in expectation, at least $\log(1/\delta)/(48\epsilon)$ samples and at least $(m/(48\epsilon))\log(1/\delta)$ setter calls on some composable interface on such a set, even if reader calls and comparisons are free.
\end{proposition}

\begin{proof}[Proof of Proposition~\ref{prop:auditlower}]
\emph{Instances.} Let $N_0=\lfloor2\epsilon N_{\delta,\epsilon}\rfloor+1$ and $\tilde\epsilon=N_0/N_{\delta,\epsilon}$. As $\log(1/\delta)\ge\log3>1$ and $N_{\delta,\epsilon}\ge\log(1/\delta)/(2\epsilon)$, $2\epsilon<\tilde\epsilon\le2\epsilon+1/N_{\delta,\epsilon}\le4\epsilon\le1/4$ and $1\le N_0\le N_{\delta,\epsilon}$. The set of states is $\{0,1\}^m\times[N_{\delta,\epsilon}]$ with cubes $\{0,1\}^m\times\{z\}$, the measure $\nu$ is uniform, and $r_i(u,z)=u_i$. The null input has $T_{i,a}(u,z)=(\upd{u}{i}{a},z)$. A negative input, indexed by a variable $j_*$ and an $N_0$-set $\cZ_0\subseteq[N_{\delta,\epsilon}]$, makes both setters of $j_*$ the identity on the cubes with $z\in\cZ_0$. On each cube the random-edit kernel averages kernels that resample one coordinate of $u$ uniformly, one of them possibly replaced by the identity, so $\nu$ is stationary for every input. Identity setters satisfy write-back, overwrite and preservation and commute with setters that keep the cube, so only success fails, on the states of the cubes with $z\in\cZ_0$ and $u_{j_*}=1-a$. At each state of such a cube, a composable interface, on the set of states or on any finite set containing it, moves the state by exactly one of its two setters of variable $j_*$, while both original setters fix the state. So the two disagreements sum to at least $\tilde\epsilon$. As a composable interface on the same states, the null input differs from the negative input only in the two setters of $j_*$, each with probability $\tilde\epsilon/2$, so $\Dist^\infty_\nu=\Dist_\nu=\tilde\epsilon/2\in(\epsilon,2\epsilon]$.

\emph{Avoidance probability.} For an integer $0\le v\le N_{\delta,\epsilon}$, let $\mathrm{Av}(v)$ be the probability that a uniformly random $N_0$-subset of $[N_{\delta,\epsilon}]$ misses a given $v$-subset, and let $\overline{\mathrm{Av}}$ be its linear interpolation on $[0,N_{\delta,\epsilon}]$, extended by zero. The decrement $\mathrm{Av}(v)-\mathrm{Av}(v+1)\ge0$ is the probability of missing $v$ given points and containing another one, which does not increase in $v$. The slope $0$ of the zero extension is at least the last slope of the interpolation. So $\overline{\mathrm{Av}}$ is nonincreasing and convex. For integers $0\le v\le\lfloor N_{\delta,\epsilon}/2\rfloor$, each factor of $\mathrm{Av}(v)=\prod_{v'<v}(1-N_0/(N_{\delta,\epsilon}-v'))$ satisfies $N_0/(N_{\delta,\epsilon}-v')\le2\tilde\epsilon\le1/2$, so $\mathrm{Av}(v)\ge(1-2\tilde\epsilon)^v\ge\eu^{-4\tilde\epsilon v}$, as $-\log(1-2\tilde\epsilon)\le4\tilde\epsilon$ for $2\tilde\epsilon\le1/2$. By convexity of the exponential, $\overline{\mathrm{Av}}(v)\ge\eu^{-4\tilde\epsilon v}$ also for real $0\le v\le\lfloor N_{\delta,\epsilon}/2\rfloor$.

\emph{Coupling.} Draw $j_*$ uniformly from $[m]$ and $\cZ_0$ uniformly among the $N_0$-subsets of $[N_{\delta,\epsilon}]$, independently of the tester's randomness and of the samples, and execute the tester on the null input; $\Pr_0$ and $\E_0$ refer to this execution. Let $n_T$ and $n_S$ be its numbers of setter calls and samples. Let $\cV_i$ be the set of cubes in which it calls a setter of variable $i$, so $\sum_i|\cV_i|\le n_T$, and let $\cV_0$ be the set of cubes that contain a sample, so $|\cV_0|\le n_S$. Couple the execution on the negative input $(j_*,\cZ_0)$, using the same randomness and samples. Setters keep the cube of their argument, so every held state lies in a cube of $\cV_0$. If $\cZ_0$ misses $\cV_{j_*}$, or if it misses $\cV_0$, then every setter call has the same output on both inputs, and, inductively, the transcripts and decisions coincide, as readers and comparisons agree on equal states. Conditional on the null execution, $\cZ_0$ misses $\cV_{j_*}$ with probability $m^{-1}\sum_i\mathrm{Av}(|\cV_i|)\ge\overline{\mathrm{Av}}(m^{-1}\sum_i|\cV_i|)\ge\overline{\mathrm{Av}}(n_T/m)$, by convexity and monotonicity of $\overline{\mathrm{Av}}$, and it misses $\cV_0$ with probability $\mathrm{Av}(|\cV_0|)\ge\overline{\mathrm{Av}}(n_S)$. Let $\mathrm{acc}$ be the event that the null input is accepted, so $\Pr_0(\mathrm{acc})\ge1-\delta$. Soundness on every negative input, averaged over $(j_*,\cZ_0)$, gives $\delta\ge\E_0[\one_{\mathrm{acc}}\,\overline{\mathrm{Av}}(n_T/m)]\ge\Pr_0(\mathrm{acc})\,\overline{\mathrm{Av}}(\bar n_T)$ for $\bar n_T=\E_0[n_T\mid\mathrm{acc}]/m$, by conditional Jensen, and in the same way $\delta\ge\Pr_0(\mathrm{acc})\,\overline{\mathrm{Av}}(\bar n_S)$ for $\bar n_S=\E_0[n_S\mid\mathrm{acc}]$. Both hold trivially if the expectations are infinite.

\emph{Extracting $\log(1/\delta)$.} We claim $\bar n_T\ge\log((1-\delta)/\delta)/(4\tilde\epsilon)$. As $1-\delta<1$ and $\tilde\epsilon>2\epsilon$, the right side is at most $\log(1/\delta)/(8\epsilon)\le(N_{\delta,\epsilon}-1)/2\le\lfloor N_{\delta,\epsilon}/2\rfloor$. So the claim holds if $\bar n_T>\lfloor N_{\delta,\epsilon}/2\rfloor$, and otherwise it follows from $\delta\ge(1-\delta)\eu^{-4\tilde\epsilon\bar n_T}$. The same argument gives $\bar n_S\ge\log((1-\delta)/\delta)/(4\tilde\epsilon)$. As $1-\delta\ge2/3$, $\tilde\epsilon\le4\epsilon$ and $\log((1-\delta)/\delta)\ge\log(1/\delta)/2$ by $(1-\delta)^2\ge\delta$, we get $\E_0n_T\ge\Pr_0(\mathrm{acc})\,m\bar n_T\ge(m/(48\epsilon))\log(1/\delta)$ and $\E_0n_S\ge\Pr_0(\mathrm{acc})\,\bar n_S\ge\log(1/\delta)/(48\epsilon)$.
\end{proof}
\section{Proof of the sufficiency in Theorem~\ref{thm:budget}}\label{app:repair}

Throughout Appendices~\ref{app:kernel}--\ref{app:repaircoeff}, $|S|=M2^m$, $\lambda\ge1$, $\mu$ is a $\lambda$-subinvariant probability measure on $S$ and $\gamma=\gamma_\mu$. Let $\cU\subseteq\cN$ be the set of states at which success, write-back, overwrite or preservation fails. For $v\ge m$, put $\Phi_v(\lambda)=h_v^{-1}\sum_{k<v}\lambda^k$, with $h_v$ as in Lemma~\ref{lem:coverage}. For a composable interface $\tilde I$, let $\cE_\mu(\tilde I)$ be the probability that some reader or setter of $I$ differs from that of $\tilde I$ at $x\sim\mu$, so $\Dist_\mu\le\cE_\mu(\tilde I)$.

\begin{theorem}[Parametric form]\label{thm:repairparam}
For $m\ge1$ and all integers $t,t'\ge m$, some composable interface $\tilde I$ on $S$ has
\[
\cE_\mu(\tilde I)\le\min\Big\{1,\ \Phi_t(\lambda)\gamma+\frac{\lambda^t}{h_t}\Phi_{t'}(\lambda)\mu(\cU)\Big\}.
\]
\end{theorem}

The proof fixes a word, takes its fibers as candidate cubes, and applies Lemma~\ref{lem:weighted}. To pay for the labels missing from a fiber, it uses a function $\kappa^\circ$ defined once from the interface alone: the capacity function of Appendix~\ref{app:capacityfn}.

\subsection{Retained steps}\label{app:kernel}

For a setter $T_{i,a}$, let $\Pk_{i,a}$ be the kernel with row zero or one defined as follows. If $a=r_i(x)$, it keeps the transition $x\to T_{i,a}x$ when $T_{i,a}x=x$. If $a\ne r_i(x)$, it keeps $x\to y=T_{i,a}x$ when $r(y)=\upd{r(x)}{i}{a}$ and $T_{i,r_i(x)}y=x$. Otherwise the row of $x$ is zero. Put $\Pk_i=(\Pk_{i,0}+\Pk_{i,1})/2$ and $\Pk=m^{-1}\sum_i\Pk_i$. Each kept transition passes~\eqref{eq:goodfull}.

Each $\Pk_i$ is symmetric: if $x\to y$ is a kept transition with $y\ne x$, then the labels of $x$ and $y$ differ exactly at $i$, and $y\to x$ is kept with the setter of the old value, so both entries are $1/2$. Entrywise, $\Pk_i\le(T_{i,0}+T_{i,1})/2$ and $\Pk\le P$. For $x\notin\cU$, write-back keeps the setter of the current value, and success, preservation, overwrite and write-back keep the other, so $\Pk_i\one(x)=1$ for all $i$. Hence
\begin{equation}\label{eq:keptrow}
0\le\one-\Pk\one\le\one_\cU.
\end{equation}

\subsection{The capacity function}\label{app:capacityfn}

Draw $t'$ independent uniform coordinates $i_1,\dots,i_{t'}$, condition on the event that they cover $[m]$, and put $\Upsilon=\E[\Pk_{i_1}\sdots\Pk_{i_{t'}}\mid\text{cover}]$ and $\kappa^\circ=\Upsilon\one$. The product applies $\Pk_{i_1}$ first. The transpose of a product reverses the word, and given coverage a uniform word has the same distribution as its reverse. So $\Upsilon$ is symmetric, and $0\le\kappa^\circ\le1$.

Fix a start state and a covering word of coordinates, and attach independent fair values to its steps. The label that these values write is uniform, as each coordinate takes its last value, and a path that survives all kept transitions ends at this label. So $\Upsilon(x,r^{-1}(b))\le2^{-m}$ for every state $x$ and label $b$. By symmetry,
\begin{equation}\label{eq:capacity}
\sum_{z:\,r(z)=b}\kappa^\circ(z)=\sum_{x\in S}\Upsilon(x,r^{-1}(b))\le|S|2^{-m}=M .
\end{equation}
By~\eqref{eq:keptrow} and $\Pk\le P$, $\mu(\one-\Pk^{t'}\one)=\sum_{k<t'}\mu\Pk^k(\one-\Pk\one)\le\sum_{k<t'}\mu P^k(\cU)\le\sum_{k<t'}\lambda^k\mu(\cU)$. Now $\Pk^{t'}\one$ is the expectation of $\Pk_{i_1}\sdots\Pk_{i_{t'}}\one$ over the coordinates. Splitting it on coverage, and bounding the product by $\one$ when the coordinates do not cover $[m]$, gives $\Pk^{t'}\one\le h_{t'}\kappa^\circ+(1-h_{t'})\one$, so $\one-\kappa^\circ\le h_{t'}^{-1}(\one-\Pk^{t'}\one)$. Integrating against $\mu$ gives
\begin{equation}\label{eq:capacitymass}
\mu(\one-\kappa^\circ)\le\Phi_{t'}(\lambda)\mu(\cU).
\end{equation}

\subsection{Proof of Theorem~\ref{thm:repairparam}}\label{app:repairproof}

\begin{proof}[Proof of Theorem~\ref{thm:repairparam}]
Fix a word $W=E_t\sdots E_1$ that writes every variable, with path $x_s=U_sx$, and let $A_W=\{x:x_j\notin\cN\text{ for }0\le j<t\}$. By Lemma~\ref{lem:insert}, each $x\in A_W$ lies in $G_W$, and each of its setters leads to a state of $G_W$ in its fiber with the updated label. On $G_W$ the map $x\mapsto(r(x),Wx)$ is injective (Lemma~\ref{lem:reversible}). Every $x\in G_W$ has the same end label, the label $b_W$ that $W$ writes. So the points $z$ with $w_z=\mu(A_W\cap W^{-1}(z))>0$ are distinct states with label $b_W$, and~\eqref{eq:capacity} gives $\sum_{z:\,w_z>0}\kappa^\circ(z)\le M$. Lemma~\ref{lem:weighted} with $G=G_W$, $A=A_W$ and $\kappa=\kappa^\circ$ gives a composable interface that agrees with all readers and setters of $I$ on $A_W\cap G'$, so
\begin{equation}\label{eq:wordcharge}
\cE_\mu(\tilde I)\le\mu(S\setminus A_W)+\sum\nolimits_zw_z(1-\kappa^\circ(z)).
\end{equation}
Now draw the steps of $W$ as $t$ independent uniform setters conditioned on writing every variable, with $x\sim\mu$ independent of them. Before conditioning, $x_k$ has distribution $\mu P^k\le\lambda^k\mu$, so
\[
\E[\mu(S\setminus A_W)\mid\text{cover}]\le h_t^{-1}\sum\nolimits_{k<t}\mu P^k(\cN)\le\Phi_t(\lambda)\gamma .
\]
The last term of~\eqref{eq:wordcharge} is the expectation of $\one_{A_W}(x)(1-\kappa^\circ(Wx))$ over $x$. Dropping $\one_{A_W}$ and the coverage from the numerator, and using~\eqref{eq:capacitymass},
\[
\E\Big[\sum\nolimits_zw_z(1-\kappa^\circ(z))\,\Big|\,\text{cover}\Big]\le h_t^{-1}\mu P^t(\one-\kappa^\circ)\le\frac{\lambda^t}{h_t}\mu(\one-\kappa^\circ)\le\frac{\lambda^t}{h_t}\Phi_{t'}(\lambda)\mu(\cU).
\]
So some word attains at most the sum of these two bounds.
\end{proof}

\subsection{The coefficients}\label{app:repaircoeff}

Let $m\ge2$, $t=t'=\tm$, so $h_t\ge1/2$ (Lemma~\ref{lem:coverage}), and recall $\mu(\cU)\le\gamma$.

\emph{The case $\lambda\le1+1/(\alpha m)$.} Such a $\mu$ is also $(1+1/(\alpha m))$-subinvariant, so take $\lambda=1+1/(\alpha m)$, and put $\bar\lambda=\lambda^t\ge1$. Then $\Phi_t(\lambda)=\alpha m(\bar\lambda-1)/h_t$, and Theorem~\ref{thm:repairparam} gives
\begin{equation}\label{eq:repairalpha}
\cE_\mu(\tilde I)\le\Phi_t(\lambda)\Big(1+\frac{\bar\lambda}{h_t}\Big)\gamma\le2\alpha m(\bar\lambda-1)(1+2\bar\lambda)\gamma\le4\alpha m\,\bar\lambda^2\gamma,\qquad \bar\lambda\le\eu^{t/(\alpha m)}\le(2\eu m)^{1/\alpha},
\end{equation}
as $t\le m\log(2m)+1$ and $1/m\le1$. This is at most $K_\alpha\gamma$ with $K_\alpha=4\alpha(2\eu)^{2/\alpha}m^{1+2/\alpha}$.

\subsection{Law checks, for every set of states}\label{app:padding}

\begin{lemma}[Padding]\label{lem:padding}
Let $I$ be an interface on a finite set $S$, with no condition on $|S|$, and $\mu$ a measure on $S$. Add $(-|S|)\bmod2^m$ states with arbitrary readings and setters that stay among the added states, and extend $\mu$ by zero to $\bar\mu$. The new interface $\bar I$ on $\bar S$ satisfies $\gamma_{\bar\mu}(\bar I)=\gamma_\mu(I)$, $\bar\mu$ is $\lambda$-subinvariant if $\mu$ is, and $\Dist^\infty_\mu(I)\le\Dist_{\bar\mu}(\bar I)$. Extending a start distribution by zero extends its sample distributions by zero.
\end{lemma}

\begin{proof}
The set $S$ is closed under the setters, and the laws at its states involve only states of $S$. Added states have mass zero. A composable interface on $\bar S$ is a candidate for $\Dist^\infty_\mu(I)$ with the same disagreements under $\mu$. Let $\bar P$ be the random-edit kernel of $\bar I$. As $S$ is closed under the setters, $\bar P$ acts on $S$ as $P$, so $\bar\mu\bar P$ is the extension of $\mu P$ by zero. Hence $\mu P\le\lambda\mu$ gives $\bar\mu\bar P\le\lambda\bar\mu$: on $S$ this is the hypothesis, and on the added states both sides vanish. By induction, if $\bar\nu$ extends a start distribution $\nu$ by zero, then $\bar\nu\bar P^s$ extends $\nu P^s$ by zero for every $s\ge0$. So, by~\eqref{eq:access}, the sample distribution of $\bar\nu$ extends $\mu_\ell$ by zero for every $\ell\ge0$.
\end{proof}

\begin{corollary}[Long runs]\label{cor:longruns}
Let $m\ge2$, $\alpha>0$, $\ell\ge\alpha m$ and $K_\alpha=4\alpha(2\eu)^{2/\alpha}m^{1+2/\alpha}$. Then $\Dist_\ell\le K_\alpha\gamma_\ell$ when $2^m\mid|S|$, and $\Dist^\infty_\ell\le K_\alpha\gamma_\ell$ for every finite $S$. So evaluating all law instances at $\lceil K_\alpha\log(1/\delta)/\epsilon\rceil$ samples of $\mu_\ell$ certifies these distances, with depth $2$ and $2m^2+2m$ setter calls per sample.
\end{corollary}

\begin{proof}
By Lemma~\ref{lem:resolvent}(i), $\mu_\ell$ is $(1+1/\ell)$-subinvariant, so~\eqref{eq:repairalpha} and Lemma~\ref{lem:padding} give the two bounds, and Appendix~\ref{app:lawcheck} gives the tester.
\end{proof}

\noindent Corollary~\ref{cor:longruns} and~\eqref{eq:repairalpha} are part~(b) of Theorem~\ref{thm:budget}.

\subsection{The coefficient at mean \texorpdfstring{$\alpha m$}{αm}}\label{app:coefficient}

For $\alpha>0$, put
\[
C_\alpha=24\cdot8^{1/(2\alpha)}\,\Gamma\big(1+\tfrac1{2\alpha}\big)+\Big(\frac{32(1+1/(2\alpha))}{\eu}\Big)^{1+1/(2\alpha)},
\]
$b_\alpha=1-\eu^{-1/(8\alpha+1)}$ and $c_\alpha=b_\alpha/(128\,C_\alpha\,16^{1/(2\alpha)})$.

\begin{proposition}[Coefficient at mean $\alpha m$]\label{prop:coefflower}
For $m\ge32$ and each $\alpha>0$, some interface and start distribution have $\Dist^\infty_{\alpha m}\ge c_\alpha m^{1/(2\alpha)}\gamma_{\alpha m}>0$.
\end{proposition}

\begin{proof}
Let $m\ge32$ and take the unmasked twisted interface $I^-$ of Appendix~\ref{app:instance} with $J=\lfloor m/8\rfloor\ge4$ and $N_c=1$, which does not depend on $\alpha$. Let $\ell=\alpha m$. By Lemma~\ref{lem:rectdistance}, $\Dist^\infty_\ell\ge((J+1)/(128J^2))(1-q_0^J)$ with $q_0=8\alpha J/(1+8\alpha J)$. As $\log(1/q_0)\ge1/(1+8\alpha J)$, $q_0^J\le\eu^{-1/(8\alpha+1)}$, so $\Dist^\infty_\ell\ge b_\alpha/(128J)$. It remains to bound $\gamma_\ell=\mu_\ell(\cN)$, where $\cN$ is given by Lemma~\ref{lem:rectlaws} and $\gamma_\ell>0$, as $\mu_\ell$ has full support: for $\ell>0$ a run reaches every label with positive probability, and by Lemma~\ref{lem:rectmeasure} its hidden bit is a fair bit independent of the final label.

Let the edits arrive at the times of a Poisson process of rate $1$ and stop at an independent time $\iota$ of density $\ell^{-1}\eu^{-t/\ell}$; this gives $\mu_\ell$ (Appendix~\ref{app:onestatehist}). Given $\iota=t$, the numbers of edits of the variables are independent and Poisson of mean $t/m$, with fair written values. So each final bit is its initial bit shifted by an independent Bernoulli variable of parameter $(1-y)/2$, where $y=\eu^{-t/m}$, and $\eu^{-\iota/m}$ has density $\alpha^{-1}y^{1/\alpha-1}$ on $(0,1)$. Given the start label and $y$, the two blocks are independent.

For a sum $Z$ of $N_1$ independent Bernoulli variables, Hoeffding's inequality gives $\Pr(Z-\E Z\ge a)\le\eu^{-2a^2/N_1}$ for $a\ge0$, and the same for the lower tail. If all parameters lie in $[1/4,3/4]$, then $\Pr(Z=h)\le\sqrt{2\pi/(3N_1)}$ for every integer $h$: Fourier inversion gives $\Pr(Z=h)\le(2\pi)^{-1}\int_{-\pi}^\pi|\E\eu^{\mathrm{i}\theta Z}|d\theta$, each factor has squared modulus $1-4p(1-p)\sin^2(\theta/2)\le\eu^{-(3/4)\sin^2(\theta/2)}$, $|\sin(\theta/2)|\ge|\theta|/\pi$ on $[-\pi,\pi]$, and a Gaussian integral finishes the bound.

Let $0<y\le1/2$. The parameters lie in $[1/4,3/4]$, so a block of $4J$ bits lies in $\{2J-1,2J\}$ with probability at most $2/\sqrt J$. A start label has a frozen block, of initial weight $J$ or $3J$, whose final weight $Z$ has mean $2J\mp Jy$. Split its bits into two parts of $2J$, with sums $Z_1$ and $Z_2$, which are independent, and let $a=|h-\E Z|$. If $Z=h$, then $Z_1$ or $Z_2$ deviates from its mean by at least $a/2$ in the direction of $h-\E Z$. By Hoeffding's inequality with $N_1=2J$, $Z_1$ does so with probability at most $\eu^{-2(a/2)^2/(2J)}=\eu^{-a^2/(4J)}$. Given such a value of $Z_1$, the event $Z=h$ fixes $Z_2$, which by the point bound with $N_1=2J$ has probability at most $\sqrt{\pi/(3J)}\le2/\sqrt J$. The same holds with the two parts exchanged, and adding the two cases gives $\Pr(Z=h)\le(4/\sqrt J)\eu^{-a^2/(4J)}$. For $h\in\{2J-1,2J\}$, $a\ge(Jy-1)_+$, and $(x-1)_+^2\ge x^2/2-1$ gives $\Pr(Z\in\{2J-1,2J\})\le(12/\sqrt J)\eu^{-Jy^2/8}$, as $\eu^{1/(4J)}<3/2$. With the other block, $\Pr(\cN\mid y)\le(24/J)\eu^{-Jy^2/8}$. For $1/2<y<1$, the frozen block has to deviate from its mean by at least $J/2-1\ge J/4$ in a fixed direction, so $\Pr(\cN\mid y)\le\eu^{-J/32}$. Integrating over $y$, with $v=Jy^2/8$,
\[
\gamma_\ell\le\frac{24}{\alpha J}\int_0^\infty y^{1/\alpha-1}\eu^{-Jy^2/8}\,dy+\eu^{-J/32}=24\cdot8^{1/(2\alpha)}\,\Gamma\big(1+\tfrac1{2\alpha}\big)J^{-1-1/(2\alpha)}+\eu^{-J/32}.
\]
As $x^{e'}\eu^{-x/32}\le(32e'/\eu)^{e'}$ for $x,e'>0$, taking $e'=1+1/(2\alpha)$ gives $\gamma_\ell\le C_\alpha J^{-1-1/(2\alpha)}$. So $\Dist^\infty_\ell/\gamma_\ell\ge(b_\alpha/(128C_\alpha))J^{1/(2\alpha)}\ge c_\alpha m^{1/(2\alpha)}$, as $J\ge m/16$.
\end{proof}
\section{The twisted rectangle}\label{app:twist}

This appendix proves the necessity in Theorems~\ref{thm:budget} and~\ref{thm:samplesonly}, part~(c) of Theorem~\ref{thm:audit}, and the depth bound at fixed thresholds of Section~\ref{sec:scope}. All of them use one family of interfaces.

\subsection{A small instance}\label{app:small}

The smallest member of the family has $J=1$. Take $m=8$, blocks $B'=\{1,2,3,4\}$ and $B''=\{5,6,7,8\}$, and states $(v,z)$ with $v\in\{0,1\}^8$ and a hidden bit $z$. Readers read $v$, and $v'$ and $v''$ are the restrictions of $v$ to the blocks. A setter of $B''$ replaces its coordinate. A setter of $B'$ replaces its coordinate, and it also toggles $z$ when it moves $|v'|$ across the threshold, between weights $1$ and $2$, while $|v''|\ge2$. Commutation between the blocks fails exactly where both weights lie in $\{1,2\}$, and all other law instances hold everywhere (Lemma~\ref{lem:rectlaws}).

Start at the label with ones at variables $1$ and $5$, so at weights $(|v'|,|v''|)=(1,1)$. The loop has four segments, with eight writes in all.
\begin{itemize}[nosep,leftmargin=1.5em]
\item Bottom: write $1$ to variables $2$ and $3$. The weights go $(1,1)\to(2,1)\to(3,1)$, and $z$ stays, as $|v''|=1$.
\item Right: write $1$ to variables $6$ and $7$. The weights go $(3,1)\to(3,2)\to(3,3)$.
\item Top: write $0$ to variables $3$ and $2$. The weights go $(3,3)\to(2,3)\to(1,3)$, and the second step toggles $z$, as $|v''|=3$.
\item Left: write $0$ to variables $7$ and $6$. The weights go $(1,3)\to(1,2)\to(1,1)$.
\end{itemize}
So the loop returns to its start label with $z$ toggled, and a composable interface that kept all eight writes would return to the start state (Lemma~\ref{lem:forcing}). The start distribution of Appendix~\ref{app:instance} draws the arguments of such loops, with random choices of the variables, so each of the $16$ setters of the blocks carries at most $8J/(J+1)=4$ times the start distribution. Every loop has a disagreement, so $1\le4\cdot16\cdot R$ for the disagreement $R$ of any candidate, and at $\ell=0$ the distance is at least $1/64$ (Lemma~\ref{lem:rectdistance}). For $J=1$ some start states meet the set where commutation fails, and for $J\ge2$ the frozen block keeps all of them outside it.

\subsection{The instances}\label{app:instance}

Let $J,N_g,N_c\ge1$ be integers and $n=8J$, with $nN_g\le m$. The states are the triples $(\sigma,v,z)\in[N_c]\times\{0,1\}^m\times\{0,1\}^{N_g}$, so there are $N_c2^{m+N_g}$ of them, and the readers read $v$. A \emph{hidden assignment} $\omega$ gives each copy $\sigma$ a \emph{mask} $e_\sigma\in\{0,1\}^m$ and $N_g$ disjoint \emph{groups} $V_{\sigma,g}\subseteq[m]$ of $n$ variables each, and it splits each group into two \emph{blocks} $B'_{\sigma,g}$ and $B''_{\sigma,g}$ of $4J$ variables each. The other variables of the copy are \emph{passive}. The \emph{relative label} of a state is $u=v\oplus e_\sigma$, with restrictions $u'_g$ and $u''_g$ to the blocks of group $g$, and $f_{g,1}=\one\{|u'_g|\ge2J\}$ and $f_{g,2}=\one\{|u''_g|\ge2J\}$ are functions of $\sigma$ and $v$. The \emph{unmasked} assignment has $N_g=1$, $e_\sigma=0^m$, $B'_{\sigma,1}=[4J]$ and $B''_{\sigma,1}=[8J]\setminus[4J]$ for all $\sigma$, and then we drop the index $g$.

The setters of $I^+$ replace one coordinate of $v$ and keep $\sigma$ and $z$, so $I^+$ is composable. In the \emph{twisted interface} $I^-_\omega$, a setter of a variable $i\in B'_{\sigma,g}$ maps $(\sigma,v,z)$ to
\begin{equation}\label{eq:twist}
\big(\sigma,\ \upd{v}{i}{a},\ z\oplus[f_{g,1}(\sigma,\upd{v}{i}{a})\oplus f_{g,1}(\sigma,v)]\,f_{g,2}(\sigma,v)\,\mathbf e_g\big),
\end{equation}
with $\mathbf e_g$ the $g$-th unit vector, and the other setters are those of $I^+$.

The start distribution $\nu_\omega$ draws $\sigma$ uniformly and then, independently for each group, one of four \emph{sides}, each with probability $1/4$: $|u'_g|$ uniform on $\{J,\dots,3J-1\}$ and $|u''_g|=J$; $|u''_g|$ uniform on $\{J,\dots,3J-1\}$ and $|u'_g|=3J$; $|u'_g|$ uniform on $\{J+1,\dots,3J\}$ and $|u''_g|=3J$; $|u''_g|$ uniform on $\{J+1,\dots,3J\}$ and $|u'_g|=J$. Given the weights, the blocks are uniform and independent. The relative labels of the passive variables and $z$ are fair bits, independent of the rest. The \emph{frozen block} $B^{\mathrm f}_g$ of a start state in group $g$ is the block whose weight the side fixes at $J$ or $3J$; the side also fixes the value of its threshold function.

\begin{lemma}[Laws]\label{lem:rectlaws}
In $I^-_\omega$, success, write-back, overwrite and preservation hold everywhere, and $\cN$ is exactly the set of states with $|u'_g|,|u''_g|\in\{2J-1,2J\}$ for some group $g$.
\end{lemma}

\begin{proof}
Every setter changes the label as written, and writing the current value changes neither $v$ nor any threshold function. Between two writes to $B'_{\sigma,g}$ the value of $f_{g,2}$ does not change, so the changes of $z$ telescope, and overwrite and commutation within this block hold. A setter of a group reads and changes only the labels of this group and the bit $z_g$, so setters of different groups commute, and so do passive setters with all others. Writes to $B''_{\sigma,g}$ commute with each other. Commutation of a variable $i$ of $B'_{\sigma,g}$ with a variable $j$ of $B''_{\sigma,g}$ fails in the bit $z_g$ by $[f_{g,1}(\upd{v}{i}{a})\oplus f_{g,1}(v)][f_{g,2}(\upd{v}{j}{b})\oplus f_{g,2}(v)]$. This vanishes unless both weights lie in $\{2J-1,2J\}$, and at those weights suitable writes cross both thresholds.
\end{proof}

Fix a copy $\sigma$ and a group $g$. The \emph{rest} of a state is its label outside $V_{\sigma,g}$ with its hidden bits other than $z_g$. Let $\nu_g$ be the distribution of the group's label and $z_g$ under $\nu_\omega$ given $\sigma$, and let $P_g$ apply a uniformly random setter of the group.

\begin{lemma}[Common sample distribution]\label{lem:rectmeasure}
Let $\ell\ge0$, $\ell_0=n\ell/m$ and $q_0=\ell_0/(1+\ell_0)$. The interfaces $I^+$ and $I^-_\omega$ have the same sample distribution $\mu_\ell$ from $\nu_\omega$, and at a sample the hidden vector is a fair vector independent of the label and of the edits of the run. The number $L_g$ of edits of group $g$ in a run satisfies $\Pr(L_g\ge j)=q_0^j$. Given $L_g=j$, these edits are independent and uniform on the setters of the group, and with $\Lambda_{\sigma,g,j}$ the conditional distribution of the final rest,
\begin{equation}\label{eq:rectaccess}
\mu_\ell(\cdot\mid\sigma)=(1-q_0)\sum\nolimits_{j\ge0}q_0^j\,(\nu_gP_g^j)\otimes\Lambda_{\sigma,g,j}.
\end{equation}
\end{lemma}

\begin{proof}
Each edit lies in the group with probability $n/m$, so for $0\le y\le1$
\[
\E y^{L_g}=\sum\nolimits_{j\ge0}(1-q)q^j(1-n/m+(n/m)y)^j=\frac{1-q}{1-q(1-n/m+(n/m)y)}=\frac1{1+\ell_0-\ell_0y}.
\]
So $L_g$ is geometric with mean $\ell_0$. A random edit can first decide whether it lies in the group and then choose its variable and value, and the choices inside the group are independent of these decisions, of the length of the run and of the choices outside the group. The setters of the group act only on its labels and $z_g$, the other setters act only on the rest, and the start state draws these two parts independently given $\sigma$. This gives~\eqref{eq:rectaccess} for each of the two interfaces. They move labels in the same way, and the hidden vector is its fair initial value plus an amount determined by the labels and the edits, so it stays a fair vector independent of the labels and the edits. So the two sample distributions are equal.
\end{proof}

\subsection{The distance}\label{app:rectdistance}

\begin{lemma}[Forcing walk]\label{lem:forcing}
Let $w_0,\dots,w_{n_1}$ satisfy $w_t=T_{i_t,a_t}w_{t-1}$, let every variable that occurs be written exactly twice, first with a value $a$ and then with $1-a$, let each first write change the state, and let $w_{n_1}\ne w_0$. Then every composable interface $\tilde I$ on a finite set $S'\supseteq S$ disagrees with one of these calls.
\end{lemma}

\begin{proof}
If $\tilde I$ kept all of them, it would follow the same states. Before the first write to a variable $i$, the reading of $\tilde I$ at $i$ is that of $w_0$. As the first write, of $a$, changes the state, this reading is $1-a$, and the last write restores it. Variables that do not occur never change, and neither does the hidden data of $\tilde I$. So $\tilde I$ would return to $w_0$, a contradiction. The argument does not use the readers of $I$.
\end{proof}

\begin{lemma}[One group]\label{lem:onegroup}
Let $\ell\ge0$, fix $\omega$, a copy $\sigma$ and a group $g$, and let $\tilde I$ be a composable interface on a finite set $S'\supseteq S$, with $\mathcal D_{i,a}$ the set of states of $S$ at which $T_{i,a}$ of $I^-_\omega$ and $\tilde T_{i,a}$ differ. Then
\begin{equation}\label{eq:onegroup}
1-q_0^J\le\frac{8J}{J+1}\sum\nolimits_{i\in V_{\sigma,g},\,a}\mu_\ell(\mathcal D_{i,a}\mid\sigma).
\end{equation}
\end{lemma}

\begin{proof}
\emph{A random loop.} Fix the rest of the state, and draw $z_g$ uniformly. Let $\pi',\pi''$ be independent uniform permutations of $B'_{\sigma,g}$ and $B''_{\sigma,g}$, and let $X_t,Y_t$ be the relative block labels with ones at the first $t$ elements. From the relative label $(X_J,Y_J)$, the word $Q$ writes the relative value $1$ to $\pi'(J+1),\dots,\pi'(3J)$, then to $\pi''(J+1),\dots,\pi''(3J)$, then the relative value $0$ to $\pi'(3J),\dots,\pi'(J+1)$, and then to $\pi''(3J),\dots,\pi''(J+1)$. A relative value $\bar a$ at $i$ is the setter $T_{i,\bar a\oplus e_{\sigma,i}}$. During the first segment $f_{g,2}=0$, the second and fourth segments do not change $z_g$, and during the third segment $f_{g,2}=1$ while $f_{g,1}$ falls from $1$ to $0$ once. So $Q$ returns to its label with $z_g$ toggled, and Lemma~\ref{lem:forcing} applies to its $8J$ calls. Let $\occ_{i,a}(X)$ be the expected number of calls of $T_{i,a}$ on the walk whose argument lies in $X$. Then $1\le\sum_{i\in V_{\sigma,g},a}\occ_{i,a}(\mathcal D_{i,a})$.

\emph{Occupation.} In the first segment, the call at weight $t$ has the argument with relative label $(X_t,Y_J)$. The other segments are alike. The current hidden bit is a fair bit independent of the permutations. So this argument has the distribution of $\nu_g$ given the first side and the weight $t$, an event of probability $1/(8J)$. Given $X_t$, the next variable written to the relative value $1$ is uniform on the $4J-t\ge J+1$ relative zeros of $X_t$, and the next one written to $0$ in the third segment is uniform on the $t\ge J+1$ relative ones. At a fixed state, $(i,a)$ is one relative value. So, state by state, $\occ_{i,a}\le(8J/(J+1))\,\nu_g$.

\emph{Noisy loops.} Fix a word $W$ of $j\le J-1$ setters of the group, and let $\mathsf L_W$ be its action on labels: a coordinate that $W$ does not write keeps its value, and a written one takes its last written value. Let $\hat{\mathsf L}_W$ apply $\mathsf L_W$ to the label and keep the hidden bit. As the hidden bit of $\nu_g$ is fair and independent of the label, and $W$ adds to it an amount determined by the label, $\nu_gW=(\hat{\mathsf L}_W)_*\nu_g$. Apply $\mathsf L_W$ to all labels of the loop $Q$, skip its calls to variables that $W$ writes, and start from a fresh fair hidden bit; call the result the \emph{noisy loop}. By induction over the positions, the label at each position is the image under $\mathsf L_W$ of the label of $Q$ there. Let $\occ^W_{i,a}$ be the occupation measure of the noisy loop. If $W$ writes $i$, then $\occ^W_{i,a}=0$. Otherwise the argument of the noisy loop has the distribution of $\hat{\mathsf L}_W$ applied to the argument of $Q$, so $\occ^W_{i,a}=(\hat{\mathsf L}_W)_*\occ_{i,a}\le(8J/(J+1))(\hat{\mathsf L}_W)_*\nu_g=(8J/(J+1))\,\nu_gW$.

As $j\le J-1$, every low end of the loop has weight at most $J+j\le2J-1$ and every high end weight at least $3J-j\ge2J+1$. Every remaining write changes the relative label by one. So in the third segment the weight of the first block falls past $2J$ exactly once while $f_{g,2}=1$, and the noisy loop returns to its label with the hidden bit toggled. Every first write of a variable changes the state. So Lemma~\ref{lem:forcing} applies again.

\emph{Summing.} Averaging over $j$ independent uniform setters of the group and over the rest drawn from $\Lambda_{\sigma,g,j}$ gives, for $0\le j<J$,
\[
1\le\frac{8J}{J+1}\sum\nolimits_{i\in V_{\sigma,g},\,a}\big[(\nu_gP_g^j)\otimes\Lambda_{\sigma,g,j}\big](\mathcal D_{i,a}).
\]
Summing with coefficients $(1-q_0)q_0^j$ and using~\eqref{eq:rectaccess} gives~\eqref{eq:onegroup}.
\end{proof}

\begin{lemma}[Distance]\label{lem:rectdistance}
For every $\omega$ and $\ell\ge0$,
\[
\Dist^\infty_{\mu_\ell}(I^-_\omega)\ge\frac{N_g(J+1)}{16mJ}(1-q_0^J).
\]
If $N_g=\lfloor m/n\rfloor$, this is at least $(1-q_0^J)/(256J)$. If $N_g=1$ and the group is the same in all copies, then $\Dist^\infty_{\mu_\ell}(I^-_\omega)\ge((J+1)/(128J^2))(1-q_0^J)\ge(1-q_0^J)/(128J)$.
\end{lemma}

\begin{proof}
Let $R=R_{\mu_\ell}(I^-_\omega,\tilde I)$ for a candidate $\tilde I$. The groups of one copy are disjoint, so summing~\eqref{eq:onegroup} over $g$ gives $N_g(1-q_0^J)\le(8J/(J+1))\sum_{i\in[m],a}\mu_\ell(\mathcal D_{i,a}\mid\sigma)$. The right side now sums over fixed setters, so we may average over the uniform copy $\sigma$, which gives $N_g(1-q_0^J)\le(8J/(J+1))\cdot2mR$. As $\lfloor m/n\rfloor\ge m/(16J)$, the second claim follows. A sum over one group alone could not be bounded by $16JR$ in this way, as the group varies with the copy. If it does not vary, averaging~\eqref{eq:onegroup} over $\sigma$ gives $1-q_0^J\le(8J/(J+1))\cdot16JR$.
\end{proof}

\subsection{Renaming}\label{app:rename}

Couple the executions of a tester on $I^+$ and $I^-_\omega$ by the same internal randomness and the same samples, drawn from the common distribution $\mu_\ell$ together with the latent sides and edits. The hidden vector of a sample is a fresh fair vector, which is legal by Lemma~\ref{lem:rectmeasure}.

\begin{lemma}[Renaming]\label{lem:rename}
Suppose that the samples drawn lie in distinct copies and that, in the execution on $I^+$, for each sample and each group $g$, one of $f_{g,1},f_{g,2}$ is constant on the labels of all states held in the copy of this sample. Then the two transcripts coincide.
\end{lemma}

\begin{proof}
Fix a sample $w_*=(\sigma,v_*,z_*)$. In the execution on $I^+$ all states held in copy $\sigma$ have hidden vector $z_*$. If $f_{g,1}$ is constant on their labels, put $b_g=0$. Otherwise let $b_g$ be the constant value of $f_{g,2}$. Put
\[
\ren(\sigma,v,z_*)=\Big(\sigma,v,z_*\oplus\sum\nolimits_gb_g[f_{g,1}(\sigma,v)\oplus f_{g,1}(\sigma,v_*)]\,\mathbf e_g\Big).
\]
Then $\ren$ fixes $w_*$, keeps labels and is injective. By~\eqref{eq:twist}, $T^-_{i,a}\ren(w)=\ren(T^+_{i,a}w)$ whenever $w$ and $T^+_{i,a}w$ are held: for $i\in B'_{\sigma,g}$ the change of $z$ in~\eqref{eq:twist} is exactly that of $\ren$, and the other setters change neither $z$ nor any $f_{g,1}$. The union of these maps over the samples is injective, as states of distinct copies differ. By induction over the steps of the tester, each state it holds on $I^-_\omega$ is the image of the one it holds on $I^+$, so all readings, comparisons, depths and choices agree.
\end{proof}

\subsection{Unmasked: a ball around each sample}\label{app:unmasked}

\begin{lemma}[Unmasked transcripts]\label{lem:unmaskedtv}
Let $\omega$ be unmasked, let $d\ge0$ and $\hat t\ge1$ be integers with $J_{\hat t}=J(1-1/n)^{\hat t}>d$, and put $\hat p=q_0^{\hat t+1}+\exp[-(J_{\hat t}-d)^2/(2\hat t)]$. The transcripts of a tester of depth at most $d$ with at most $s$ samples of $\mu_\ell$ on $I^+$ and on $I^-_\omega$ differ in total variation by at most $s\hat p+s(s-1)/(2N_c)$, whatever its number of setter calls.
\end{lemma}

\begin{proof}
Fix a side, condition on the complete initial label of the frozen block, and let $Z_j$ be the weight of this block after $j$ edits of the group. A coordinate is untouched with probability $(1-1/n)^j$, and otherwise its last value is a fair bit, so $\E Z_j=2J\mp J(1-1/n)^j$ on a side with initial weight $J$ or $3J$. Changing one edit changes $Z_j$ by at most two, so the bounded-difference inequality \citep{vanhandel2016probability} gives $\Pr(Z_j-\E Z_j\ge y)\le\eu^{-y^2/(2j)}$, and the same for the lower tail. Call a sample good if $Z_j\le2J-d-1$ on the low side and $Z_j\ge2J+d$ on the high side. Then the threshold function of the frozen block is constant on the Hamming ball of radius $d$ around the label of the sample. For $1\le j\le\hat t$, a failure needs a deviation of at least $J_{\hat t}-d$ in a fixed direction. For $j=0$ the sample is good, as $J\ge J_{\hat t}>d$. So a sample is good except with probability $\hat p$. A held state on $I^+$ differs from its sample in at most $d$ coordinates, as a write changes one coordinate. Lemma~\ref{lem:rename} finishes the proof, and two of $s$ samples share a copy with probability at most $s(s-1)/(2N_c)$.
\end{proof}

\subsection{Masked: every query separately}\label{app:masked}

For polynomial budgets the tester queries few labels, so a mask replaces the union over the whole ball by a union over the queries. In this subsection $N_g=\lfloor m/n\rfloor$, and $\omega$ is \emph{uniform}: for each copy, independently, a uniform mask on $\{0,1\}^m$, uniform disjoint groups and uniform splits into blocks. We let the tester learn the length of each run, which only helps it. Write each edit $T_{i,a}$ of a sample in copy $\sigma$ as the pair $(i,\bar a)$ with relative value $\bar a=a\oplus e_{\sigma,i}$.

The \emph{latent data} $\mathcal K$ of a sample in a new copy are the groups and blocks of this copy, the sides, the initial relative label and the list of pairs of its edits. The tester sees the visible label $v^\star$ of the sample and the length of its run, and it holds the hidden vector only behind a handle. Hidden from it are $\mathcal K$ and the mask.

\begin{lemma}[Independence]\label{lem:independence}
Generate the copies of the first $s$ potential samples in advance, condition on the event that they are distinct, and condition on the table of their run lengths.
\begin{enumerate}[label=(\roman*),nosep,leftmargin=*]
\item The visible label of each sample is uniform and independent of its latent data, and its hidden vector is a fair vector independent of both. The data of different samples are independent.
\item The whole execution of the tester on $I^+$ is a function of the visible labels of the samples, the table and the internal randomness, and it does not depend on the mask or the latent data.
\item So, for each held state, its sample and the set $Y\subseteq[m]$ of coordinates where its label differs from the label of this sample are independent of the latent data of all samples. If the depth is at most $d$, then $|Y|\le d$.
\end{enumerate}
\end{lemma}

\begin{proof}
(i) Fix the length of a run in a new copy $\sigma$. The groups, the blocks, the sides and the initial relative label do not depend on $e_\sigma$. Given these and $e_\sigma$, the pairs are independent and uniform on $[m]\times\{0,1\}$, as $a$ is a fair bit independent of $i$. So $\mathcal K$ is independent of $e_\sigma$. The relative end label $u^\star$ is a function of $\mathcal K$, as each coordinate keeps its initial relative value or takes the last relative value written to it. The visible label is $v^\star=e_\sigma\oplus u^\star$. So, for each $v$ and each event $E$, $\Pr(v^\star=v,\mathcal K\in E)=\sum_y\Pr(\mathcal K\in E,u^\star=y)\Pr(e_\sigma=v\oplus y)=2^{-m}\Pr(\mathcal K\in E)$: $v^\star$ is uniform and independent of $\mathcal K$. The hidden vector is a fair vector independent of $\mathcal K$ and $e_\sigma$ by Lemma~\ref{lem:rectmeasure}. Different samples lie in different copies, whose data are independent.

(ii) In a copy of $I^+$, a setter replaces one coordinate of the label and keeps the hidden vector. So the end points of paths and their equalities depend only on the root label and the setters called. By induction over the steps of the tester, each reading, each comparison and each choice is a function of the visible labels, the table and the internal randomness.

(iii) This follows from (i) and (ii). A write changes one coordinate of the label, so a state of depth at most $d$ has $|Y|\le d$.
\end{proof}

\begin{lemma}[Masked transcripts]\label{lem:maskedtv}
Let $0\le d<m/2$ and $\hat t\ge0$ be integers, $\eta=1-2d/m$, and
\begin{equation}\label{eq:zeta}
\zeta=\exp[-J\eta^2/2]+\exp\big[-J^2\eta^2(1-1/n)^{2\hat t}/(8(4J+\hat t))\big].
\end{equation}
A tester of depth at most $d$ with at most $s$ samples of $\mu_\ell$ and at most $s_{\mathrm h}$ samples and setter calls in total has transcripts on $I^+$ and on $I^-_\omega$ that differ in total variation by at most $sN_gq_0^{\hat t+1}+s_{\mathrm h}N_g\zeta+s(s-1)/(2N_c)$. This holds with free reader calls, comparisons and computation, and also if the tester learns the length of each run.
\end{lemma}

\begin{proof}
\emph{One query, one group.} Fix $Y$ with $|Y|\le d$ and a group $g$, and condition on $L_g=j\le\hat t$. This keeps the law of the groups, as the number of edits of a set of $n$ variables has the same distribution for each such set. So the frozen block $B^{\mathrm f}_g$ is a uniform $4J$-subset of $[m]$. Let $\ov=|Y\cap B^{\mathrm f}_g|$, so $\E\ov\le4Jd/m=2J(1-\eta)$. Reveal the $4J$ elements of $B^{\mathrm f}_g$ one at a time. Each step has conditional range at most one, as swapping two elements changes $\ov$ by at most one, so the Azuma--Hoeffding inequality \citep{vanhandel2016probability} gives $\Pr(\ov-\E\ov\ge y)\le\eu^{-2y^2/(4J)}$. With $y=J\eta$, $\Pr(\ov>2J(1-\eta/2))\le\eu^{-J\eta^2/2}$, and otherwise $1-\ov/(2J)\ge\eta/2$.

Let $Z$ be the relative weight of $B^{\mathrm f}_g$ at the queried label, and let $\E Z$ be its conditional expectation given $B^{\mathrm f}_g$, the side, $Y$ and $L_g=j$. Given $L_g=j$, the edits of the group are independent and uniform on its pairs. Each initial bit has mean $1/4$ or $3/4$, a coordinate is untouched with probability $\theta_j=(1-1/n)^j$, and otherwise its last relative value is fair. Flipping the $\ov$ coordinates of $Y$ in $B^{\mathrm f}_g$ gives
\[
\E Z=2J\mp J(1-\ov/(2J))\theta_j,
\]
with the sign of the side. Now fix $B^{\mathrm f}_g$ and the side. The initial label is the first $J$ or $3J$ elements of a uniform permutation of $B^{\mathrm f}_g$. Swapping two of them changes at most two initial bits, and changing one pair at most two final bits. So, given these data, $Z$ is a function with at most $4J+j$ martingale increments of conditional range at most two, and $\Pr(Z-\E Z\ge y)\le\eu^{-y^2/(2(4J+j))}$, also for the lower tail. If $1-\ov/(2J)\ge\eta/2$, crossing the threshold $2J$ needs a deviation of at least $J\eta\theta_{\hat t}/2$ in a fixed direction. Averaging over $B^{\mathrm f}_g$ and the side, the threshold function of $B^{\mathrm f}_g$ at the queried label differs from its value on the side with probability at most $\zeta$.

\emph{All queries and groups.} Let the execution on $I^+$ be \emph{good} if the first $s$ potential samples lie in distinct copies, each drawn sample has $L_g\le\hat t$ for every group, and at each of the at most $s_{\mathrm h}$ held states and for every group, the threshold function of the frozen block of its sample has its value on the side. The union bound runs over the unconditioned execution on $I^+$ and uses only the distribution of each $L_g$, not a joint law of the groups. By Lemma~\ref{lem:independence}(iii), the query bound applies to each held state and each group. This gives the stated bound on the probability that the execution is not good, and Lemma~\ref{lem:rename} applies to good executions.
\end{proof}

\begin{lemma}[Expected cost]\label{lem:expectedcost}
Let a tester accept a first input with probability at least $2/3$ and a second input with probability at most $1/3$, where an input may be a mixture of instances. Fix the operations that it counts. The \emph{truncation at $t$} of an execution is the whole execution before its $(t+1)$-st counted operation, with the free operations in between and the outcome if it stops there. Let $\hat p\ge0$, and suppose that for every $t$ the truncations at $t$ on the two inputs differ in total variation by at most $t\hat p+t(t-1)/(2N_c)$. Then the tester uses at least $\lfloor\min\{1/(12\hat p),\sqrt{N_c/6}\}\rfloor/6$ counted operations in expectation on the first input, with $1/0=\infty$.
\end{lemma}

\begin{proof}
Let $t=\lfloor\min\{1/(12\hat p),\sqrt{N_c/6}\}\rfloor$, so the truncations at $t$ differ by at most $1/6$. Let $E$ be the event that the tester accepts before its $(t+1)$-st counted operation, which the truncation at $t$ determines, and let $E'$ be the event that it makes more than $t$ of them. With $\Pr_+$ and $\Pr_-$ for the first and the second input, $\Pr_-(E)\le1/3$, so $\Pr_+(E)\le1/2$, and $\Pr_+(E)\ge2/3-\Pr_+(E')$. So $\Pr_+(E')\ge1/6$, and the expected number is at least $t/6$.
\end{proof}

\noindent Lemmas~\ref{lem:unmaskedtv} and~\ref{lem:maskedtv} supply this hypothesis. The tester stopped before its $(t+1)$-st counted operation has at most $t$ counted operations, and by Lemma~\ref{lem:rename} the coupling keeps its free operations and its outcome.

\subsection{Necessity}\label{app:necessity}

\begin{proposition}[Necessity]\label{prop:necessity}
Fix either distance. Consider testers that certify this distance at $\epsilon_m$ with confidence $2/3$ on samples of $\mu_{\ell_m}$, for all interfaces and start distributions.
\begin{enumerate}[label=(\alph*),nosep,leftmargin=*]
\item Let $\epsilon_m=1/(32m)$, $2\le d_m\le\lfloor m/16\rfloor$, $\ell^*_m=m\log(m/d_m)/\log m$ and $\ell_m=o(\ell^*_m)$. Every such tester of depth at most $d_m$ uses a superpolynomial expected number of samples on $I^+$, for the unmasked start distribution.
\item Fix $C>1$ and $0<c<1/2$. Let $1/(32m)\le\epsilon_m\le(\log m)^{-C}$, $2\le d_m\le cm$ and $\ell_m=o(m\log(1/\epsilon_m)/\log m)$. Every such tester of depth at most $d_m$, even if it learns the length of each run, uses a superpolynomial expected number of samples and setter calls on $I^+$, for some start distribution $\nu_\omega$.
\item Let $1/(32m)\le\epsilon_m=o(1/\log m)$, $J=\lfloor1/(512\epsilon_m)\rfloor$, $\ell_m=0$ and $2\le d_m\le\lfloor m/2-w_m\rfloor$ with $w_m>0$ and $Jw_m^2/(m^2\log m)\to\infty$. The conclusion of~(b) holds.
\end{enumerate}
\end{proposition}

\begin{proof}
(a)~Take the unmasked assignment with $J=\lfloor m/8\rfloor$, $n=8J$ and $N_c=2^m$. As $\ell^*_m\le m$, $\ell_m=o(m)$, so $J/(1+\ell_0)\to\infty$ and $q_0^J\to0$. Lemma~\ref{lem:rectdistance} gives $\Dist^\infty_{\ell_m}\ge(1-o(1))/(128J)\ge(1-o(1))/(16m)>\epsilon_m$ for large $m$ and every $\omega$. So a tester as in the statement accepts $(I^+,\nu_\omega)$ with probability at least $2/3$ and $(I^-_\omega,\nu_\omega)$ with probability at most $1/3$.

Put $d^+=\max\{d_m,n^{3/4}\}$ and $\hat t=\lfloor(n-1)\log(J/(3d^+/2))\rfloor$. As $d_m\le J/2=n/16$ and $n^{3/4}\le n/16$ for large $m$, the logarithm is at least $\log(4/3)$, so $\hat t\ge1$. As $\log(1-1/n)\ge-1/(n-1)$, $J_{\hat t}\ge3d^+/2$, so $J_{\hat t}-d_m\ge d^+/2$. As $\hat t\le n\log n/4$ and $d^+\ge n^{3/4}$, the second term of $\hat p$ in Lemma~\ref{lem:unmaskedtv} is at most $\eu^{-(d^+)^2/(8\hat t)}\le\eu^{-\sqrt n/(2\log n)}$. Let $a_0=1-\log12/\log16>0$. If $d_m\ge n^{3/4}$, the logarithm in $\hat t$ is $\log(n/(12d_m))\ge a_0\log(n/d_m)$, as $n/d_m\ge16$. Otherwise it is at least $(\log n)/8\ge\log(n/d_m)/8$ for $n\ge12^8$. So $\hat t$ is at least a positive constant times $m\log(m/d_m)$, as $n/m\to1$. This gives $(\hat t+1)/((1+\ell_0)\log m)\to\infty$, since $\ell_0\le\ell_m=o(\ell^*_m)$ and $\ell^*_m\to\infty$. As $q_0\le\eu^{-1/(1+\ell_0)}$, $\hat p=m^{-\omega(1)}$. Lemma~\ref{lem:expectedcost}, counting samples, gives the claim.

(b)~Take $J=\lfloor1/(512\epsilon_m)\rfloor$, $n=8J$, $N_g=\lfloor m/n\rfloor$ and $N_c=2^m$. For large $m$, $J\ge(\log m)^C/1024$ and $\log J\sim\log(1/\epsilon_m)$, and $J\le m/16$ as $\epsilon_m\ge1/(32m)$, so $N_g\ge2$. As $\log(1/\epsilon_m)\le\log(32m)$, $\ell_m=o(m)$, so
\[
q_0^J\le\exp\Big[-\frac{J}{1+8J\ell_m/m}\Big]=\exp\Big[-\frac1{1/J+8\ell_m/m}\Big]\to0.
\]
Lemma~\ref{lem:rectdistance} gives $\Dist^\infty_{\ell_m}\ge(1-o(1))/(256J)\ge(2-o(1))\epsilon_m>\epsilon_m$ for large $m$ and every $\omega$. So a tester as in the statement accepts $(I^+,\nu_\omega)$ with probability at least $2/3$ and $(I^-_\omega,\nu_\omega)$ with probability at most $1/3$.

Let $a_C=(C-1)/(4C)>0$ and $\hat t=\lfloor a_C(n-1)\log J\rfloor$ in Lemma~\ref{lem:maskedtv}. As $\log(1-1/n)\ge-1/(n-1)$, $(1-1/n)^{\hat t}\ge J^{-a_C}$. With $\eta_0=1-2c>0$ and $\eta\ge\eta_0$, the exponent of the second term of~\eqref{eq:zeta} is at least $\eta_0^2J^{1-2a_C}/(8(4+8a_C\log J))$. Here $C(1-2a_C)=(C+1)/2>1$, and $y^{1-2a_C}/(4+8a_C\log y)$ increases for large $y$. So, with $J\ge(\log m)^C/1024$, this exponent divided by $\log m$ is at least a positive constant times $(\log m)^{(C-1)/2}/\log\log m$, which tends to infinity. The first term of~\eqref{eq:zeta} is superpolynomially small as well, so $\zeta=m^{-\omega(1)}$. Next,
\[
\frac{n\log J}{(1+n\ell_m/m)\log m}=\Big(\frac{\log m}{n\log J}+\frac{\ell_m\log m}{m\log J}\Big)^{-1}\to\infty,
\]
so $q_0^{\hat t+1}\le\exp[-(\hat t+1)/(1+\ell_0)]=m^{-\omega(1)}$. As $N_g\le m$, $\hat p=N_g(q_0^{\hat t+1}+\zeta)=m^{-\omega(1)}$. Lemma~\ref{lem:expectedcost}, counting samples and setter calls, with the mixtures over $\omega$ of the instances $(I^+,\nu_\omega)$ and of the instances $(I^-_\omega,\nu_\omega)$ as the two inputs, gives the bound on the first mixture, hence for some $\omega$.

(c)~Take $n=8J$, $N_g=\lfloor m/n\rfloor$ and $N_c=2^m$, so $J\le m/16$. As $q_0=0$, Lemma~\ref{lem:rectdistance} gives $\Dist^\infty_\nu\ge1/(256J)\ge2\epsilon_m$. As $d_m\le m/2-w_m$, $\eta=1-2d_m/m\ge2w_m/m>0$, so $d_m<m/2$ and Lemma~\ref{lem:maskedtv} applies. Take $\hat t=0$ there, so $\zeta\le2\exp[-J\eta^2/32]$. Squaring $\eta\ge2w_m/m$, whose two sides are positive, gives $\zeta\le2\exp[-Jw_m^2/(8m^2)]=m^{-\omega(1)}$. Conclude as in~(b).
\end{proof}

\begin{proof}[Proof of Theorem~\ref{thm:budget}]
Part~(b) is Corollary~\ref{cor:longruns} and~\eqref{eq:repairalpha}. For part~(a), $\epsilon_m=m^{-\beta}$ lies in the range of Proposition~\ref{prop:necessity}(b) for $C=2$ and large $m$, and $\log(1/\epsilon_m)/\log m=\beta$, so its hypothesis is $\ell_m=o(m)$. If $\ell_m\ge\alpha m$ for all large $m$, part~(b) gives a polynomial budget at depth $2\le d_m$. If $\ell_m$ is not $\Omega(m)$, some subsequence has $\ell_m/m\to0$, and part~(a) along it excludes a polynomial budget.

For the statement after the theorem, let $(\log m)^C\le\ell_m=o(m)$ and $\epsilon_m=1/\ell_m$, which lies in the range of Proposition~\ref{prop:necessity}(b) for large $m$. Fix $y>0$. As $t/\log t$ increases for $t\ge\eu$ and $\ell_m\le ym$ for large $m$, $\ell_m\log m/(m\log\ell_m)\le y\log m/\log(ym)\to y$. So $\ell_m=o(m\log(1/\epsilon_m)/\log m)$.
\end{proof}

\begin{proof}[Proof of Theorem~\ref{thm:audit}(c)]
\emph{Upper bound.} By Hoeffding's inequality, $\tau_2\le\exp[-2(d-2-m/2)^2/m]$ for $d\ge2+m/2$. So $d=2+\lceil m/2+\sqrt{(m/2)\log(2(m+2)/\epsilon_m)}\rceil$ gives $(m+2)\tau_2\le\epsilon_m/2$, and $d=m/2+O(\sqrt{m\log m})$. If this value exceeds $m+1$, take $d=m+1$ and part~(a). Part~(b) gives certification of both distances with $\lceil2\log3/\epsilon_m\rceil$ rounds of at most $2m^2+6m$ setter calls, which is a polynomial budget.

\emph{Lower bound.} Take $w_m=m^{1-\beta/2}\log m$, which is positive and $o(m)$, in Proposition~\ref{prop:necessity}(c). For large $m$, $J\ge m^\beta/1024$, so $Jw_m^2/(m^2\log m)\ge\log m/1024\to\infty$. So depths at most $m/2-w_m$ have no polynomial budget, for both distances.
\end{proof}

\subsection{Samples from the start distribution}\label{app:depthsource}

Let $\ell=0$, so that a tester draws samples of the start distribution $\mu=\nu$. A tester is an \emph{$(\epsilon,\delta)$-tester} if it certifies the distance under $\mu$ at $\epsilon$ with confidence $1-\delta$ for every interface on a finite set and every start distribution $\mu$.

\begin{theorem}[Samples from the start distribution]\label{thm:depthsource}
For both distances, at $\ell=0$: let $0<\delta<1/2$, $m\ge32$ and $0<\epsilon\le1/512$. No $(\epsilon,\delta)$-tester has a bound on its number of samples, or on its expected number, that depends only on $m$, $\epsilon$ and $\delta$, if its depth is at most $\lfloor\min\{m/8,1/(128\epsilon)\}\rfloor$, which is at least $\min\{m,1/\epsilon\}/256$.
\end{theorem}

\begin{proof}[Proof of Theorem~\ref{thm:depthsource}]
Let $y=\min\{m/8,1/(128\epsilon)\}$. The ranges of $m$ and $\epsilon$ give $y\ge4$, so $J=\lfloor y\rfloor\ge y/2\ge\min\{m,1/\epsilon\}/256$, and $m\ge8J$. Take the unmasked instance with $\ell=0$ and any $N_c$. As $J\le1/(128\epsilon)$, Lemma~\ref{lem:rectdistance} gives $\Dist_\nu(I^-)\ge\Dist^\infty_\nu(I^-)\ge1/(128J)\ge\epsilon$.

Fix a start state with block labels $u'_0,u''_0$, hidden bit $z_0$ and $b=f_2(u''_0)$, and on the states of its copy with hidden bit $z_0$ whose block labels lie within Hamming distance $J$ of $(u'_0,u''_0)$ put $\ren(\sigma,v,z_0)=(\sigma,v,z_0\oplus b[f_1(u')\oplus f_1(u'_0)])$. This map fixes the start state, keeps the readers and is injective. Setters of $B''_\sigma$ and of passive variables do not change $f_1$, and $I^-$ treats them as $I^+$ does. A setter of $B'_\sigma$ whose two labels lie in this ball changes $f_1$ only while $f_2=b$. Indeed, if $|u''_0|\in\{J,3J\}$ and $f_2(u'')\ne b$, then $u''$ is at Hamming distance at least $J$ from $u''_0$, so $u'=u'_0$ at both ends and $f_1$ does not change. If $|u'_0|\in\{J,3J\}$ and $f_1$ changes, one end is at Hamming distance at least $J$ from $u'_0$, so $u''=u''_0$ and $f_2=b$. Every start state has such a block. So $T^-_{i,a}\ren(w)=\ren(T^+_{i,a}w)$ for every state $w$ of this domain whose block labels lie within Hamming distance $J-1$ of $(u'_0,u''_0)$. A setter call of depth at most $J$ from the start state acts on a state reached by at most $J-1$ calls. So, as in Lemma~\ref{lem:rename}, the two transcripts of a tester of depth at most $J$ coincide when its samples lie in distinct copies, and two of $s$ samples share a copy with probability at most $s(s-1)/(2N_c)$.

Let $t=\lfloor\sqrt{N_c(1-2\delta)}\rfloor$, let $E$ be the event that the tester accepts before it draws a $(t+1)$-st sample, and let $E'$ be the event that it draws more than $t$ samples. Then $\Pr_-(E)\le\delta$ and $\Pr_+(E)\ge1-\delta-\Pr_+(E')$. The coupling then gives $\Pr_+(E')\ge1-2\delta-t(t-1)/(2N_c)\ge(1-2\delta)/2$. Hence the expected number of samples under $I^+$ is at least $t(1-2\delta)/2$, which grows without bound with $N_c$. A bound on the number of samples that depends only on $m$, $\epsilon$ and $\delta$ would bound this expectation.
\end{proof}
\section{Proof of Theorem~\ref{thm:samplesonly}}\label{app:partial}

The necessity in Theorem~\ref{thm:samplesonly} follows from Proposition~\ref{prop:necessity}(a). If $\ell_m$ is not $\Omega(\ell^*_m)$, some subsequence has $\ell_m/\ell^*_m\to0$, and the proposition, applied along it, excludes a polynomial budget of samples. This appendix proves the sufficiency. A random word need not write every variable: short words from its end can stand in for the few variables that it misses.

For an integer $\bar d\ge0$, let $\cN_{\bar d}$ be the set of states from which some word of length at most $\bar d$ reaches $\cN$, so $\cN_0=\cN$. Let $H_0=0$ and $H_j=\sum_{v\le j}1/v$.

\begin{theorem}[Partial coverage]\label{thm:partial}
Let $m\ge2$, $2\le d\le m$, $\bar d=d-2$, $\lambda\ge1$, let $\mu$ be $\lambda$-subinvariant and $t=\lceil2m(H_m-H_{\bar d})\rceil$. Then $\Dist^\infty_\mu\le\min\{1,2(t+1)\lambda^t\mu(\cN_{\bar d})\}$, and if $2^m\mid|S|$, then $\Dist_\mu\le\min\{1,(t+1)(t+4)\lambda^t\mu(\cN_{\bar d})\}$.
\end{theorem}

\paragraph{The tester.} Exhausting all words of length at most $\bar d$ from a sample and evaluating all law instances at their ends has depth at most $\bar d+2=d$. It rejects with probability $\mu(\cN_{\bar d})$ and accepts every composable interface. So, if $K$ is the coefficient of Theorem~\ref{thm:partial}, $\lceil K\log(1/\delta)/\epsilon\rceil$ samples certify as in Appendix~\ref{app:lawcheck}. The number of setter calls of a sample can be exponential in $d$.

\begin{proof}[Proof of Theorem~\ref{thm:partial}]
\emph{The word.} Draw $t$ independent uniform setters $E_1,\dots,E_t$, with prefixes $U_k=E_k\sdots E_1$. Let $\mathcal H$ be the event that at most $\bar d$ variables are not written, and let $R_0$ be the set of these variables. On $\mathcal H$, let $W_0$ write $0$ once to each variable of $R_0$ in increasing order, and put $W=W_0U_t$, which writes every variable. The number of draws needed to collect $m-\bar d$ distinct variables has mean $m(H_m-H_{\bar d})$, so Markov's inequality gives $\Pr(\mathcal H)\ge1/2$.

\emph{The distance with added states.} Let $A_W$ be the set of states with $U_kx\notin\cN$ for $k<t$ and $U_tx\notin\cN_{\bar d}$. Every step of $W_0$ acts on a state within $\bar d-1$ writes of $U_tx$, so the path of $x\in A_W$ under $W$ avoids $\cN$ before its last step. By Lemma~\ref{lem:insert}, Lemma~\ref{lem:structure} applies to $A_W$, and $\Dist^\infty_\mu\le\mu(S\setminus A_W)$. For $x\sim\mu$ independent of the setters, $U_kx$ has distribution $\mu P^k\le\lambda^k\mu$, and $\cN\subseteq\cN_{\bar d}$. So, dropping $\mathcal H$ from the numerator and dividing by $\Pr(\mathcal H)\ge1/2$,
\begin{equation}\label{eq:partialA}
\E[\mu(S\setminus A_W)\mid\mathcal H]\le2\Big[\sum\nolimits_{k<t}\lambda^k+\lambda^t\Big]\mu(\cN_{\bar d})\le2(t+1)\lambda^t\mu(\cN_{\bar d}).
\end{equation}

\emph{A local rule.} Let $z\notin\cN_{\bar d}$ and $|R_0|\le\bar d$. Let $W_g$ write the values $g\in\{0,1\}^{R_0}$ to the variables of $R_0$, once each, in increasing order. Each suffix of $W_g$ acts on a state within $\bar d-1$ writes of $z$, so these states lie outside $\cN$. Commuting a setter $T_{i,e}$ with $i\in R_0$ through $W_g$ and absorbing it by overwrite gives $T_{i,e}W_gz=W_{g'}z$, where $g'$ is $g$ with coordinate $i$ replaced by $e$. In particular $W_{g''}W_gz=W_{g''}z$ for all $g,g''$. Commuting a setter $E$ of a variable outside $R_0$ through $W_g$ gives $EW_gz=W_gEz$.

\emph{A second word.} Let $g_1,\dots,g_t$ be fair bits and $g_0\in\{0,1\}^{R_0}$ a fair vector, independent of the rest. Put $V_k=T_{i_k,g_k}$ and $V=V_1\sdots V_t$, which applies $V_t$ first. For $1\le a\le k\le t$ put $z_{a,k}=U_{a-1}V_{k+1}\sdots V_tx$, and for $0\le j\le t$ put $x'_j=U_jV_{j+1}\sdots V_tx$. The positions used by $z_{a,k}$ are disjoint, so before conditioning $z_{a,k}$ has distribution $\mu P^{t-k+a-1}$, and $x'_j$ has distribution $\mu P^t$. Consider the event $\mathcal G$ that all $z_{a,k}\notin\cN$ and all $x'_j\notin\cN_{\bar d}$. Summing the $t(t+1)/2$ and the $t+1$ terms,
\begin{equation}\label{eq:partialG}
\Pr(\mathcal G^c\mid\mathcal H)\le2\Big[\frac{t(t+1)}2+t+1\Big]\lambda^t\mu(\cN_{\bar d}).
\end{equation}

\emph{A witness.} Assume $\mathcal G$. Fix $k$ and let $j_k$ be the first position with $i_{j_k}=i_k$. Commutation at $z_{a,k}$ for $a<j_k$ and overwrite at $z_{j_k,k}$ give, for $j\ge k$, $U_jV_kV_{k+1}\sdots V_tx=U_jV_{k+1}\sdots V_tx$. Removing $V_1,\dots,V_j$ in turn gives $U_jVx=x'_j$. Let $y=W_{g_0}Vx$, and let $\hat u$ be the label predicted by applying the writes of $V$ and $W_{g_0}$ to $r(x)$. Given $W$ and $x$, $\hat u$ is uniform: $V$ applies $V_1$ last, so each written variable takes the bit $g_k$ of $V$ at its first position $k$ in $W$, and each variable of $R_0$ a coordinate of $g_0$. The steps of $V$ act at $z_{1,k}\notin\cN$ and those of $W_{g_0}$ within $\bar d-1$ writes of $x'_0=Vx\notin\cN_{\bar d}$, so $r(y)=\hat u$. The steps of $U_t$ avoid $R_0$ and act at $U_{j-1}Vx=x'_{j-1}\notin\cN_{\bar d}$, so the local rule moves $W_{g_0}$ through them: $U_ty=W_{g_0}U_tVx=W_{g_0}U_tx$. As $x'_t=U_tx\notin\cN_{\bar d}$, the local rule gives $W_0W_{g_0}U_tx=W_0U_tx$. Hence $Wy=Wx$ and $r(y)=\hat u$.

\emph{The distance on the same states.} By Lemma~\ref{lem:structure}, $\Dist_\mu\le\mu(S\setminus A_W)+\Delta_{A_W}$. Given $W$, the term $\Delta_{A_W}$ is at most the probability that $\hat u$ occurs nowhere in the fiber of $x$, which needs $\mathcal G^c$. Averaging given $\mathcal H$ and adding~\eqref{eq:partialA} and~\eqref{eq:partialG} gives $2[t(t+1)/2+2t+2]\lambda^t\mu(\cN_{\bar d})=(t+1)(t+4)\lambda^t\mu(\cN_{\bar d})$.
\end{proof}

\begin{proof}[Proof of the sufficiency in Theorem~\ref{thm:samplesonly}]
Let $2\le d\le m/16$, $\bar d=d-2$, and $\ell\ge b\ell^*_m$ with $\ell^*_m=m\log(m/d)/\log m$ and a constant $b>0$. For $\bar d\ge1$, integral comparison gives $H_m-H_{\bar d}\le\log(m/\bar d)$, and $\bar d\ge d/3$ for $d\ge3$, so $\log(m/\bar d)\le\log3+\log(m/d)\le2\log(m/d)$. For $\bar d=0$, $H_m\le1+\log m\le2\log(m/2)$ for $m\ge32$. So $t\le5m\log(m/d)$, and with $\lambda=1+1/\ell$, $\lambda^t\le\eu^{t/\ell}\le m^{5/b}$. As $(t+1)(t+4)$ is polynomial in $m$, the tester above uses a polynomial number of samples, for both distances; for $\Dist^\infty$, no condition on $|S|$ is needed.

\emph{The explicit bound.} Let $m\ge96$, $d=\lfloor m/16\rfloor$ and $\ell\ge\varsigma m/\log m$ with $\varsigma>0$. Then $\bar d\ge m/32\ge1$, so $H_m-H_{\bar d}\le\log32$ and $t\le2m\log32+1\le8m$, and $(t+1)(t+4)\le140m^2$. Also $\lambda^t\le\eu^{t/\ell}\le m^{8/\varsigma}$. So $\lceil140m^{2+8/\varsigma}\log(1/\delta)/\epsilon\rceil$ samples certify both distances.
\end{proof}
\section{Proof of Theorem~\ref{thm:onestate}}\label{app:onestate}

\paragraph{Route.} Part~(a) follows the four steps of Section~\ref{sec:onestate}, and each subsection gives one ingredient.
\begin{itemize}[nosep,leftmargin=1.5em]
\item Appendix~\ref{app:onestatehist} sets up continuous time and predicted labels, and gives two events on which the history of a variable matters only through $B_j$:~\eqref{eq:os-front} moves the last write to the front, and strongly good words allow a whole history to be replaced. Their failure probabilities are charged to $\gamma$ in~\eqref{eq:os-last} and~\eqref{eq:os-strong}.
\item Appendix~\ref{app:os-birth} couples the histories given $B$ as the time grows, which bounds the derivative in time of the end state given $B$ in~\eqref{eq:os-deriv1}.
\item Appendix~\ref{app:os-kernel} proves a bound on the density of the time given a label (Lemma~\ref{lem:os-kernel}) and turns a derivative into a variance bound~\eqref{eq:os-L1}.
\item Appendix~\ref{app:os-weight} introduces the weight $\varpi$, which is what a setter pays when it resamples a history. It bounds the error from resampling one history at a time~\eqref{eq:os-local} and from the time~\eqref{eq:os-time}, and it chooses a state $f(b)$ for each label $b\ne r(o)$.
\item Appendix~\ref{app:os-initial} treats the label $r(o)$, where the weight is not integrable at time $0$ and $f(r(o))=o$.
\item Appendix~\ref{app:os-cube} places the chosen states in one cube and proves the bound~\eqref{eq:os-param} for $\ell\ge2$.
\item Appendix~\ref{app:os-short} treats means at most $2$, where a Stein kernel replaces the coupling and the integral kernel.
\item Appendix~\ref{app:os-assemble} assembles the cases, with means of at least $2m$ taken from~\eqref{eq:repairalpha}, and Appendix~\ref{app:fixed} proves part~(b).
\end{itemize}

Throughout Appendices~\ref{app:onestatehist}--\ref{app:os-cube}, $m\ge2$, the start distribution is $\nu=\delta_o$, $\ell\ge2$, $q=\ell/(1+\ell)$, $\mu=\mu_\ell$, $\gamma=\gamma_\mu$, $\alpha=\ell/m$ (this ratio, not the free constant $\alpha$ of Theorem~\ref{thm:budget}), and all law instances hold at $o$. Each law instance then fails with probability at most $\gamma$ at $x\sim\mu$, and this is the only use of $\gamma$.

\subsection{Histories and two good events}\label{app:onestatehist}

\paragraph{Continuous time.} Let the edits arrive at the times of a Poisson process of rate $1$, each a uniformly random setter, and let $\rho_t=\eu^{-t}\sum_{n\ge0}(t^n/n!)\,\nu P^n$ be the distribution of the state at time $t$. Stop at an independent time $\iota$ with density $\ell^{-1}\eu^{-t/\ell}$. The number $L$ of edits before $\iota$ satisfies
\[
\Pr(L=n)=\frac1{\ell\,n!}\int_0^\infty\eu^{-(1+1/\ell)t}t^n\,dt=\frac{\ell^n}{(1+\ell)^{n+1}}=(1-q)q^n,
\]
by integration by parts and induction on $n$. So the state $X$ at time $\iota$ has distribution $\mu$; this holds for every start distribution and every $\ell>0$. For a nonnegative function $\varphi$ on states, this gives $\E\int_0^\iota\rho_s\varphi\,ds=\int_0^\infty\eu^{-s/\ell}\rho_s\varphi\,ds=\ell\mu\varphi$.

\paragraph{Predicted labels.} Let $b^0=r(o)$. Applying the same edits to the vector $b^0$ as coordinate writes gives the \emph{predicted label} $B$. Split by variable, the edits form independent Poisson processes of rate $1/m$, with independent fair written values. So, with $w=|b\oplus b^0|$,
\begin{equation}\label{eq:os-label}
\Pr(B=b\mid\iota=t)=2^{-m}(1-\eu^{-t/m})^w(1+\eu^{-t/m})^{m-w},
\end{equation}
which is positive for $t>0$. Given $B=b$ and $\iota=t$, the histories of the variables, with their arrival times and written values, are still independent, as the event $B=b$ is the intersection of one event for each variable.

\paragraph{Moving the last write to the front.} Fix a variable $j$. For an edit $E=T_{i,a}$, let $I_j(x,E)$ indicate that, for $b=0$ or $b=1$, the equality $T_{j,b}Ex=ET_{j,b}x$ fails if $i\ne j$, or $T_{j,b}T_{j,a}x=T_{j,b}x$ fails if $i=j$. Let $\bar I_j(x)=(2m)^{-1}\sum_{i,a}I_j(x,T_{i,a})$, so $\mu\bar I_j\le2\gamma$. Let $W=E_n\sdots E_1$ be a word, $U_s=E_s\sdots E_1$ and $U_0=\Id$, and let $W_{-j}$ delete the writes to $j$ from $W$. If $I_j(U_{s-1}o,E_s)=0$ for every $s$ before the last write to $j$, then
\begin{equation}\label{eq:os-front}
Wo=W_{-j}T_{j,B_j}o.
\end{equation}
Indeed, if $W$ writes $j$, let $s_*$ be its last write to $j$, with value $b=B_j$, and put $T_*=T_{j,b}$. For $s<s_*$, induction gives $T_*U_so=U^{(-j)}_sT_*o$, where $U^{(-j)}_s$ deletes the writes to $j$: an edit of another variable commutes with $T_*$ at $U_{s-1}o$, and a write to $j$ is overwritten by $T_*$. So $U_{s_*}o=T_*U_{s_*-1}o=U^{(-j)}_{s_*-1}T_*o$, and the later edits do not write $j$. If $W$ does not write $j$, then $B_j=b^0_j$, and write-back at $o$ gives $T_{j,b^0_j}o=o$. If $I_j(U_{s-1}o,E_s)=0$ for all $s$, the same induction gives $T_{j,a}Wo=W_{-j}T_{j,a}o$ for $a=0,1$.

Let $\bar p_j(t)$ be the probability that the word of the edits up to time $t$ fails the condition of~\eqref{eq:os-front}, and $p_j(t)$ the probability that it fails the stronger condition for all $s$. An edit arriving at time $s$ acts at a state of distribution $\rho_s$, and a later write to $j$ occurs with probability $1-\eu^{-(t-s)/m}$, independently of this state and this edit. So
\begin{equation}\label{eq:os-last}
\bar p_j(t)\le\int_0^t\rho_s\bar I_j\,[1-\eu^{-(t-s)/m}]\,ds,\qquad p_j(t)\le\int_0^t\rho_s\bar I_j\,ds.
\end{equation}

\paragraph{Strongly good words.} Call a word \emph{strongly good for $j$} if the four overwrite equalities $T_{j,a}T_{j,b}x=T_{j,a}x$ hold at every prefix state, including the end state, and every edit of another variable commutes with $T_{j,0}$ and $T_{j,1}$ at the state where it acts. If a word is strongly good for $j$, then replacing the whole history of $j$ by another one with the same value $B_j$ keeps the end state. To see this, scan the old writes to $j$, the new writes to $j$ and the common other edits in their merged order, with $x$ the old prefix state, $y$ the new prefix state and $e$ the new predicted value of $j$. The invariant $y=T_{j,e}x$ holds at the start by write-back at $o$. An old write to $j$ changes only $x$, and overwrite at $x$ keeps the invariant. A new write to $j$ changes $y$ and $e$, and overwrite at $x$ keeps it. Another edit commutes with $T_{j,e}$ at $x$. At the end, $y=T_{j,B_j}X$. Induction along the old word, with commutation for other edits and $T_{j,a}T_{j,a}x=T_{j,a}x$ for writes to $j$, gives $T_{j,B_j}X=X$. So $y=X$. A new write may come after all old edits, which is why the end state is included.

Let $I^\circ_j$ indicate that one of the four overwrite equalities fails at a state, and let $\bar I^*_j$ be $I^\circ_j$ plus the average over the edit of the indicator that this edit, if it does not write $j$, fails to commute with $T_{j,0}$ or $T_{j,1}$. Then $\mu I^\circ_j\le4\gamma$ and $\mu\bar I^*_j\le6\gamma$. Let $p^*_j(t)$ be the probability that the word up to time $t$ is not strongly good for $j$, and $p^*_j(t\mid b)$ the same given $B=b$. Then $p^*_j(t)\le\int_0^t\rho_s\bar I^*_j\,ds+\rho_tI^\circ_j$, so
\begin{equation}\label{eq:os-strong}
\E\,p^*_j(\iota)\le(6\ell+4)\gamma.
\end{equation}

\subsection{A pure-birth coupling given the label}\label{app:os-birth}

Fix a variable, let $y=t/m$, $p_n=\eu^{-y}y^n/n!$ and $R_n=\Pr\{\mathrm{Pois}(y)>n\}$. Given its predicted value and $t$, the number of writes to this variable has the masses $p_n/(1-p_0)$ for $n\ge1$ if the value differs from the initial one, and $2p_0/(1+p_0)$ for $n=0$ and $p_n/(1+p_0)$ for $n\ge1$ otherwise. Differentiating the distribution functions $1-R_n/(1\mp p_0)$ in $t$, using $dp_0/dy=-p_0$ and $dR_n/dy=p_n$, shows that these masses are the marginals of a pure-birth process with rates
\[
\frac1m\Big[1-\frac{p_0R_n}{(1-p_0)p_n}\Big],\qquad\frac1m\Big[1+\frac{p_0R_n}{(1+p_0)p_n}\Big]\ (n\ge1),\qquad\frac1{m(1+p_0)}\ (n=0),
\]
in the two cases. Let $N(t)$ be the state of this process at time $t$. Its entrance law is $N(0+)=1$ when the predicted value differs from the initial one, and $N(0+)=0$ when they are equal. As $R_n/p_n=\sum_{v\ge1}y^vn!/(n+v)!\le\eu^y-1$, all rates lie in $[0,2/m]$. Given $n\ge1$ writes, the normalized arrival times are independent uniform points of $[0,1]$, the last value is the given one, and the earlier values are fair. At a birth, insert an independent uniform point. If it is not the new last point, give it a fair value. If it is, redraw the old last value as a fair bit and give the new point the given value. For $n=0$ there is no old last value, so the first birth skips the redraw. This keeps the law of the history. Multiplying by $t$ does not change the order of the normalized points. So running this process independently for each variable couples the histories given $B=b$ for all $t>0$, and a birth changes one history only. If the word before the birth is strongly good for that variable, the end state does not change.

Let $g_{b,x}(t)=\Pr(X=x\mid B=b,\iota=t)$ and $v_b(t)=1-\sum_xg_{b,x}(t)^2$. As the rates are at most $2/m$, the probability that the end state changes in $(t,t+h]$ is at most $(2/m)\int_t^{t+h}\sum_jp^*_j(s\mid b)\,ds$. These conditional probabilities are quotients of power series that converge uniformly with their derivatives on compact subsets of $(0,\infty)$. So
\begin{equation}\label{eq:os-deriv1}
\sum\nolimits_x|g'_{b,x}(t)|\le\frac4m\sum\nolimits_jp^*_j(t\mid b).
\end{equation}

\subsection{An integral kernel of the time given the label}\label{app:os-kernel}

\begin{lemma}[Integral kernel]\label{lem:os-kernel}
Let $n\ge1$ and $0\le\bar k\le n$ be integers and $\ell'>0$. Let $\psi$ be the probability density on $(0,\infty)$ proportional to $\eu^{-t/\ell'}(1-\eu^{-t/n})^{\bar k}(1+\eu^{-t/n})^{n-\bar k}$, with distribution function $\Psi$. Then $\Psi(t)(1-\Psi(t))/\psi(t)\le200(1+\ell'/n)\min\{t,n\}$ for all $t>0$.
\end{lemma}

\begin{proof}
Let $\mathcal S=(\log\psi)'$, so
\[
\mathcal S(t)=-\frac1{\ell'}+\frac{\bar k}{n(\eu^{t/n}-1)}-\frac{n-\bar k}{n(\eu^{t/n}+1)},\qquad \mathcal S(t)\ge-\frac1{\ell'}-\frac12 .
\]
With $a'=n/\ell'$ and $y=\eu^{-t/n}$, $n(1-y^2)\mathcal S(t)=(a'+n)y^2+(2\bar k-n)y-a'$. The right side has roots $y_+\in(0,1]$ and $-y_-<0$, with $(a'+n)y_+y_-=a'$; $y_+=1$ exactly when $\bar k=0$. So $\psi$ has a unique mode at $t_*=-n\log y_+$: an interior maximum if $\bar k\ge1$, and the boundary point $t_*=0$ if $\bar k=0$, when $\psi$ decreases on $(0,\infty)$. With $v=\eu^{-(t-t_*)/n}$,
\[
\mathcal S(t)=\frac1{\ell'}(v-1)\frac{1+(y_+/y_-)v}{1-y_+^2v^2},
\]
where the last fraction is at least $1$. Hence $\mathcal S(t_*+z)\le-(1-\eu^{-z/n})/\ell'$ for $z>0$, and $\mathcal S(t_*-z)\ge(\eu^{z/n}-1)/\ell'$ for $0<z<t_*$.

\emph{The scale $n$.} For $t\le t_*$, moving left by at most $n$ the density ratio is at most $1$, and beyond that the logarithmic derivative is at least $(\eu-1)/\ell'\ge1/\ell'$. So $\Psi(t)/\psi(t)\le n+\ell'$. For $t\ge t_*$, moving right by at most $n$ the ratio is at most $1$, and beyond that the logarithmic derivative is at most $-(1-\eu^{-1})/\ell'\le-1/(2\ell')$. So $(1-\Psi(t))/\psi(t)\le n+2\ell'$. As $\Psi(1-\Psi)\le\min\{\Psi,1-\Psi\}$, always $\Psi(1-\Psi)/\psi\le n+2\ell'\le2(1+\ell'/n)n$. This gives the claim for $t\ge n/64$, as then $n\le64\min\{t,n\}$.

\emph{The scale $t$.} Let $t<n/64$. If $t\le t_*$, $\psi$ does not decrease on $(0,t]$, so $\Psi(t)/\psi(t)\le t$. Let $t>t_*$, and let $q'=\ell'/(1+\ell')$. If $t\le q'$, then for $0<z<t$, $\psi(z)/\psi(t)\le\eu^{(1/\ell'+1/2)t}\le\eu$, so $\Psi(t)/\psi(t)\le3t$.

Let $t>q'$ and $\ell'\ge2$, so $t\ge2/3$. For $0<z\le\bar k/2\le n/2$, $\eu^{z/n}-1<2z/n$ gives $\mathcal S(z)>-1/\ell'-1/2+\bar k/(2z)\ge0$. So $t_*\ge\bar k/2$, and $\bar k\le2t$. Let $z_0=32t<n/2$. For $z_0\le z\le n$, $\eu^{z/n}-1\ge z/n$ and $\eu^{z/n}+1<4$ give $\mathcal S(z)\le\bar k/z-(1-\bar k/n)/4\le1/16-31/128<-1/16$. So $\psi$ does not increase on $[t,z_0]$, it falls at rate at least $1/16$ on $[z_0,n]$, and $\psi(n)/\psi(t)\le\eu^{-n/32}$. With the bound $n+2\ell'$ at $n$ and $n\eu^{-n/32}\le32$,
\[
\frac{1-\Psi(t)}{\psi(t)}\le32t+16+32(1+2\ell'/n)\le(104+96\ell'/n)t.
\]

Let $t>q'$ and $\ell'<2$. For $0<z\le\bar k\ell'/(2+\ell')<n/2$, the same estimate gives $\mathcal S(z)>0$, so $\bar k\le(2/\ell'+1)t$. For $z\ge8t$, $\mathcal S(z)\le-1/\ell'+\bar k/z\le-(6-\ell')/(8\ell')\le-1/(2\ell')$. So $(1-\Psi(t))/\psi(t)\le8t+2\ell'\le14t$, as $\ell'\le(1+\ell')t<3t$.
\end{proof}

For a positive probability density $\psi$ with distribution function $\Psi$ and a locally absolutely continuous vector $g$ of probabilities, the inequality $\|y-y'\|_2^2\le\|y-y'\|_1$ for probability vectors gives
\begin{equation}\label{eq:os-L1}
\sum\nolimits_x\mathrm{Var}_\psi\,g_x=\int_{s<t}\|g(t)-g(s)\|_2^2\,\psi(s)\psi(t)\,ds\,dt\le\int_0^\infty\|g'(z)\|_1\Psi(z)(1-\Psi(z))\,dz.
\end{equation}

\subsection{Weighted errors and the chosen states}\label{app:os-weight}

Let $\varpi(t)=1/(1-\eu^{-t/m})$, which is at least $1$, and $\Omega_\ell=\ell+(m/q)\log(1+\ell)$. For every nonnegative $\varphi$ on states,
\begin{equation}\label{eq:os-weight}
\E\Big[\varpi(\iota)\int_0^\iota\rho_s\varphi\,ds\Big]\le\Omega_\ell\,\mu\varphi .
\end{equation}
Indeed, $\varpi(t)\le1+m/t$. The constant part gives $\ell\mu\varphi$. For the other part, expand $\rho_s\varphi$ over the number $k$ of edits. The coefficient of $\nu P^k\varphi$ is $(m/\ell)\int_{1/\ell}^\infty[v(1+v)^{k+1}]^{-1}dv$, after writing $\eu^{-t/\ell}/t=\int_{1/\ell}^\infty\eu^{-tv}dv$ and exchanging integrals. The substitution $(1+v)^{-1}=y$ turns it into $m(1-q)q^k\sum_{v\ge0}q^v/(k+v+1)\le(1-q)q^k(m/q)\log(1+\ell)$, as $q/\ell=1-q$.

\paragraph{One history at a time.} Let $V(t)=\sum_b\Pr(B=b\mid\iota=t)\,v_b(t)$. For independent variables $\omega_1,\dots,\omega_m$ and a vector-valued function $f$ of them,
\begin{equation}\label{eq:os-tensor}
\E\|f-\E f\|^2\le\sum\nolimits_j\E\,\mathrm{Var}_{\omega_j}(f\mid\omega_{-j})=\tfrac12\sum\nolimits_j\E\|f-f^{(j)}\|^2,
\end{equation}
where $f^{(j)}$ resamples $\omega_j$. This is the tensorization of the variance, also called the Efron--Stein inequality \citep{vanhandel2016probability}. Take $f=(\one_{X=x})_x$ and resample the history of $j$ given $B_j$. By~\eqref{eq:os-front} the two end states agree unless one of the two words fails the condition there, and the two words have the same distribution. So $V(t)\le2\sum_j\bar p_j(t)$. As $\varpi(t)[1-\eu^{-(t-s)/m}]\le1$, \eqref{eq:os-last} gives
\begin{equation}\label{eq:os-local}
\E[\varpi(\iota)V(\iota)]\le4m\ell\gamma.
\end{equation}
Along a run, the reading of $j$ stays equal to $B_j$ unless success or preservation fails. So, by~\eqref{eq:os-weight},
\begin{equation}\label{eq:os-read}
\E[\varpi(\iota)\one_{r(X)\ne B}]\le m\Omega_\ell\gamma.
\end{equation}

\paragraph{The time given the label.} Let $b\ne b^0$. By~\eqref{eq:os-label}, the density of $\iota$ given $B=b$ under the measure $\varpi(\iota)\,d\Pr$ is proportional to $\eu^{-t/\ell}(1-\eu^{-t/m})^{w-1}(1+\eu^{-t/m})^{m-w}$; call it $\psi_b$, with distribution function $\Psi_b$. The substitution $n=m-1$, $\bar k=w-1$, $z=nt/m$, $\ell'=n\ell/m$ reduces it to Lemma~\ref{lem:os-kernel}, and gives $\Psi_b(1-\Psi_b)/\psi_b\le200(1+\alpha)\min\{t,m\}$. Let $Z_b=\E[\varpi(\iota)\one_{B=b}]$. With~\eqref{eq:os-deriv1}, \eqref{eq:os-L1}, $\varpi(t)\min\{t,m\}/m\le2$ and~\eqref{eq:os-strong},
\begin{equation}\label{eq:os-time}
\begin{aligned}
\sum\nolimits_{b\ne b^0}Z_b\sum\nolimits_x\mathrm{Var}_{\psi_b}g_{b,x}&\le\frac{800(1+\alpha)}m\E\Big[\varpi(\iota)\min\{\iota,m\}\sum\nolimits_jp^*_j(\iota\mid B)\Big]\\
&\le1600(1+\alpha)m(6\ell+4)\gamma.
\end{aligned}
\end{equation}

Put $\bar g_{b,x}=\E_{\psi_b}g_{b,x}$. The law of total variance gives $1-\sum_x\bar g_{b,x}^2=\E_{\psi_b}v_b+\sum_x\mathrm{Var}_{\psi_b}g_{b,x}$. For $b\ne b^0$ with $S_b=r^{-1}(b)$ nonempty, let $f(b)\in S_b$ have the largest $\bar g_{b,x}$. The largest $\bar g_{b,x}$ over all $x$ is at least $\sum_x\bar g_{b,x}^2$ and exceeds $\bar g_{b,f(b)}$ by at most the mass of $\bar g_b$ outside $S_b$. So $1-\bar g_{b,f(b)}\le1-\sum_x\bar g_{b,x}^2+\sum_{x\notin S_b}\bar g_{b,x}$, also when $S_b$ is empty, with the convention $\bar g_{b,f(b)}=0$.

\subsection{The initial label}\label{app:os-initial}

Put $f(b^0)=o$. Let $v_0(t)=\ell^{-1}\eu^{-t/\ell}\Pr(B=b^0\mid t)$ and $\varpi_0=\varpi v_0$, which is not integrable at $0$. Given $B=b^0$ and $t$, the probability of no edits tends to $1$ as $t\downarrow0$, so $g_{b^0,o}(t)\to1$. Let $e_0=\int_0^\infty\varpi_0(t)[1-g_{b^0,o}(t)]\,dt$. The bound of Appendix~\ref{app:os-birth} on the probability that the end state changes gives $1-g_{b^0,o}(t)\le(2/m)\int_0^t\sum_jp^*_j(s\mid b^0)\,ds$. So, by an exchange of integrals,
\[
e_0\le\frac2m\int_0^\infty\sum\nolimits_jp^*_j(s\mid b^0)\int_s^\infty\varpi_0(t)\,dt\,ds.
\]
For $t\le m$, the logarithmic derivative of $v_0$ is at most $-1/(\eu^{t/m}+1)\le-1/4$, and beyond $m$ at most $-1/\ell$. With $\varpi(t)\le1+m/t$ and $\varpi(t)\le2$ for $t\ge m$, for $s\le m$,
\[
\frac1{v_0(s)}\int_s^\infty\varpi_0(t)\,dt\le4+m[1+\log(1+4/s)]+2\ell\le m[6+2\alpha+\log^+(1/s)],
\]
using $\int_0^\infty\eu^{-z}/(z+s/4)\,dz\le\log(1+4/s)+1$, $4\le2m$ and $\log5<2$. For $s\ge m$ the left side is at most $2\ell$.

Also $\E[\log^+(1/\iota)\,p^*_j(\iota)]\le18\gamma$. For the path term, $\int_s^1\eu^{-t/\ell}\log(1/t)\,dt\le1\le2\eu^{-s/\ell}$ for $0<s<1$, so its expectation is at most $2\mu\bar I^*_j\le12\gamma$. For the end state, given $k$ edits the density of $\iota$ is proportional to $\eu^{-\lambda't}t^k$ with $\lambda'=1+1/\ell\le3/2$, and $\iota$ is independent of the word. So $\E[\log^+(1/\iota)\mid L=k]\le\lambda'^{\,k+1}/(k!(k+1)^2)\le3/2$, the bound being largest at $k=0$, and this term is at most $(3/2)\mu I^\circ_j\le6\gamma$. With~\eqref{eq:os-strong},
\begin{equation}\label{eq:os-zero}
e_0\le(12+4\alpha)m(6\ell+4)\gamma+36m\gamma.
\end{equation}

\subsection{One cube}\label{app:os-cube}

The states $f(b)$ have distinct labels. Place them in one cube, each at its label, fill the other positions of this cube with other states, which exist as $2^m$ divides $|S|$, and split the remaining states into cubes in any way. This defines a composable interface $\tilde I$ on $S$ that keeps the readers at every state $f(b)$. Let $\mathcal G=\{(f(b),b):S_b\ne\varnothing\}$ and $e_*=\E[\varpi(\iota)\one_{(X,B)\notin\mathcal G}]$. By the last paragraph of Appendix~\ref{app:os-weight} and~\eqref{eq:os-local}, \eqref{eq:os-time}, \eqref{eq:os-read} and~\eqref{eq:os-zero},
\begin{equation}\label{eq:os-graph}
e_*\le m\{4\ell+[1612+1604\alpha](6\ell+4)+36+\Omega_\ell\}\gamma.
\end{equation}

Fix a setter $T_{j,a}$. Given $\iota=t$, keep the histories of the other variables and resample the history of $j$ conditioned on its last value being $a$. The new history has density at most $2\varpi(t)$ with respect to the old one, by~\eqref{eq:os-label}. If both words satisfy the stronger condition of Appendix~\ref{app:onestatehist}, then the new end state is $Y=W_{-j}T_{j,a}o=T_{j,a}X$. If moreover $(X,B)$ and $(Y,\upd{B}{j}{a})$ lie in $\mathcal G$, then $T_{j,a}X=f(\upd{B}{j}{a})=\tilde T_{j,a}X$. By~\eqref{eq:os-weight} and~\eqref{eq:os-last}, $\E[\varpi(\iota)p_j(\iota)]\le2\Omega_\ell\gamma$. So each setter disagrees with probability at most $e_*+2e_*+2\Omega_\ell\gamma+4\Omega_\ell\gamma$, and each reader with probability at most $e_*$. With~\eqref{eq:os-graph}, $6\ell+4\le8\ell$, $108\le54\ell$, $q\ge2/3$ and $3m+6\le6m$,
\begin{equation}\label{eq:os-param}
\begin{aligned}
\Dist_\mu&\le m\{12\ell+[4836+4812\alpha](6\ell+4)+108\}\gamma+(3m+6)\Omega_\ell\gamma\\
&\le m^2[38760\alpha+38496\alpha^2+9\log(1+\ell)]\gamma.
\end{aligned}
\end{equation}

\subsection{Short means}\label{app:os-short}

Let $0<\ell<4$, and keep the other notation. For an integer $0\le w\le m$, let $\psi^\circ_w$ be the density on $(0,\infty)$ proportional to $\eu^{-t/\ell}(1-\eu^{-t/m})^w(1+\eu^{-t/m})^{m-w}$, the density of $\iota$ given $|B\oplus b^0|=w$, with mean $\bar t_w$. Define $\vartheta_w$, the Stein kernel of $\psi^\circ_w$ \citep{saumard2019weighted}, by $\vartheta_w(t)\psi^\circ_w(t)=\int_0^t(\bar t_w-z)\psi^\circ_w(z)\,dz$.

\begin{lemma}[Short means]\label{lem:os-short}
For $0<\ell<4$, $\vartheta_w(t)\le\xi_\ell\,t$ for all $t>0$, with $\xi_\ell=\ell/(1-\ell/4)$.
\end{lemma}

\begin{proof}
Fix $w$ and drop it from the notation, and write $\xi=\xi_\ell$. \emph{The mean.} Write $\psi^\circ(t)$ as a positive constant times $t^w\eu^{-t/\ell}g(t)$ with $g(t)=((1-\eu^{-t/m})/t)^w(1+\eu^{-t/m})^{m-w}$. Both factors of $g$ do not increase: the derivative of $(1-\eu^{-t/m})/t$ has the numerator $(1+t/m)\eu^{-t/m}-1\le0$. Let $\tau,\tau'$ be independent Gamma variables with shape $w+1$ and scale $\ell$, of mean $(w+1)\ell$. Then $2\,\mathrm{Cov}(\tau,g(\tau))=\E[(\tau-\tau')(g(\tau)-g(\tau'))]\le0$, and $g$ is bounded and positive, so $\bar t=\E[\tau g(\tau)]/\E g(\tau)\le(w+1)\ell$.

\emph{A sign change.} Let $\Psi^\circ(t)=\xi t\,\psi^\circ(t)-\int_0^t(\bar t-z)\psi^\circ(z)\,dz$ and $\beta_0(t)=\xi+t-\bar t+\xi t\mathcal S(t)$, with $\mathcal S$ the logarithmic derivative of $\psi^\circ$. Then $(\Psi^\circ)'=\beta_0\psi^\circ$. Let $\varphi_0(y)=y/(\eu^y-1)$ and $\varphi_1(y)=y/(\eu^y+1)$, so that $t\mathcal S(t)=-t/\ell+w\varphi_0(t/m)-(m-w)\varphi_1(t/m)$. The function $\varphi_0$ does not increase, as the numerator $\eu^y(1-y)-1$ of its derivative vanishes at $0$ and has derivative $-y\eu^y\le0$. Also $-\varphi_1'(y)\le1/4$ for $y>0$: this is equivalent to $4y\le\eu^y+6+5\eu^{-y}$, and $\eu^y+6-4y\ge7-3y+y^2/2=(y-3)^2/2+5/2>0$. Hence
\[
\beta_0'(t)=1-\frac\xi\ell+\frac\xi m\big[w\,\varphi_0'(t/m)-(m-w)\varphi_1'(t/m)\big]\le1-\frac\xi\ell+\frac\xi4=0 .
\]
As $t\mathcal S(t)\to w$ for $t\downarrow0$, $\beta_0(0+)=\xi(w+1)-\bar t\ge0$, by the bound on the mean and $\xi>\ell$. As $\mathcal S(t)\to-1/\ell$ and $t(\mathcal S(t)+1/\ell)\to0$ for $t\to\infty$, $\beta_0(t)=\xi-\bar t+(1-\xi/\ell)t+o(1)\to-\infty$. So $\beta_0$ is first nonnegative and then nonpositive.

\emph{Conclusion.} As $t\psi^\circ(t)\to0$ at $0$ and at $\infty$, and $\int_0^\infty(\bar t-z)\psi^\circ(z)\,dz=0$, $\Psi^\circ(0+)=\Psi^\circ(\infty)=0$. As $\Psi^\circ$ first increases and then decreases, $\Psi^\circ\ge0$, which is $\vartheta(t)\le\xi t$.
\end{proof}

\begin{lemma}[Bound for short means]\label{lem:os-shortbound}
For $0<\ell\le2$ and $m\ge2$, if all law instances hold at $o$, then $\Dist_\mu\le296m^2\gamma$.
\end{lemma}

\begin{proof}
Write $\psi=\psi^\circ_w$, $\vartheta=\vartheta_w$ and $\xi=\xi_\ell$, with $\xi\le4$ as $\ell\le2$. For every bounded smooth function $\varphi$ on $(0,\infty)$,
\begin{equation}\label{eq:os-poincare}
\begin{aligned}
\mathrm{Var}_\psi\,\varphi&=\int_{y<y'}(\varphi(y')-\varphi(y))^2\psi(y)\psi(y')\,dy\,dy'\\
&\le\int_{y<y'}(y'-y)\int_y^{y'}\varphi'(t)^2\,dt\,\psi(y)\psi(y')\,dy\,dy'=\int_0^\infty\varphi'(t)^2\vartheta(t)\psi(t)\,dt,
\end{aligned}
\end{equation}
by the Cauchy--Schwarz inequality and an exchange of integrals, with $\bar t$ the mean of $\psi$ and
\[
\int_0^t\int_t^\infty(y'-y)\psi(y)\psi(y')\,dy'\,dy=\int_0^t(\bar t-y)\psi(y)\,dy.
\] This is the weighted Poincar\'e inequality \citep{klaassen1985inequality,saumard2019weighted}. By Lemma~\ref{lem:os-short}, the right side is at most $\xi\int_0^\infty t\varphi'(t)^2\psi(t)\,dt$.

Fix $b$ and $t$, and put $y=t/m$. Let $L_i$ be the number of edits of variable $i$. If $b_i\ne b^0_i$, then $L_i$ is Poisson with mean $y$, conditioned to be positive, and its variance is $y/(1-\eu^{-y})-y^2\eu^{-y}/(1-\eu^{-y})^2\le y$. If $b_i=b^0_i$, its variance is $y/(1+\eu^{-y})+y^2\eu^{-y}/(1+\eu^{-y})^2\le y+y^2\eu^{-y}\le2y$. These numbers are conditionally independent, so $L_t=\sum_iL_i$ has $\mathrm{Var}(L_t\mid b,t)\le2t$. Given $\iota=t$, $L_t$ is Poisson with mean $t$, and given the numbers of edits and $B$, the end state does not depend on $t$. So $\Pr(X=x,B=b\mid\iota=t)=\eu^{-t}\sum_{n\ge0}\chi_{b,x,n}t^n/n!$ with numbers $\chi_{b,x,n}$ that do not depend on $t$, namely the probabilities of $X=x$ and $B=b$ given $n$ edits. As $t\,(t^n/n!)'=n\,t^n/n!$, the quotient rule for $g_{b,x}$ gives $g'_{b,x}(t)=t^{-1}\mathrm{Cov}(\one_{X=x},L_t\mid b,t)$. By the Cauchy--Schwarz inequality,
\[
\sum\nolimits_xg'_{b,x}(t)^2\le\frac{\mathrm{Var}(L_t\mid b,t)}{t^2}\sum\nolimits_xg_{b,x}(t)(1-g_{b,x}(t))\le\frac2t\,v_b(t).
\]
Let $p_{b,x}=\Pr(X=x\mid B=b)$. By the law of total variance, $1-\sum_xp_{b,x}^2=\E[v_b(\iota)\mid B=b]+\sum_x\mathrm{Var}(g_{b,x}(\iota)\mid B=b)\le(1+2\xi)\E[v_b(\iota)\mid B=b]$. Averaging over $b$, $V\le2\sum_j\bar p_j$ and $\E\bar p_j(\iota)\le2\ell\gamma$ give
\[
\sum\nolimits_b\Pr(B=b)\Big(1-\sum\nolimits_xp_{b,x}^2\Big)\le4m\ell(1+2\xi)\gamma.
\]
For each label $b$ with nonempty $S_b$, choose $f(b)\in S_b$ with the largest $p_{b,x}$, and let $\pi_0$ be the joint distribution of $(X,B)$ and $\mathcal G=\{(f(b),b):S_b\ne\varnothing\}$. The inequality at the end of Appendix~\ref{app:os-weight}, with $p_{b,x}$ in place of $\bar g_{b,x}$, gives $1-p_{b,f(b)}\le1-\sum_xp_{b,x}^2+\sum_{x\notin S_b}p_{b,x}$, also when $S_b$ is empty, with the convention $p_{b,f(b)}=0$. Average this over $b$. The bound above takes care of the first two terms, and the last term averages to $\Pr(r(X)\ne B)\le m\ell\gamma$. So $\pi_0(\mathcal G^c)\le m\ell(5+8\xi)\gamma$. Place the states $f(b)$ in one cube as in Appendix~\ref{app:os-cube}. Let $\mathcal T_{i,a}(x,b)=(T_{i,a}x,\upd{b}{i}{a})$. The geometric history gives $\pi_0=(1-q)\delta_{(o,b^0)}+(q/(2m))\sum_{i,a}\pi_0\mathcal T_{i,a}$, so $\pi_0\mathcal T_{i,a}\le(2m/q)\pi_0$. If $(X,B)\in\mathcal G$ and $\mathcal T_{i,a}(X,B)\in\mathcal G$, then $T_{i,a}X=\tilde T_{i,a}X$. So each setter disagrees with probability at most $(1+2m/q)\pi_0(\mathcal G^c)$. With $\ell/q=1+\ell$, $\Dist_\mu\le m[2m+(2m+1)\ell](5+8\xi)\gamma$. As $\ell\le2$ and $\xi\le4$, this is at most $8m^2\cdot37\gamma=296m^2\gamma$.
\end{proof}

\subsection{Proof of Theorem~\ref{thm:onestate}(a)}\label{app:os-assemble}

The proof treats four cases: means $\ell\ge2m$ by~\eqref{eq:repairalpha}, means $\ell<2m$ with a law instance failing at $o$ by Lemma~\ref{lem:resolvent}(iii), and, with all law instances holding at $o$, means $\ell\le2$ by Lemma~\ref{lem:os-shortbound} and means $2\le\ell\le2m$ by~\eqref{eq:os-param}.

Let $K=m^2[231504+9\log(1+2m)]$.

\begin{proof}[Proof of Theorem~\ref{thm:onestate}(a)]
As $m\ge2$, the intervals $(0,2]$, $[2,2m]$ and $[2m,\infty)$ cover every mean $\ell>0$. We show $\Dist_\ell\le K\gamma_\ell$ in four cases. With $\Dist_\ell\le1$, this gives $\Dist_\ell\le\min\{1,K\gamma_\ell\}$.

\emph{Means $\ell\ge2m$.} By Lemma~\ref{lem:resolvent}(i), $\mu_\ell$ is $(1+1/\ell)$-subinvariant, hence $(1+1/(2m))$-subinvariant, whether or not the laws hold at $o$. So~\eqref{eq:repairalpha} with $\alpha=2$ gives $\bar\lambda\le(2\eu m)^{1/2}$ and $\Dist_\ell\le8m\bar\lambda^2\gamma_\ell\le16\eu m^2\gamma_\ell$. As $16\eu<48$, this coefficient is at most $K$.

\emph{Means $\ell<2m$ with a law instance failing at $o$.} Then $o\in\cN$, and Lemma~\ref{lem:resolvent}(iii) gives $\gamma_\ell\ge\mu_\ell(\{o\})\ge1/(1+\ell)$. So $\Dist_\ell\le1\le(1+\ell)\gamma_\ell\le(1+2m)\gamma_\ell$, and $1+2m\le2m^2\le K$ as $m\ge2$.

\emph{Means $\ell\le2$ with all law instances holding at $o$.} Lemma~\ref{lem:os-shortbound} gives $\Dist_\ell\le296m^2\gamma_\ell$, and $296m^2\le K$.

\emph{Means $2\le\ell\le2m$ with all law instances holding at $o$.} These are the assumptions of Appendices~\ref{app:onestatehist}--\ref{app:os-cube}, with $\alpha=\ell/m\le2$ and $\log(1+\ell)\le\log(1+2m)$. So~\eqref{eq:os-param} gives
\[
\Dist_\ell\le m^2[2\cdot38760+4\cdot38496+9\log(1+2m)]\gamma_\ell=K\gamma_\ell .
\]

For $\Dist^\infty$ on every finite $S$, apply this to the interface of Lemma~\ref{lem:padding}, which keeps the start state, the law instances at it and $\gamma_\ell$. The tester follows from Appendix~\ref{app:lawcheck}.
\end{proof}

\subsection{Proof of Theorem~\ref{thm:onestate}(b)}\label{app:fixed}

\begin{proof}[Proof of Theorem~\ref{thm:onestate}(b)]
\emph{A core.} Let $S^\circ$ consist of eight distinct states $x,w,y,v',v,\bar x,\bar y,\bar w$. For $m=2$, the table in Figure~\ref{fig:core} gives the readers and setters at $x$, $w$, $y$, $v'$ and $v$. The states $\bar x,\bar y,\bar w$ read $(0,0)$, and every setter fixes them. Let $\nu^\circ=(\delta_x+\delta_w)/2$, and call $x$ and $w$ its \emph{atoms}.

\emph{Four entries that no composable interface keeps.} We claim that no composable interface $\tilde I$, on $S^\circ$ or on a finite set containing $S^\circ$, has $\tilde T_{1,1}x=y$, $\tilde T_{2,1}x=v'$, $\tilde T_{2,0}w=y$ and $\tilde T_{1,0}w=v$. In a composable interface, every setter keeps the cube of its argument. So the first and third entries put $x$, $y$ and $w$ in one cube, and the other two entries put $v'$ and $v$ in it as well. By the first and third entries, the label of $y$ in $\tilde I$ starts with $(1,0)$. As $x\ne y$, the first entry changes $x$, so the label of $x$ starts with $(0,0)$. As $w\ne y$, the label of $w$ starts with $(1,1)$ in the same way. So $v'=\tilde T_{2,1}x$ and $v=\tilde T_{1,0}w$ both have labels that start with $(0,1)$. The four setters keep all other coordinates, so the labels of $v'$ and $v$ agree, and $v'=v$, a contradiction. Hence every composable interface, on $S^\circ$ or on a larger set, disagrees with $I$ at $x$ or at $w$, and $\Dist^\infty_{\nu^\circ}\ge1/2$.

\emph{The laws hold at $x$ and $w$.} Success, write-back and preservation at $x$ and $w$ follow from the table. Overwrite involves two writes to one variable. At $x$, these stay in $\{x,y\}$ for variable~$1$ and in $\{x,v'\}$ for variable~$2$, and at $w$, in $\{v,w\}$ and in $\{y,w\}$. On each of these pairs, the two setters of the variable act as an exact binary write, so overwrite holds. For commutation, both orders of $T_{1,a}$ and $T_{2,b}$ give the same state:
\[
\begin{array}{c|cc}
x&b=0&b=1\\\hline
a=0&x&v'\\
a=1&y&w
\end{array}
\qquad
\begin{array}{c|cc}
w&b=0&b=1\\\hline
a=0&x&v\\
a=1&y&w
\end{array}.
\]
So all $4m^2+3m=22$ law instances hold at $x$ and $w$, and $\gamma_{\nu^\circ}=0$.

\emph{A composable interface at distance $1/2$.} Give $x,y,v',w$ the labels $00,10,01,11$ in one cube, and $\bar x,\bar y,v,\bar w$ the same labels in a second cube. This composable interface keeps the readers at $x$ and $w$. At $x$ and $w$, its setters differ from those of $I$ only in $T_{1,0}w=v'\ne v$. So $\Dist_{\nu^\circ}=1/2$.

\emph{More variables.} For $m\ge2$, let $S^\circ_m=S^\circ\times\{0,1\}^{m-2}$. The first two variables act on the first component as above, the other variables are coordinate readers and setters of the second component, and labels concatenate. Let $\nu^\circ=(\delta_{(x,0^{m-2})}+\delta_{(w,0^{m-2})})/2$. The laws that involve only the first two variables hold as for $m=2$, and those that involve only the others hold for coordinate readers and setters. Setters of the first two variables and of the others act on different components, so they commute and keep each other's readings. Hence $\gamma_{\nu^\circ}=0$. The four entries at the atoms keep the second component, so the argument above gives $\Dist^\infty_{\nu^\circ}\ge1/2$. The product of the composable interface above with the interface that reads and replaces the coordinates of $\{0,1\}^{m-2}$ changes only $T_{1,0}$ at $(w,0^{m-2})$ on the support of $\nu^\circ$, so $\Dist_{\nu^\circ}=1/2$. The core $S^\circ_m$ has $2^{m+1}$ states.

\emph{One start state.} Let $k\ge1$, let $o_0=(x,0^{m-2})$ and $o_1=(w,0^{m-2})$ be the atoms of $\nu^\circ$, and add $k$ extra states $o'_0,\dots,o'_{k-1}$ with arbitrary readings. Every setter maps $o'_j$ to $o'_{j+1}$ for $j<k-1$, and $T_{i,a}$ maps $o'_{k-1}$ to $o_a$. With $\nu$ the point mass at $o'_0$, the last of $k$ random edits reaches $o_a$ when it writes $a$, so $\nu P^k=\nu^\circ$. Add $(-k)\bmod2^m$ further states, with arbitrary readings and fixed by every setter, so that $2^m$ divides $|S|$. As $S^\circ_m$ is closed under the setters, the law instances at the atoms involve only states of $S^\circ_m$, and $\gamma_{\nu P^k}=\gamma_{\nu^\circ}=0$. The lower bound concerns only the entries at the atoms, so it holds as before. The added states form full cubes, so the composable interface above extends to $S$ with disagreement $1/2$. Thus $\Dist^\infty_{\nu P^k}=\Dist_{\nu P^k}=1/2$.
\end{proof}
\section{Fixed thresholds: reduction to one quantity}\label{app:modulus}

For an integer $d\ge1$, a state $x$ is \emph{locally exact to depth $d$} if every word of length at most $d$ that starts at $x$ ends at a state whose label is the one predicted by coordinate writes from $r(x)$, and two such words end at the same state exactly when their predicted labels agree. Let $A_d$ be the set of these states, and let
\[
a_d=\sup\{\Dist^\infty_\nu(I):I\text{ a finite interface of any number of variables, }\nu(A_d)=1\}.
\]
So $a_d\in[0,1]$ does not increase with $d$. The \emph{local exactness check} of depth $d$ reads the end label of every word of length at most $d$ from a sample, checks it against the predicted label, and checks for every two of these words that their ends are equal exactly when their predicted labels agree. It rejects with probability $\mu(S\setminus A_d)$.

\begin{theorem}[Reduction]\label{thm:modulus}
For every interface and $\mu$, $\Dist^\infty_\mu\le\mu(S\setminus A_d)+\mu(A_d)a_d$. The following are equivalent: (1)~$a_d\to0$; (2)~for every fixed $\ell>0$, every sequence of interfaces and start distributions with $m\to\infty$ and $\gamma_\ell=m^{-\omega(1)}$ has $\Dist^\infty_\ell\to0$; (3)~the same for one fixed $\ell>0$; (4)~for all fixed $\ell>0$ and $0<\eta,\delta<1$, law checks at samples of $\mu_\ell$ certify $\Dist^\infty_{\mu_\ell}<\eta$ from a number of samples polynomial in $m$, with degree and coefficient depending on $\ell,\eta,\delta$ only.
\end{theorem}

\begin{lemma}[Passive embedding]\label{lem:embedding}
Let $I$ be an interface with $n_1$ variables on $S$, and let $\nu(A_d)=1$ with $d\ge2$. For all $m\ge n_1$ and $\ell>0$ there is an interface $\bar I$ with $m$ variables and a start distribution with $\Dist^\infty_\ell(\bar I)\ge\Dist^\infty_\nu(I)/(1+\ell)$ and $\gamma_\ell(\bar I)\le(\ell n_1/(m+\ell n_1))^{d-1}$.
\end{lemma}

\begin{proof}
Let $\bar S=S\times\{0,1\}^{m-n_1}$, $\bar r(s,v)=(r(s),v)$ and $\bar\nu=\nu\otimes\delta_{0^{m-n_1}}$. The first $n_1$ setters act on $s$, the others replace coordinates of $v$. Then $\bar\mu_\ell\ge\bar\nu/(1+\ell)$. A composable interface with $m$ variables, with the last $m-n_1$ readers and setters forgotten, is a composable interface with $n_1$ variables: each $m$-dimensional cube splits into $2^{m-n_1}$ cubes by the last coordinates. Its set of states contains the slice $S\times\{0^{m-n_1}\}$, so it is a candidate for $\Dist^\infty_\nu(I)$. This gives the distance bound. The number $L_a$ of edits of the first $n_1$ variables is geometric with mean $\ell n_1/m$, as in Lemma~\ref{lem:rectmeasure}. If $L_a\le d-2$, both sides of each law instance on these variables are words of length at most $L_a+2\le d$ from the start state, which lies in $A_d$, so the law holds at the end of the run. Passive coordinates are exact and commute with all writes. So $\gamma_\ell\le\Pr(L_a\ge d-1)$.
\end{proof}

\begin{proof}[Proof of Theorem~\ref{thm:modulus}]
\emph{The inequality.} If $\mu(A_d)=0$, the right side is $1$ and each disagreement probability is at most $1$. Otherwise, let $\nu$ be $\mu$ conditioned on $A_d$. For every composable interface and every disagreement set $X$, $\mu(X)\le\mu(S\setminus A_d)+\mu(A_d)\nu(X)$. Taking the largest $X$ and the infimum over candidates gives the inequality. As $A_{d+1}\subseteq A_d$, $a_d$ does not increase.

\emph{From laws to local exactness.} If no state within $d-1$ writes of $x$ lies in $\cN$, then $x\in A_d$. Indeed, along a word of length at most $d$, success and preservation give the predicted labels. Commutation sorts the writes by variable, overwrite deletes all but the last write of each variable, and write-back at $x$ deletes a write of the current value of $x$. Each commutation or overwrite applies to a state reached by at most $d-2$ writes, and write-back applies at $x$ after commutation moves the write to the front. So two words with the same predicted label reduce to the same word, and words with different predicted labels end at different labels. If $\mu$ is $\lambda$-subinvariant, $\mu(S\setminus A_d)\le\gamma_\mu\sum_{k<d}(2m\lambda)^k$, as the push-forwards of $\mu$ under all $(2m)^k$ words of length $k$ sum to $(2m)^k\mu P^k\le(2m\lambda)^k\mu$.

\emph{(1) implies (2).} Fix $d$ and $\ell$. As $\mu_\ell$ is $(1+1/\ell)$-subinvariant, $\gamma_\ell=m^{-\omega(1)}$ gives $\mu_\ell(S\setminus A_d)\to0$, so $\limsup\Dist^\infty_\ell\le a_d$. Let $d\to\infty$. Choosing a fixed $\ell>0$ in (2) gives (3).

\emph{(3) implies (1).} Suppose $a_\infty=\lim_da_d>0$ and put $\eta=a_\infty/2$. For each $j\ge2$, choose an interface $I_j$ with $n_j$ variables and $\nu_j$ with $\nu_j(A_j)=1$ and $\Dist^\infty_{\nu_j}(I_j)\ge\eta$. Fix $\ell>0$ and choose increasing integers $m_j\ge\max\{j,n_j^2,(\ell n_j)^2,4\}$. For $m\ge m_2$, let $j(m)=\max\{j:m_j\le m\}$, which tends to infinity, and embed $I_{j(m)}$ by Lemma~\ref{lem:embedding}. Then $\Dist^\infty_\ell\ge\eta/(1+\ell)$ and $\gamma_\ell\le(\ell n_{j(m)}/m)^{j(m)-1}\le m^{-(j(m)-1)/2}=m^{-\omega(1)}$, against (3).

\emph{(1) implies (4).} Fix $\ell,\eta,\delta$, choose $d$ with $a_d\le\eta/2$ and put $\lambda=1+1/\ell$. If $\Dist^\infty_{\mu_\ell}\ge\eta$, the inequality gives $\mu_\ell(S\setminus A_d)\ge\eta/2$, so $\gamma_\ell\ge\eta/(2\sum_{k<d}(2m\lambda)^k)$. So $\lceil(2/\eta)\sum_{k<d}(2m\lambda)^k\log(1/\delta)\rceil$ samples suffice.

\emph{(4) implies (2).} If (2) fails, there are a fixed $\ell>0$, a fixed $\eta>0$ and a sequence with $\Dist^\infty_\ell\ge\eta$ and $\gamma_\ell=m^{-\omega(1)}$. A law-check tester with polynomially many samples rejects these with probability $o(1)$.
\end{proof}
\section{Reads, measures and small supports}\label{app:reads}

This appendix proves Propositions~\ref{prop:reads}--\ref{prop:support}.

\subsection{Reads along random writes}\label{app:readrisk}

Let $\ell>0$ and let $k\ge1$ be an integer. Draw $X_0\sim\mu_\ell$ and independent uniformly random setters $(i_1,a_1),\dots,(i_k,a_k)$, independent of $X_0$. Put $X_s=T_{i_s,a_s}X_{s-1}$, $B_0=r(X_0)$ and $B_s=\upd{B_{s-1}}{i_s}{a_s}$, the label that a record predicts.

\begin{proposition}[Reads]\label{prop:reads}
For every interface and start distribution, the probability that $r(X_s)\ne B_s$ for some $s\le k$ is at most $\ell(q^{-k}-1)\gamma_\ell$, which is at most $\eu k\gamma_\ell$ for $k\le\ell$.
\end{proposition}

\begin{proof}
Take the first $s$ with $r(X_s)\ne B_s$. As $r(X_{s-1})=B_{s-1}$, the setter does not update the label as written at $X_{s-1}$, so success or preservation fails there, and $X_{s-1}\in\cN$. Here $X_{s-1}$ has distribution $\mu_\ell P^{s-1}\le q^{-(s-1)}\mu_\ell$ by Lemma~\ref{lem:resolvent}(ii). Summing over $s$ with $\sum_{s<k}q^{-s}=\ell(q^{-k}-1)$ gives the bound. The last estimate of Corollary~\ref{cor:workload} gives $\ell(q^{-k}-1)\le\eu k$ for $k\le\ell$.
\end{proof}

\noindent Each step checks a whole label at once, so no union over the $m$ variables appears.

\subsection{Correct reads at constant distance}\label{app:eight}

For $m\ge2$, let $S=\{0,1\}^m\times\{0,1\}$ and $r(u,z)=u$, and let
\begin{equation}\label{eq:eight}
T_{1,v}(u,z)=(\upd{u}{1}{v},\,z\oplus(v\oplus u_1)u_2),\qquad T_{i,v}(u,z)=(\upd{u}{i}{v},\,z)\quad(i\ne1).
\end{equation}
For $m=2$ this is the interface of Figure~\ref{fig:eight}.

\begin{proposition}[Eight states]\label{prop:eight}
Let $\mu$ be the uniform measure on $S$. Success, preservation, write-back and overwrite hold at every state, so after any sequence of writes the label is the one that coordinate writes predict. Commutation fails at every state. Also $\mu P=\mu$ and $\Dist^\infty_\mu\ge1/8$.
\end{proposition}

\begin{proof}
Every setter updates the label as written, which gives success and preservation. For $v=u_1$ the change of $z$ in~\eqref{eq:eight} vanishes, which gives write-back. Writing $w$ and then $v$ to variable $1$ changes $z$ by $(w\oplus u_1)u_2\oplus(v\oplus w)u_2=(v\oplus u_1)u_2$, the change of writing $v$ alone, which gives overwrite for this variable; for the other variables it is direct. At $(u,z)$, the two orders of $T_{1,1-u_1}$ and $T_{2,1-u_2}$ give the same label, and their values of $z$ differ by $(1\oplus u_1\oplus u_1)(u_2\oplus1\oplus u_2)=1$.

Each flip is an involution, as $F_1(u,z)=(u\oplus\mathbf e_1,z\oplus u_2)$ and the other flips change one coordinate of $u$. So each flip preserves $\mu$. By write-back, the two setters of variable $i$ average to $(\Id+F_i)/2$, so $\mu P=\mu$.

From $x=(u,z)$, apply $F_1$, $F_2$, $F_1$, $F_2$. Each of the two coordinates is first written with the opposite value and then with its initial value. The label returns to $u$, and $z$ changes by $u_2\oplus(1-u_2)=1$. Let $\tilde I$ be a composable interface on a finite set $S'\supseteq S$. By Lemma~\ref{lem:forcing}, it disagrees with one of the four calls. Draw $x\sim\mu$. As the flips are involutions, the argument of each call is uniform. A call of variable $i$ uses one of its two setters, so it disagrees with probability at most the sum of their disagreements, at most $2R_\mu(I,\tilde I)$. So $1\le8R_\mu(I,\tilde I)$.
\end{proof}

\subsection{An audit at mean zero certifies the start distribution}\label{app:twostate}

Let $m=1$, $S=\{x,y\}$, $\nu=\delta_x$, $r(x)=0$ and $r(y)=1$, and let $T_0x=x$, $T_1x=y$, $T_0y=x$ and $T_1y=x$.

\begin{proposition}[Two states]\label{prop:twostate}
Every audit round with target label $\hat a=0$ and depth $m+1=2$ passes, with or without its bridge check, and $\Dist_\nu=\Dist^\infty_\nu=0$. After one random write, $\Dist_{\nu P}=\Dist^\infty_{\nu P}=1/2$.
\end{proposition}

\begin{proof}
\emph{The round.} The path check has $T_0x=x$, with the correct label, and the comparison with $x$ passes. Write-back at $x$ is $T_0x=x$. The flip is $y_1=T_1x=y$, of label $1$. Its path has $T_0y=x$, of label $0$, which equals $x_1=x$, and the undo $T_1x=y$ returns to $y$. For the bridge check, $u^{(1)}=a=0$, and $w_0$ is $x$ if $u=0$ and $y$ if $u=1$, with the correct label. In both cases $T_0w_0=x=w_1$.

\emph{The distances.} The record with labels $r(x)=0$ and $r(y)=1$, whose setters $T_0$ and $T_1$ always output $x$ and $y$, agrees with the interface at $x$, so both distances under $\nu$ are $0$. Now $\nu P=(\delta_x+\delta_y)/2$. This record differs only in $T_1y$, which gives the upper bound $1/2$. A candidate with disagreement below $1/2$ would keep all readers and setters at both atoms. As $T_1y=x$ and $r(x)=0$, it would violate success, on the same states or with added states.
\end{proof}

\subsection{Lengthening the runs}\label{app:lengthen}

\begin{proposition}[Longer runs]\label{prop:lengthen}
Let $m\ge2$, $\ell\ge0$ and $L\ge\ell$ with $L>0$. Then $\mu_\ell\le\frac{1+L}{1+\ell}\mu_L$, so $\Dist^\infty_{\mu_\ell}\le\frac{1+L}{1+\ell}\Dist^\infty_{\mu_L}$, and also $\Dist_{\mu_\ell}\le\frac{1+L}{1+\ell}\Dist_{\mu_L}$ when $2^m\mid|S|$. Let $L=\max\{\ell,m\}$. Then $\frac{1+L}{1+\ell}\le1+m$, and law checks at depth $2$ at $\lceil16\eu^2m^3\frac{1+L}{1+\ell}\log(1/\delta)/\epsilon\rceil$ samples of $\mu_L$ certify $\Dist_{\mu_\ell}<\epsilon$ when $2^m\mid|S|$, and $\Dist^\infty_{\mu_\ell}<\epsilon$ for every finite $S$.
\end{proposition}

\begin{proof}
Put $q_\ell=\ell/(1+\ell)$. As $q_\ell\le q_L$, comparing~\eqref{eq:access} term by term gives $(1-q_\ell)q_\ell^s\le((1-q_\ell)/(1-q_L))(1-q_L)q_L^s$, and $(1-q_\ell)/(1-q_L)=(1+L)/(1+\ell)$. The bound on the distances follows for each candidate and each disagreement set. For the tester, Corollary~\ref{cor:longruns} with $\alpha=1$ gives $\Dist_{\mu_L}\le16\eu^2m^3\gamma_{\mu_L}$ when $2^m\mid|S|$, and $\Dist^\infty_{\mu_L}\le16\eu^2m^3\gamma_{\mu_L}$ for every finite $S$. So a distance of at least $\epsilon$ under $\mu_\ell$ gives $\gamma_{\mu_L}\ge\epsilon(1+\ell)/(16\eu^2m^3(1+L))$, and law checks reject as in Appendix~\ref{app:lawcheck}. If $\ell\ge m$ then $L=\ell$, and otherwise $L=m$, which gives the bound on the ratio.
\end{proof}

\subsection{Small supports with added states}\label{app:support}

\begin{proposition}[Small support]\label{prop:support}
Let $m,N\ge1$, $0<\epsilon,\delta<1$, and let $\mu$ have support of size at most $N$. A tester of depth at most one with one-sided error, $\lceil(N/\epsilon)\log(N/\delta)\rceil$ samples and at most $2mN$ setter calls certifies $\Dist^\infty_\mu<\epsilon$ with confidence $1-\delta$. It needs only the bound $N$.
\end{proposition}

\begin{proof}
\emph{The tester.} Draw the samples and remove repetitions by comparison, which leaves a set $B_N$. For each $x\in B_N$, call all $T_{i,a}x$ and read the labels of their outputs. Reject if $T_{i,r_i(x)}x\ne x$ or $r(T_{i,a}x)\ne\upd{r(x)}{i}{a}$ for some $i,a$. Among all held states, join $x$ and $T_{i,a}x$ by an edge, and reject if a connected component has two distinct states of one label. The calls act only on samples, so the depth is at most one, and there are at most $2mN$ of them. A composable interface passes, as each setter keeps the cube and a cube has one state of each label.

\emph{A pass gives a composable interface.} Suppose the tester passes. Each component has distinct labels. Give each component its own cube, with its states at their labels, and fill the other positions with added states. Give each state of $S$ that is not held a cube of its own in the same way. This is a composable interface on a finite set containing $S$. It keeps the readers on $S$, and for $x\in B_N$ all setters at $x$, as $T_{i,a}x$ lies in the cube of $x$ with the updated label. So $\Dist^\infty_\mu\le\mu(S\setminus B_N)$.

\emph{Heavy atoms are sampled.} Let $A_N$ be the support of $\mu$ and $H_N=\{x\in A_N:\mu(x)\ge\epsilon/N\}$. With $s=\lceil(N/\epsilon)\log(N/\delta)\rceil$ samples, a state of $H_N$ is missed with probability at most $(1-\epsilon/N)^s\le\delta/N$, so $H_N\subseteq B_N$ except with probability at most $\delta$. On this event, $\mu(S\setminus B_N)\le\mu(A_N\setminus H_N)<\epsilon$. So an interface with $\Dist^\infty_\mu\ge\epsilon$ passes with probability at most $\delta$.
\end{proof}

\end{document}